\documentclass[nofootinbib,twocolumn,aps,superscriptaddress,prl,10pt]{revtex4-2}
\usepackage{amsmath}
\usepackage{amssymb}
\usepackage{amsfonts}
\usepackage{dsfont}
\usepackage{soul}
\usepackage{hyperref}
\usepackage{color}
\usepackage{braket}
\usepackage{comment}
\usepackage{titlesec}
\usepackage[utf8]{inputenc}

\newcommand{\bigsection}[1]{%
\begin{center}
{\huge #1}
\end{center}
}
\newcounter{appendixsection}

\usepackage{tikz,xcolor}
\definecolor{lime}{HTML}{A6CE39}
\DeclareRobustCommand{\orcidicon}{%
	\begin{tikzpicture}
	\draw[lime, fill=lime] (0,0) 
	circle [radius=0.16] 
	node[white] {{\fontfamily{qag}\selectfont \tiny ID}};
	\draw[white, fill=white] (-0.0625,0.095) 
	circle [radius=0.007];
	\end{tikzpicture}
	\hspace{-2mm}
}
\foreach \x in {A, ..., Z}{%
	\expandafter\xdef\csname orcid\x\endcsname{\noexpand\href{https://orcid.org/\csname orcidauthor\x\endcsname}{\noexpand\orcidicon}}
}

\begin{document}
\date{\today}
\title{Self gravity decoheres quantum systems}

\author{Alessio Lapponi\orcidA{}}
\affiliation{Scuola Superiore Meridionale, Largo San Marcellino 10, 80138 Napoli, Italy}
\affiliation{Istituto Nazionale di Fisica Nucleare (INFN), Sezione di Napoli,
Complesso Universitario di Monte S. Angelo, Via Cintia Edificio 6, 80126 Napoli, Italy}
\affiliation{Institute for Quantum Computing Analytics (PGI-12), Forschungszentrum J\"ulich, 52425 J\"ulich, Germany}
\author{Stefano Mancini\orcidB{}}
\email{stefano.mancini@unicam.it}
\affiliation{School of Science and Technology, University of Camerino, Via Madonna delle Carceri 9, 62032 Camerino, Italy.}
\affiliation{Istituto Nazionale di Fisica Nucleare (INFN), Sezione di Perugia, Via A.~Pascoli, 06123 Perugia, Italy.}
\author{Frank K. Wilhelm\orcidC{}}
\affiliation{Institute for Quantum Computing Analytics (PGI-12), Forschungszentrum J\"ulich, 52425 J\"ulich, Germany}
\affiliation{Theoretical Physics, Universit\"at des Saarlandes, 66123 Saarbr\"ucken, Germany}
\author{David Edward Bruschi\orcidZ{}}
\email{david.edward.bruschi@posteo.net}
\affiliation{Institute for Quantum Computing Analytics (PGI-12), Forschungszentrum J\"ulich, 52425 J\"ulich, Germany}
\affiliation{Theoretical Physics, Universit\"at des Saarlandes, 66123 Saarbr\"ucken, Germany}

\begin{abstract}
We study the effects of self gravity on the quantum state of a massive and static particle that initially contains quantum coherence between two positions.
We employ linearized quantum gravity to obtain the self-interacting dynamics of the particle mediated by gravitons, and find that the effective evolution of the particle's state can be viewed as a quantum channel composed of a unitary, dephasing, depolarizing, and erasure part. Depolarization drives the state towards maximal mixedness while depolarization and dephasing decrease its quantum coherence. Crucially, the intrinsic diffusion and dephasing timescales of the problem determine a  relation between the mass and size of the particle that naturally identifies the transition between its classical and quantum regime. Our work therefore provides an explanation for the observational difference between the quantum behavior of small and light systems and the classical behavior of larger and heavier ones.
\end{abstract}

\maketitle

%------------------------%
%\section{Introduction}
%------------------------%
A key open question at the overlap of relativistic and quantum physics is \emph{how do quantum systems gravitate}. 
While a definitive answer would most likely require unifying general relativity and quantum mechanics into a full theory of quantum gravity \cite{Kiefer2012_QuantumGravity3rd,Ashtekar2021_ShortReviewLQG,vanDongen2021_StringTheoryEinsteinIdentity,Bose:Fuentes:2025}, tackling this question already in the low energy and weak curvature regimes would have the advantage of potentially providing new insights while circumventing such difficult task. Extensive work has been dedicated in the past decades on many topics at the interplay of relativity and quantum mechanics, such as quantum state reduction and spontaneous collapse of the wave function due to gravity \cite{Karolyhazy:1966,Diosi:1984,Bassi:Lochan:13}, gravitational bound states \cite{Nesvizhevsky:Boerner:2002,Jenke:Cronenberg:2014}, stochastic gravity \cite{Brahim:Remy:2006,Hu:Verdaguer:08}, and advanced theories of (post and quantum) gravity \cite{Page:Geilker:1981,Rovelli:2008,Khoury:2013,Oppenheim:2023,Bianconi:2025,Pietsch:Petruzziello:2026}. Furthermore, experiments to test the quantum nature of gravity have also been recently proposed \cite{Marletto:Vedral:2017,Carney2019}. Nevertheless, to date there is no definitive proof that quantum features, such as coherence and entanglement, affect the gravitational field of physical systems.

Decoherence of quantum states is naturally expected in presence of external forces or in presence of interaction with an environment. Thus, it is logical to expect that \emph{gravitational decoherence} can provide a natural mechanism for spatial configurations of (large-enough) massive objects to exhibit a classical behavior \cite{Bassi:Lochan:13,Anastopoulos:Hu:2013,Grossardt2017}.  Multiple approaches have been proposed to tackle this question, predicting a wide spectrum of phenomenological implications and magnitudes of the effects. For example, semiclassical gravity has been employed to show that \emph{intrinsic decoherence}, i.e., decoherence induced by the quantum nature of the state itself, arises from the fluctuations of the gravitational field induced by the state to itself \cite{Anastopoulos:Hu:2013,Grossardt2017,Hu:Verdaguer:08,Toros:Mazumdar:2024}. This approach predicts a complete decoherence of particles initially in a superposition of states that occurs at a time proportional to the difference of the gravitational energy of the states. Remarkably, the result is identical to the one predicted by the Penrose-Di\'osi mechanism, which considers the collapse as a fundamental and stochastic phenomenon \cite{Diosi:89,Penrose:96,Bassi:Lochan:13,Nandi:Petruccione:2026}.
Quantum theories of gravity, instead, predict the existence of a \textit{environmental decoherence}, where the entropy of a quantum state increases due to interaction with the quantized gravitational field that acts as an environment. Differently from the semiclassical gravity induced decoherence, the collapse rate depends on the graviton spectrum constituting the initial bath, and there is no decoherence when no gravitons are present \cite{Blencowe:2013,Oniga:Wang:2016}. In the absence of experimental evidence, only bounds on the magnitude of such decoherence can be provided \cite{Pfister2016,Carlesso2016}.

Here we tackle the following concrete question: \emph{how do spatial quantum superpositions of massive particles evolve in time when self-interacting gravitationally}? To answer our question we consider a particle that is initially found in a superposition of two possible locations, and study the evolution of system while it self-interacts via graviton-exchange. We find that the evolution of the system can be viewed as a quantum channel, where the state undergoes a depolarizing and dephasing process without the need of an initial bath of gravitons. The decoherence obtained this way is thus intrinsic to the system, and it allows us to place a fundamental constraints on the mass and size any quantum particle can have.

%------------------------%
%\section{Setup}
%------------------------%
We model particles as an excitations of a massive scalar quantum field $\hat{\Phi}$ with mass $m$. In flat spacetime $\hat{\Phi}$ can be expanded as $\hat{\Phi}  =\int \textrm{d}^3k[\hat{a}_{\boldsymbol{k}} u_{\boldsymbol{k}} + \hat{a}^{\dagger}_{\boldsymbol{k}} u^*_{\boldsymbol{k}}]$, where $u_{\boldsymbol{k}} (x^{\mu})=(2(2\pi)^3\omega_{\boldsymbol{k}})^{-1/2}\,\exp[i\,k_{\mu}\,x^{\mu}]$ are plane wave modes, while $\omega_{\boldsymbol{k}}:=\sqrt{|\boldsymbol{k}|^2+m^2}$ are the mode frequencies \cite{Srednicki:2007}. The annihilation operators $\hat{a}_{\boldsymbol{k}}$ define the vacuum state $|0\rangle$ through $\hat{a}_{\boldsymbol{k}} |0\rangle=0\,\,\,\forall\, \boldsymbol{k}$ and satisfy the canonical commutation relations $[\hat{a}_{\boldsymbol{k}},\hat{a}^{\dagger}_{\boldsymbol{k}'}]=\delta^3({\boldsymbol{k}}-{\boldsymbol{k}}')$. Here we employ natural units $c=\hbar=1$ unless stated and use Einstein's summation convention.

If gravity is weak we can employ the framework of \textit{linearized gravity} \cite{Sathyaprakash:Schutz:09,Anastopoulos:Hu:2014}, where the spatial extension of the particle is (much) larger than its own Schwarzschild radius $r_\textrm{S}:=\frac{2G m}{c^2}$ to guarantee that it does not collapse into a black hole. In this regime the metric tensor $g_{\mu\nu}$ can be decomposed as $g_{\mu \nu}=\eta_{\mu \nu}+h_{\mu \nu}$, where $\eta_{\mu \nu}=\textrm{diag}(-1,1,1,1)$ is the Minkowski metric and $|h_{\mu\nu}|\ll1$. The metric perturbation $h_{\mu\nu}$ obeys the linearized Einstein equations
$\Box \bar{h}_{\mu\nu}=0$ in the absence of matter, which have wave-like solutions \cite{Maggiore2007}, where
$\bar{h}_{\mu\nu}=h_{\mu\nu}-\frac{1}{2}\eta_{\mu\nu}h$ is the trace-reversed perturbation in the Lorenz gauge
$\partial^\mu \bar{h}_{\mu\nu}=0$ while $h=h^\mu{}_\mu$.

Weak gravity also allows us to circumvent the necessity of a full theory of quantum gravity by employing the well-established framework of \textit{linearized quantum gravity} \cite{Gupta:1952}, where the metric perturbation $h_{\mu\nu}$ is promoted to an operator $\hat{h}_{\mu\nu}$. This approach provides an effective low-energy field theory where matter fields interact in flat spacetimes with excitations of the quantized linearized gravitational field, called \textit{gravitons}. Quantization of the wave-solution to Einstein equations using the field variable $\hat{h}_{\mu\nu}=\hat{\gamma}_{\mu\nu}-\frac{1}{2}\hat{\gamma}\eta_{\mu\nu}$ gives the spin-$2$ and spin-$0$ components
$\hat{\gamma}_{\mu\nu}=\frac{1}{\pi m_\textrm{P}}\int \frac{\textrm{d}^3k}{\sqrt{|\boldsymbol{k}|}}\bigl(\hat{P}_{\mu\nu}(\boldsymbol{k})e^{i\boldsymbol{k}\cdot\boldsymbol{x}}+\hat{P}_{\mu\nu}^\dag(\boldsymbol{k}) e^{-i\boldsymbol{k}\cdot\boldsymbol{x}}\bigr)$ and $\hat{\gamma}=\frac{2}{\pi m_\textrm{P}}\int \frac{\textrm{d}^3k}{\sqrt{|\boldsymbol{k}|}} \bigl(\hat{P}(\boldsymbol{k}) e^{i\boldsymbol{k}\cdot\boldsymbol{x}}+\hat{P}^\dagger(\boldsymbol{k}) e^{-i\boldsymbol{k}\cdot\boldsymbol{x}}\bigr)$ respectively. We have denoted $m_\textrm{P}=\sqrt{\hbar\,c/G}$ as Planck's mass. The graviton operators $\hat{P}_{\mu\nu}(\boldsymbol{k})$ and $\hat{P}(\boldsymbol{k})$ satisfy the the commutation relations $[\hat{P}_{\mu\nu}(\boldsymbol{k}),\hat{P}_{\mu'\nu'}^\dag(\boldsymbol{k}')]=(\eta_{\mu\mu'}\eta_{\nu\nu'}+\eta_{\mu\nu'}\eta_{\mu'\nu})\delta^3(\boldsymbol{k}-\boldsymbol{k}')$ and $[\hat{P}(\boldsymbol{k}),\hat{P}^\dagger(\boldsymbol{k}')]=-\delta^3(\boldsymbol{k}-\boldsymbol{k}')$, while all others vanish (see Supp. Mat.~\ref{suppl_sec_1}). 

Selecting the initial state of a localized particle requires us to provide a method for localizing field excitations. This can be achieved by defining the extended operator 
\begin{equation}\label{wave packet annihilation operator}
    \hat{a}_K=\int \textrm{d}^3k\, F(\boldsymbol{k})e^{i\boldsymbol{k}\cdot\boldsymbol{x}_K }\hat{a}_{\boldsymbol{k}},
\end{equation}
 where the profile function $F(\boldsymbol{k})$ is chosen such that $[\hat{a}_K,\hat{a}_K^\dag]=\int \textrm{d}^3k\,|F(\boldsymbol{k})|^2=1$.  Here, $\boldsymbol{x}_K$ determines the location of the particle, and $K$ labels the two possible configurations of the position degree of freedom that we denote left (L) and right (R). We can then create localized orthogonal one-particle states via $\ket{1_K}:=\hat{a}_K ^\dagger\ket{0}$, such that $\langle 1_K|1_{K'}\rangle=\delta_{KK'}$.
 
\begin{figure}
    \centering
    \includegraphics[width=\linewidth]{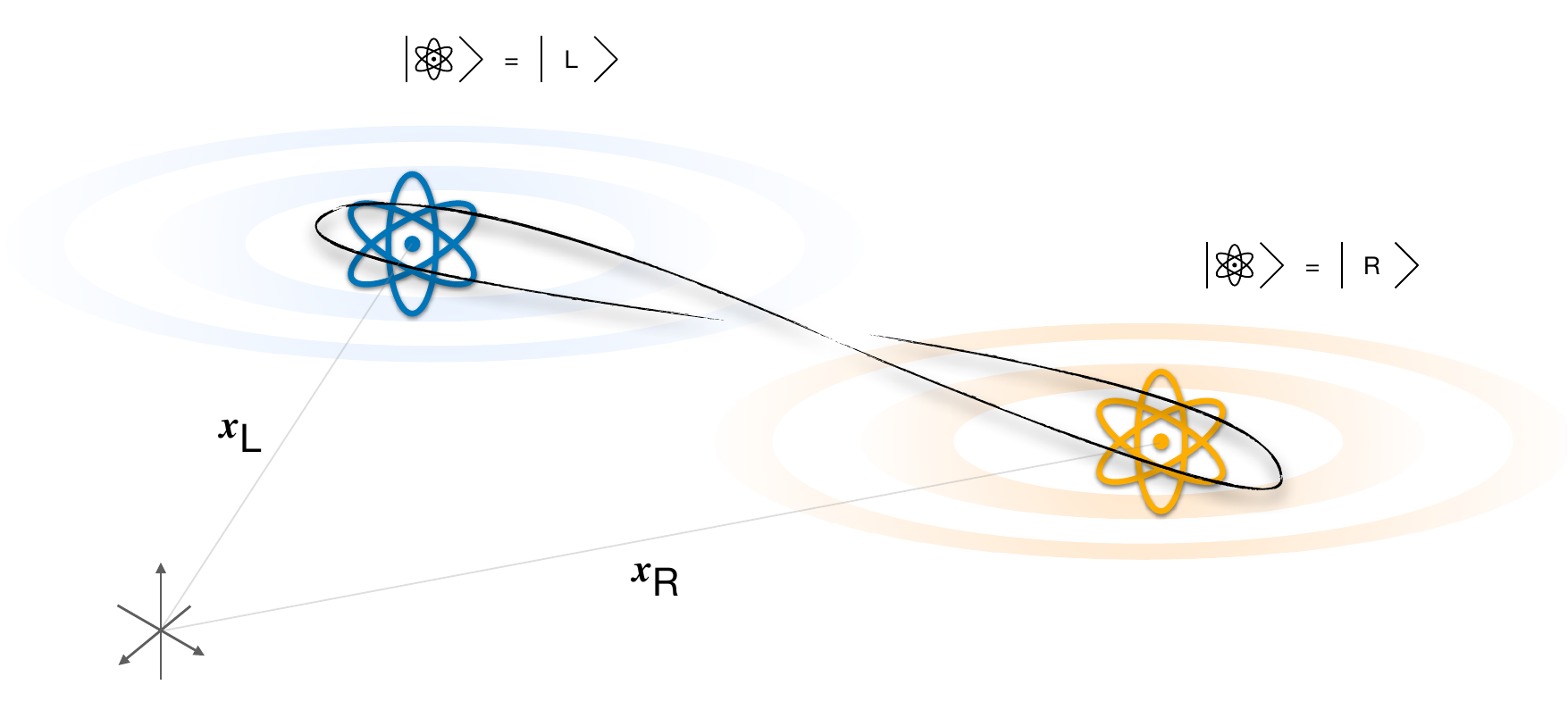}
    \caption{Pictorial representation of a particle in a superposition of two different locations $|\textrm{L}\rangle$ and $|\textrm{R}\rangle$, represented by the black lines $x_\textrm{L}$ and $x_\textrm{R}$, that (self) interacts with gravitons, loosely represented by the waves emanating from the particle.}
    \label{fig: setup}
\end{figure}

 We consider particles that are static, have characteristic size $\ell_0$, and are localized at positions $\boldsymbol{x}_K$, as depicted in Fig. \ref{fig: setup}. Thus, the momentum distribution functions $F$ are centered around $\boldsymbol{k}_0=\mathbf{0}$ with width $\sigma\propto l_0^{-1}$. We then require that $\frac{\sigma}{m}\ll1$, which guarantees that the particle is static since $F$ has mostly support when $\frac{|\boldsymbol{k}|}{\sigma}\leq\mathcal{O}(1)$. We also assume that the distance $L:=|\boldsymbol{x}_\textrm{R}-\boldsymbol{x}_\textrm{L}|$ between the two positions is always larger than the particle's size, i.e., $\ell_0/L\ll 1$. Importantly, the static regime is valid only for times $t$ smaller than the diffusion timescale $t_\textrm{diff}$ (discussed at the end), after which the wave-packet diffuses \cite{Srednicki:2007}. For such times the states $\ket{1_K}$ are eigenstates of the free Hamiltonian $\hat{H}_{0,\textrm{S}}$ with eigenvalue $m$, i.e.  $\hat{H}_{0,\textrm{S}}\ket{1_K}=m\ket{1_K}$. Details are given in Supp. Mat.~\ref{Supp:Mat:Initial:State}.

The initial state of the full system is $\hat{\rho}(0)=\hat{\rho}_\textrm{S}(0)\otimes\hat{\rho}_\textrm{E}(0)\otimes\hat{\rho}_\textrm{G}(0)$, where 
\begin{align}
    \hat{\rho}_\textrm{S}(0)=\alpha\ket{1_\textrm{L}}&\bra{1_\textrm{L}}+(1-\alpha)\ket{1_\textrm{R}}\bra{1_\textrm{R}}\nonumber\\
    &+\beta\ket{1_\textrm{L}}\bra{1_\textrm{R}}+\beta^*\ket{1_\textrm{R}}\bra{1_\textrm{L}}\,,\label{initial particle state}
\end{align}
is the initial reduced state of the particle, and $\hat{\rho}_\textrm{G}(0)=|0_\textrm{G}\rangle\langle0_\textrm{G}|$ is the initial vacuum state of the gravitons. Here, $\hat{\rho}_\textrm{E}(0)=\bigotimes_{\underline{\lambda}\in\textrm{E}}|0_{\underline{\lambda}}\rangle\langle 0_{\underline{\lambda}}|$ is the reduced state of the \emph{system environment}, i.e., the subsystem determined by the localized particle modes $\{F_{\underline{\lambda}}\}$ that, together with the modes $F_\textrm{L}$ and $F_\textrm{R}$, form an orthonormal basis $\mathcal{B}_\textrm{ext}=\{F_\textrm{L},F_\textrm{R},\{F_{\underline{\lambda}}\}\}$ of extended modes. Note that $\mathcal{B}_\textrm{ext}$ is infinite dimensional since the original basis $\mathcal{B}=\{u_{\boldsymbol{k}} (x^{\mu})\}$ of plane waves is infinite-dimensional. In order to have a valid physcial state we require $0\leq\alpha\leq1$, that $-1/2\leq|\beta|\leq1/2$ and $(\alpha-1/2)^2+|\beta|^2\leq1/4$.

In the Schr\"odinger picture, the state of the whole system evolves via the von Neumann equation $\hat{\rho}(t)=\hat{U}(t)\hat{\rho}(0)\hat{U}^\dag(t)$, where $\hat{U}(t)=e^{-i\hat{H}t}$ and $\hat{H}=\hat{H}_0+\hat{H}_\textrm{I}$. Here, $\hat{H}_0=\int \textrm{d}^3k\, \omega_{\boldsymbol{k}} \hat{a}_{\boldsymbol{k}}^\dag\hat{a}_{\boldsymbol{k}}+\frac{1}{2}\int \textrm{d}^3k\,\omega^{\textrm{G}}_{\boldsymbol{k}}\,\bigl[\hat{P}_{\mu\nu}^\dag(\boldsymbol{k})\hat{P}^{\mu\nu}(\boldsymbol{k})-2\hat{P}^\dag(\boldsymbol{k})\hat{P}(\boldsymbol{k})\bigr]$ and $\hat{H}_\textrm{I} $ are the free and interaction Hamiltonian respectively. The latter is determined at time $t=0$ by the matter-graviton interaction \cite{Gupta:1952}, given by
\begin{equation}\label{eq: Interaction Hamiltonian}
    \hat{H}_I =-\frac{1}{2}\int \textrm{d}^3x\,\hat{h}_{\mu\nu}(\boldsymbol{x}):\hat{T}^\Phi{}^{\mu\nu}(\boldsymbol{x}):,
\end{equation}
where we have introduced the stress-energy tensor field $\hat{T}_{\mu\nu}:=\partial_{\mu}\hat{\phi}\partial_{\nu}\hat{\phi}-\frac{1}{2}\,\eta_{\mu \nu}\bigl[\partial^\lambda\hat{\phi}\partial_{\lambda}\hat{\phi}+m^2\hat{\phi}^2\bigr]$ of the matter field, and $:\hat{A}:$ denotes normal ordering of $\hat{A}$.

The initial state 
$\hat{\rho}(0)=\hat{\rho}_{\textrm{SE}}(0)\otimes\ket{0_{\textrm{G}}}\bra{0_{\textrm{G}}}$ contains no correlations between matter and gravity, which not only allows the gravitational degrees of freedom to be treated as an environment, but also to attribute all effects of interest to the self-gravity dynamics alone. The state $\hat{\rho}_\textrm{SE}(t)=\textrm{Tr}_{\textrm{G}}\bigl(\hat{U}(t)\hat{\rho}(0)\hat{U}^\dagger(t)\bigr)$ of the matter sector at time $t$ is then obtained by tracing out the gravitational degrees of freedom. Lengthy algebra found in Supp. Mat.~\ref{suppl_sec_3} gives
\begin{align}\label{time:evolution:matrix:representation}
    \hat{\rho}_\textrm{SE}(t)
    &=\hat{U}_2(t)\hat{\rho}_\textrm{SE}(0)\hat{U}_2^\dag(t)\nonumber\\
    +&\frac{2}{\pi}\frac{m^2}{m_\textrm{P}^2}\int_0^t\textrm{d}t'\textrm{d}t''\int\frac{\textrm{d}^3x\textrm{d}^3y}{m^2}D_{\mu\nu,\mu'\nu'}(x-y)\nonumber\\
    \times&\left[\hat{T}^{\mu\nu}_x\hat{\rho}_\textrm{SE}(0)\hat{T}^{\mu'\nu'}_y-\frac{1}{2}\left\{\hat{T}^{\mu'\nu'}_y\hat{T}^{\mu\nu}_x,\hat{\rho}_\textrm{SE}(0)\right\}\right],
\end{align}
where $x\equiv(x^\mu)$ for convenience, $\hat{U}_2(t):=\exp[-\frac{1}{2}\int_0^t\textrm{d}t'\int_0^{t'}\textrm{d}t''\bra{0_\textrm{G}}[\hat{H}_\textrm{I}(t'-t),\hat{H}_\textrm{I}(t''-t)]\ket{0_\textrm{G}}]$, $\hat{T}^{\mu\nu}_x\equiv:\hat{T}^{\mu\nu}(x):$ for the sake of simplicity, and $D_{\mu\nu,\mu'\nu'}(x-y)$ is a graviton propagator-like quantity.

We can find the reduced state $\hat{\rho}_\textrm{S}(t)$ of the system starting from  \eqref{time:evolution:matrix:representation} and by 
first invoking the rotating wave approximation for the stress-energy tensor, which reduces to a number-preserving multimode-mixer operator, subsequently projecting $\hat{\rho}_\textrm{SE}(t)$ onto the full one-particle space of modes $F_\textrm{L},F_\textrm{R},\{F_{\underline{\lambda}}\}$, then tracing over $E$, and finally computing the coefficients $\rho_{\textrm{S},KK'}:=\bra{1_K}\hat{\rho}_\textrm{SE}(t)\ket{1_{K'}}$ and $\rho_{\textrm{S},00}:=\bra{0_\textrm{S}}\hat{\rho}_\textrm{S}(t)\ket{0_\textrm{S}}$, see Supp. Mat~\ref{sec_state_elements}. Straightforward algebra gives us the perturbative form $\hat{\rho}_\textrm{S}(t)=\hat{\rho}_\textrm{S}(0)+\epsilon^2\hat{\rho}^{(2)}_\textrm{S}(t)$, where the small control parameter $\epsilon:=\frac{m}{m_\textrm{P}}\ll1$ determines the magnitude of the effects in terms of the relevant physical scales \cite{Marletto:Vedral:2017,Aziz:Howl:2025}, and the matrix $\hat{\rho}^{(2)}_\textrm{S}(t)$ has the form
\begin{equation}
    \hat{\rho}^{(2)}_\textrm{S}(t)
    =\rho^{(2)}_{00}(t)\ket{0_\textrm{S}}\bra{0_\textrm{S}}\oplus
    \begin{pmatrix}
        \rho^{(2)}_{\textrm{LL}}(t) & \rho^{(2)}_{\textrm{LR}}(t) \\
        \rho^{(2)*}_{\textrm{LR}}(t) & \rho^{(2)}_{\textrm{RR}}(t)
    \end{pmatrix}
\end{equation}
in the $\ket{0_\textrm{S}},|1_\textrm{L}\rangle$, $|1_\textrm{R}\rangle$ basis. Note that the first-order term $\hat{\rho}^{(1)}_\textrm{S}(t)$ vanishes identically in the absence of external sources of gravitons (e.g., a background bath of gravitons or the presence of an additional massive object). 

The evolution \eqref{time:evolution:matrix:representation} is not unitary. The second term in particular signals the occurrence of open quantum system dynamics. We can therefore formally write
\begin{equation}\label{eq: Channel representation}
    \hat{\rho}_\textrm{S}(t)=\textrm{Tr}_{\textrm{G,E}}\left(\hat{U}(t)\hat{\rho}(0)\hat{U}^\dagger(t)\right)\equiv\mathcal{N}(\hat{\rho}_\textrm{S}(0)),
\end{equation}
which establishes the evolution $\hat{\rho}_\textrm{S}(0)\overset{\mathcal{N}}{\rightarrow}\hat{\rho}_\textrm{S}(t)$ as a \textit{quantum channel}. Stinespring's theorem guarantees that $\mathcal{N}$ is complete-positive and trace preserving map \cite{Stinespring1955}. 
The use of the quantum communication framework allows us to characterize the evolution using well-known methods and tools through the lens of quantum information.
\begin{figure}
    \centering
    \includegraphics[width=0.9\linewidth]{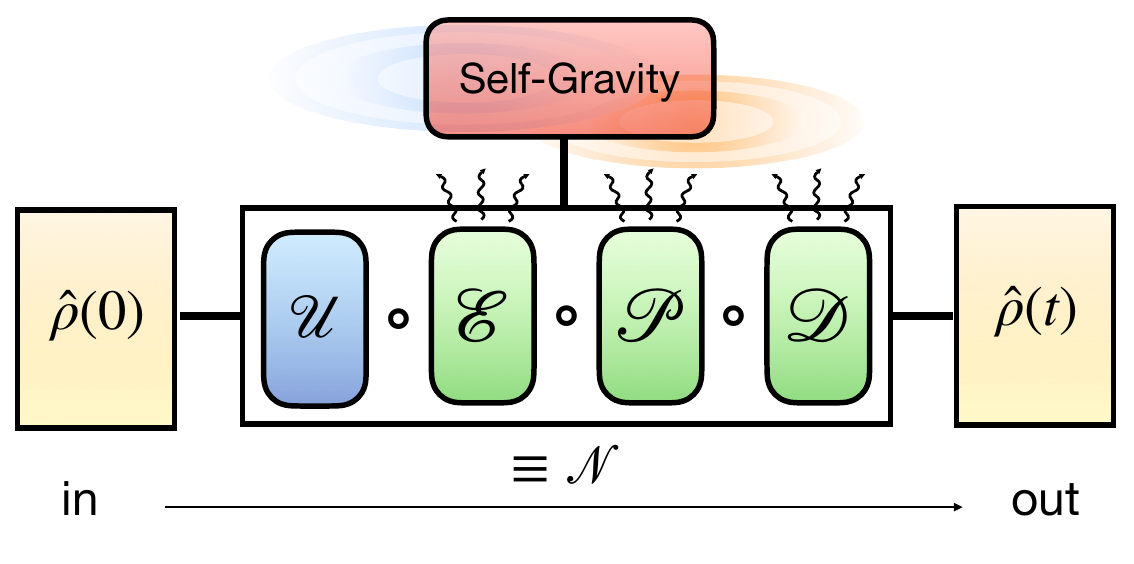}
    \caption{\textbf{Self-gravity quantum channel:} Time evolution of the system due to self-gravity can be viewed a quantum channel $\mathcal{N}=\mathcal{U}\circ\mathcal{E}\circ\mathcal{P}\circ\mathcal{D}$ with unitary $\mathcal{U}$, erasure $\mathcal{E}$, depolarizing $\mathcal{P}$, and dephasing $\mathcal{D}$ components.}
    \label{fig:QC}
\end{figure}
The quantum channel $\mathcal{N}$ arising from \eqref{eq: Channel representation} can be decomposed as $\mathcal{N}=\mathcal{E}\circ\mathcal{P}\circ\mathcal{D}$, where $\mathcal{E}$ is an erasure channel, $\mathcal{P}$ is a depolarization channel, and $\mathcal{D}$ is a dephasing channel, see Supp. Mat.~\ref{suppl_sec_channel}. A depiction of the channel and its decomposition is given in Figure~\ref{fig:QC}. Since they all occur at the same perturbative order $\mathcal{O}(\epsilon^2)$, they commute. In the following we identify the contribution of each channel.

\emph{Erasure channel}---Conservation of the trace requires $\textrm{Tr}(\hat{\rho}_\textrm{S}(t))=1$. However, in general we have $\rho^{(2)}_{\textrm{LL}}(t)+\rho^{(2)}_{\textrm{RR}}(t)<0$ for $t>0$ due to the interaction with gravitons. Therefore, we must have $\rho^{(2)}_{00}(t)=-\rho^{(2)}_{\textrm{LL}}(t)-\rho^{(2)}_{\textrm{RR}}(t)$ since it is the only other nonzero coefficient that contributes to the trace. In turn, this informs us on the presence of the \emph{erasure channel} $\mathcal{E}$, which has general action $\mathcal{E}:\hat{\rho}_\textrm{S}(0)\rightarrow\mathcal{E}(\hat{\rho}_\textrm{S}(0))=(1-p_\textrm{e}(t))\hat{\rho}(0)+ p_\textrm{e}(t)|\textrm{e}\rangle\langle\textrm{e}|$ on the initial state $\hat{\rho}_\textrm{S}(0)$ of a 2-dimensional system \cite{Zhong:Oh:2023}. Here $|\textrm{e}\rangle$ is a flag state orthogonal to the state $\hat{\rho}(0)$, and in our case we have $|\textrm{e}\rangle\equiv\ket{0_\textrm{S}}$.
Thus, the channel has erasure probability $p_\textrm{e}(t)\equiv-(\rho^{(2)}_{\textrm{LL}}(t)+\rho^{(2)}_{\textrm{RR}}(t))$.

For the remainder of this work we condition all analysis to the presence of a particle which, interestingly, is directly related to the renormalization of vacuum and divergent terms. This effectively allows us to set $p_\textrm{e}(t)=0$ and focus on the left-right single-particle subsector.

\emph{Depolarization channel}---The diagonal elements now have corrections that are equal in magnitude but opposite in sign. This motivates us to consider a \emph{depolarization channel} $\mathcal{P}$, which is defined by the concrete action $\mathcal{P}:\hat{\rho}(0)\rightarrow\mathcal{P}(\hat{\rho}(0))=\frac{\lambda}{d}\mathds{1}+(1-\lambda)\hat{\rho}(0)$, where $d$ is the dimension of $\hat{\rho}(0)$ and $\lambda$ is the depolarization parameter. In our case, the state is effectively 2-dimensional, and therefore we set $d=2$. We thus immediately obtain the depolarization parameter $\lambda(t)=\frac{2}{1-2\alpha}((1-\alpha)\rho^{(2)}_{\textrm{LL}}(t)-\alpha\rho^{(2)}_{\textrm{RR}}(t))$.

\emph{Dephasing channel}---We are left with a final contribution of the form $\rho_{\textrm{LR}}(t)=(1-\varphi(t))\rho_{\textrm{LR}}(0)$ on the off-diagonal elements, which motivates us to consider the \emph{dephasing channel} $\mathcal{D}:\hat{\rho}_\textrm{S}(0)\rightarrow\mathcal{D}(\hat{\rho}_\textrm{S}(0))$ with action
\begin{equation}
    \mathcal{D}:\left(\begin{matrix}
        \rho_{00}&\rho_{01}\\
        \rho_{01}^*&\rho_{11}
    \end{matrix}\right)
    \mapsto
    \left(\begin{matrix}
        \rho_{00}&(1-\varphi)\rho_{01}\\
        (1-\varphi)\rho_{01}^*&\rho_{11}
    \end{matrix}\right).
\end{equation}
We find $\varphi(t)=-(\frac{1}{\beta}\rho^{(2)}_{\textrm{LR}}(t)+\frac{1}{1-2\alpha}(\rho^{(2)}_{\textrm{LL}}(t)-\rho^{(2)}_{\textrm{RR}}(t)))$.

The three channels obtained above account for all effects to $\mathcal{O}(\epsilon^2)$. In Supp. Mat.~\ref{suppl_sec_channel} we show that \emph{all} quantities above are well defined for all values of $\alpha,\beta$. We have therefore achieved a full characterisation of the dynamics.

We now specify the momentum distribution $F(\boldsymbol{k})$ to seek an explicit solution to the problem. We employ a Gaussian profile $F(\boldsymbol{k})=(2\pi\sqrt{2\pi}\sigma^3)^{-1/2}\exp[-|\boldsymbol{k}|^2/4\sigma^2]$, which is symmetrically peaked around the origin $\boldsymbol{k}=0$ and is normalized by $\int \textrm{d}^3k |F(\boldsymbol{k})|^2=1$.
With this choice of momentum distribution we obtain the renormalized correction coefficients of the state
{\small
\begin{subequations}\label{coefficeints:for:final:state:main}
  \begin{align}
    \rho^{(2)}_{\textrm{S},\textrm{LL}}(t)=&(1-2\alpha)\,Q(t)\,e^{-\sigma^2L^2},\\
    \rho^{(2)}_{\textrm{S},\textrm{RR}}(t)=&-(1-2\alpha)\,Q(t)\,e^{-\sigma^2L^2},\\
    \rho^{(2)}_{\textrm{S},\textrm{LR}}(t)=& \beta(Q_{2L}(t)-Q(t))+i(1-2\alpha)\frac{\pi\sigma t}{128\sigma L}e^{-\frac{\sigma^2L^2}{2}},
\end{align}
\end{subequations}
}

\noindent where 
$Q_{x}(t):=\frac{1}{32}\int_0^{+\infty}\textrm{d}v
\frac{\sin(\sigma\,x\,v)}{\sigma\,x}\frac{\sin^2(\sigma\,t\,v)}{v^2}e^{-v^2}$,  $Q(t)\equiv Q_{x\to0}(t)=(\frac{\sigma t}{8})^2 {}_2 F_2(1,1;\frac{3}{2},2;-(\sigma t)^2)$, and ${}_2F_2(a,b;c,d;z)$ is a generalized hypergeometric function.

The renormalized channel parameters then read
\begin{subequations}\label{channel:parameters:main}
    \begin{align}
        \lambda(t)=&2\epsilon^2\,Q(t)\,e^{-\sigma^2L^2},\label{depolarizaion:parameter:main}\\
         \varphi(t)=&\epsilon^2 \bigl((1-2e^{-\sigma^2L^2})\,Q(t)-Q_{2L}(t)\bigr).
    \end{align}
\end{subequations}
We are now in the position of extracting all relevant physical information of the evolution of the system under self gravity. We start by considering the probability $P_K(t):=\bra{K}\hat{\rho}_\textrm{S}(t)\ket{K}$ of finding the particle at either position $K=\textrm{L}$ or $K=\textrm{R}$. 
We have
\begin{subequations}
    \begin{align}
        P_\textrm{L}(t):=&\alpha+(1-2\alpha)\epsilon^2\,Q(t)\,e^{-\sigma^2L^2},\\
        P_\textrm{R}(t):=&1-\alpha-(1-2\alpha)\epsilon^2\,Q(t)\,e^{-\sigma^2L^2}.
    \end{align}
\end{subequations}
As expected, $P_\textrm{L}(t)+P_\textrm{R}(t)=1$ in the renormalized state.

We then quantify the quantum coherence in the state. In general, a quantum state of the form $\hat{\rho}=\sum_{jk} \rho_{jk}\ket{j}\bra{k}$, where $\ket{j}$ form an orthonormal basis, contains quantum coherence. This can be seen by using the \emph{ relative entropy of coherence} $C (\hat{\rho}):=S(\hat{\rho}_\textrm{diag})-S(\hat{\rho})$, where $S(\hat{\rho}):=-\textrm{Tr}(\hat{\rho}\ln\hat{\rho})$ is the von Neumann entropy of state $\hat{\rho}$ and $\hat{\rho}_\textrm{diag}$ is the state obtained from $\hat{\rho}$ by removing all off-diagonal terms \cite{Zhang:Shao:2016}. In our case we have 
\begin{align}
    C (t)=-\sum_{\sigma=\pm}\left[\lambda^\textrm{diag}_\sigma\ln\lambda^\textrm{diag}_\sigma-\lambda_\sigma\ln\lambda_\sigma\right],
    \label{coherence def}
\end{align}
where $C (t)\equiv C (\hat{\rho}_\textrm{S}(t))$, and it is supplemented by 
{\small
\begin{subequations}\label{lambda terms}
    \begin{align}
        2\lambda^\textrm{diag}_\pm=&1\pm|1-2\alpha|\left(1-\lambda/2\right),\\
        2\lambda_\pm=&1\pm\sqrt{(1-2\alpha)^2\left(1-\lambda\right)+4\beta^2\left(1-2\lambda-2\varphi\right)}
    \end{align}
\end{subequations}
}

\noindent given in terms of the depolarizing and dephasing parameters $\lambda$ and $\varphi$ respectively  \eqref{channel:parameters:main}. 

The expressions above allow us to conclude the following. First, we know that depolarization and dephasing are responsible for ``mixing the state''. We therefore study the evolution of the quantity $Q(t)$ appearing in \eqref{channel:parameters:main}, which increases monotonically in time. Thus $P_\textrm{L}(t)$ decreases and $P_\textrm{R}(t)$ increases when $\alpha>1/2$, while the opposite occurs when $\alpha<1/2$. Together, these statements imply that $|P_\textrm{R}(t)-P_\textrm{L}(t)|=|1-2\alpha|(1-2\epsilon^2\,Q(t)\,e^{-\sigma^2L^2})$, which monotonically decreases in time, and thus the state evolves towards the maximal mixed scenario $P_\textrm{L}(t)=P_\textrm{R}(t)=1/2$ since the coherence decreases as well, see below. Such state is the most classical state possible for the system, but the effects are exponentially suppressed as a function of the distance-to-particle-size ratio. Interestingly, when $\alpha=\frac{1}{2}$ the probabilities $P_K(t)$ do not change as a function of time, and thus the diagonal elements $\rho_{\textrm{S},KK}(t)=\frac{1}{2}$ remain constant. Second, it is easy to verify that the amount of coherence $C (t)$ is a monotonic function of the combination $\beta(1-\lambda-\varphi)$ of relevant parameters for all time $t$, where crucially $C (t)=0$ for $\beta=0$ regardless of the interaction. This means that there is no
coherence change, including no generation of coherence,
when none is initially present (i.e., the state is mixed and
diagonal), which is consistent with the observational evidence that an initially classical system will never enjoy quantum features due to gravitational self-interaction. To get a better insight we fix $\alpha=\frac{1}{2}$ and take $\beta$ as a free parameter. The initial state is maximally mixed (and has no coherence) for $\beta=0$, and is maximally entangled for $\beta=\frac{1}{2}$. Then, Eqs. \eqref{coherence def} and \eqref{lambda terms} imply that the coherence $C (t)$ is a monotonically increasing function of $\beta$ that is bounded by $0\leq C (t)\leq\ln2+ \frac{\lambda+\varphi}{2}\ln\frac{\lambda+\varphi}{2}$ and decreases in time. Crucially, when the distance $L$ is much larger than the size of the particle $\ell_0$ we have that $\lambda\approx0$ and $\varphi\approx\epsilon^2 Q(t)$, and thus the coherence decreases as a function of time \textit{independently of the distance itself}. 

Finally, we note that there are two timescales intrinsic to the problem: the \textit{diffusion} time $t_\textrm{diff}$ and \textit{dephasing} time $t_\textrm{deph}$ determined by
{\small
\begin{subequations}
    \begin{align}
        t_\textrm{diff}:=\kappa_\textrm{diff}\frac{\ell_0^2 m}{\hbar}=&\kappa_\textrm{diff}\,t_\textrm{P}\frac{\ell_0^2}{\ell_\textrm{P}^2}\frac{m}{m_P},\\
        f_\textrm{deph}\left(\frac{t_\textrm{deph}}{t_\textrm{P}}\frac{\ell_\textrm{P}}{\ell_0}\right)=&\kappa_\textrm{deph}\frac{m_\textrm{P}^2}{m^2},
    \end{align}
\end{subequations}
}

\noindent where $t_\textrm{P}:=\sqrt{\frac{\hbar\,G}{c^5}}$ and $\ell_\textrm{P}:=\sqrt{\frac{\hbar\,G}{c^3}}$ are Planck's time and length. Here, $\kappa_\textrm{diff}$ and $\kappa_\textrm{deph}$ are constants that depend only on the specific details of the setup (such as the shape of the particle), while $f_\textrm{deph}$ is a setup-dependent function that can be obtained explicitly once the setup has been given. The diffusion timescale governs the validity of the massive static regime, beyond which the wave-packet diffuses as predicted by quantum field theory \cite{Ford:Oconnell:2002,Edery:2021}, while the dephasing timescale governs the validity of the weak-field regime (up to $\mathcal{O}(\epsilon^2)$), and thus the time at which decoherence become relevant. 

In our case, $f_\textrm{deph}(x)\approx \ln x$ for large $x$. We then have 
\begin{equation}
    t_\textrm{deph}=t_\textrm{P}\frac{\ell_0}{\ell_\textrm{P}}e^{\kappa_\textrm{deph}\frac{m_\textrm{P}^2}{m^2}}.
\end{equation}
We expect that heavy objects behave classically, while light ones can exhibit quantum features. This must occur in particular for diffusion and dephasing times much larger than Planck's time $t_\textrm{P}=5\times10^{-44}\textrm{s}$, i.e., $\frac{t_\textrm{diff}}{t_\textrm{P}},\frac{t_\textrm{deph}}{t_\textrm{P}}\gg1$, as evidenced by observation. 
This is guaranteed by the massive static regime, which reads $\frac{\ell_0}{\ell_\textrm{P}}\frac{m}{m_\textrm{P}}\gg1$. Thus, our work applies to objects with characteristic size that largely exceeds the Planck length, since $m<m_\textrm{P}$ for the perturbative regime to apply. 

We are motivated to state that a \emph{particle is classical} when $t_\textrm{deph}\ll t_\textrm{diff}$, i.e., in the case of large and heavy objects, while a \emph{particle exhibits quantum behavior} when $t_\textrm{diff}\ll t_\textrm{deph}$, i.e., when it is light and small. These regimes imply  $\kappa_\textrm{diff}\frac{\ell_0}{\ell_\textrm{P}}\gg\frac{m_\textrm{P}}{m}\exp[\kappa_\textrm{deph}\frac{m_\textrm{P}^2}{m^2}]>1$  and  $\kappa_\textrm{diff}\frac{\ell_0}{\ell_\textrm{P}}\ll\frac{m_\textrm{P}}{m}\exp[\kappa_\textrm{deph}\frac{m_\textrm{P}^2}{m^2}]$ respectively for our case.

We then note that the two timecales $t_\textrm{diff}$ and $t_\textrm{deph}$ have, in general, different functional dependencies on mass and size of the particle. Thus, we identify a separation regime $(\ell_0(m),m)$ in the particle parameter space where the two timescales coincide. It is given by the implicit relation $\frac{m^2}{m_\textrm{P}^2}f_\textrm{deph}\bigl(\kappa_\textrm{diff}\frac{\ell_0}{\ell_\textrm{P}}\frac{m}{m_\textrm{P}}\bigr)=\kappa_\textrm{deph}$, which in our present case reads $\ell_0(m)=\kappa_\textrm{diff}^{-1}\ell_\textrm{P}\frac{m_\textrm{P}}{m}\exp[\kappa_\textrm{deph}\frac{m_\textrm{P}^2}{m^2}]$. We use this relation to define the regime across which a transition between classical and quantum behaviour occurs. A visual summary is given in Table~\ref{tab:my_label}.

\begin{table}[ht!]
    \begin{tabular}{|l|c|}
                  \hline
        \multicolumn{1}{|c|}{\textbf{Physical regime}} & \textbf{Parameter constraint} \\
       \hline
       \hline
       Quantum  & $t_\textrm{diff}\ll t_\textrm{deph}$ \\
       \hline
       Macroscopic quantum  & $t_\textrm{diff}\approx t_\textrm{deph}$\\
       \hline
       Classical & $t_\textrm{diff}\gg t_\textrm{deph}$\\
       \hline
    \end{tabular}
    \caption{Parameter constraints for the quantum-to-classical regime transition of self-gravitating physical systems. }
    \label{tab:my_label}
\end{table}

It is easy to see that for very light particles, such as electrons, for which $m=10^{-30}\textrm{kg}$ and $\ell_0\leq10^{-20}\textrm{m}$, we have that $t_\textrm{diff}\approx10^{-36}\textrm{s}\ll t_\textrm{deph}\approx\infty$, which means that quantum coherence can be established and maintained for such small and light systems. On the other hand, in the case of heavier and more macroscopic objects, for example a small rod of the size $\ell_0=10^{-6}\textrm{m}$ and mass $m=10^{-9}\textrm{kg}$, we have $t_\textrm{deph}\approx2\times10^{-6}\textrm{s}\ll t_\textrm{diff}\approx10^{13}\textrm{s}$, which corroborates the observation that heavier objects are never detected in a quantum state.
In the present setup, a transition regime effectively occurs when $m$ become close enough to Planck's mass $m_\textrm{P}$, i.e., within a few orders of magnitude $m\approx 10^{-1}-10^{-2}m_\textrm{P}$. For smaller masses, the coherence in the state can be maintained at will. The transitional regime is currently beyond those found, for example, within proposals employing Bose-Einstein condensates (BECs) \cite{Cirac1998,Howl:Penrose:2019}, but is closer to those at the centre of experimental proposals to measure gravitational effects on macroscopic quantum system \cite{Bose:Mazumdar:2017,Marletto:Vedral:2017,Carney2019,Strasser:Christodoulou:2025}. A nonperturbative analysis of composite systems where the full time-evolution is taken into account is needed to make reliable predictions for masses that are close to or larger than Planck's mass.

Our work provides a mechanism, and therefore a potential explanation, for the difference in gravitational behavior of large and heavy objects as compared to small and light ones. We identify the transition regime, defined as the area in parameter space where the diffusion and dephasing timescales are of the same order of magnitude. Our results complement the ongoing efforts to understand gravitation of quantum systems \cite{Anastopoulos2021, Belenchia2019,Bassi2017}, as well as the transition between classical-to-quantum regimes \cite{Zurek2003,Schlosshauer2007} and the verification of the quantumness of gravity \cite{Anastopoulos2021,Belenchia2019,Bassi2017}. In this regard, our work directly depends on the quantum nature of the interaction mediated by gravitons, and cannot be explained classically \cite{Kafri2014,Anastopoulos2021,Belenchia2019,Aziz:Howl:2025,Marletto:Oppenheim:2025,Biswas:Bose:2026}. 

\vspace{0.2cm}

\textit{Acknowledgments}---We thank Andreas Wolfgang Schell for useful comments and discussions. AL acknowledges the hospitality of PGI-12 at Forschungzentrum J\"ulich where the overhaul of this work was conducted.

\vspace{0.2cm}

\textit{Author contribution statement}---DEB conceived the idea, led the project, and produced the first version of this work. DEB and AL overhauled the first version to include the covariant methods developed in the literature~\cite{Gupta:1952}, which allowed DEB to obtain the results in current form. FWM and SM aided in the analysis and interpretation of the results. All authors contributed to writing the manuscript.

\bibliographystyle{apsrev4-2}
\bibliography{WeightPaper}

\clearpage

\onecolumngrid

\appendix
\setcounter{appendixsection}{0}

\appendix
\setcounter{section}{0}
\renewcommand{\thesection}{\Alph{section}}

\bigsection{SUPPLEMENTAL MATERIAL 
\\for
\\``Self gravity decoheres quantum systems''}

\vspace{1cm}

%---------------------------%
\section{PRELIMINARIES}\label{suppl_sec_1}
%---------------------------%
In this work we extensively use the results obtained in the literature \cite{Gupta:1952,Bruschi:2016,Bose:Mazumdar:2017,Bose:Mazumdar:2022}, which we do not reprint in full here. We leave it to the interested reader to peruse the foundational works. Importantly for our study, we work in the Schr\"odinger picture, thus we will seek the effective Hamiltonian of the system to evolve its initial quantum state. We will also employ perturbation theory with the perturbative parameter $\epsilon\ll1$, and all quantities of interest will be expanded as $A=A^{(0)}+\epsilon\,A^{(1)}+\epsilon^2\,A^{(2)}$. We work in natural units $c=\hbar=1$, such that $G=m_\textrm{P}^{-2}$, $t_\textrm{P}:=\sqrt{G}$, and $\ell_\textrm{P}:=\sqrt{G}$, where $G$ stands for Newton's constant, while $m_\textrm{P}$, $t_\textrm{P}$ and $\ell_\textrm{P}$ are Planck's mass, time, and length respectively.

Throughout this work we employ the following conventions:
\begin{subequations}
    \begin{align}
        (2\pi)^3\,\delta^3(\boldsymbol{k})=&\int \textrm{d}^3x e^{i\boldsymbol{k}\cdot\boldsymbol{x}},\\
        F(\boldsymbol{k})=&\int \textrm{d}^3x \tilde{F}(\boldsymbol{x})e^{i\boldsymbol{k}\cdot\boldsymbol{x}},\\
        \tilde{F}(\boldsymbol{x})=&\int \frac{\textrm{d}^3k}{(2\pi)^3}\,F(\boldsymbol{k})e^{-i\boldsymbol{k}\cdot\boldsymbol{x}},
    \end{align}
\end{subequations}
which are consistent with the fact that the spatial Fourier transform of $\tilde{F}(\boldsymbol{x})=1$ is $F(\boldsymbol{k})=(2\pi)^3\,\delta^3(\boldsymbol{k})$.
For later reference, this also means that
\begin{align*}
    \int \textrm{d}^3k |F(\boldsymbol{k})|^2=1
    \qquad
    \Rightarrow
    \qquad
    (2\pi)^3\int \textrm{d}^3x |\tilde{F}(\boldsymbol{x})|^2=1.
\end{align*}

%---------------------------%
\subsection{Field quantization}\label{field:quantization:appendix}
%---------------------------%
In this work matter is modelled by employing the conventional theory of field quantization and we refer the reader to the literature for an in depth introduction to the quantum field theory in curved spacetime \cite{BandD,Jacobson:2005}.  
Elementary particles, in particular, are modelled as excitations of an uncharged massive scalar quantum field $\hat{\phi}(x^\mu)$ with mass $m$ that propagates on a curved background spacetime $\mathcal{M}$, which is endowed by a metric $g_{\mu\nu}$ and is parametrized by coordinates $x^\mu$. This simplification allows us to gain insights in the qualitative behavior without the need of dealing with all complications that arise using Dirac fields to model matter \cite{Srednicki:2007}.
Given the approach taken here, field quantization can be done in a standard fashion and it allows us to find the expression for the field in terms of the Fourier coefficients, later to be promoted to creation and annihilation operators \cite{Srednicki:2007}. Note that we consider an action of the form $S=\int d^4x\mathcal{L}$, where $\mathcal{L}=-\frac{1}{2}\partial_\mu\hat{\phi}\partial^\mu\hat{\phi}-\frac{1}{2}m^2\hat{\phi}^2$ for matter (i.e., the field of choice here), which determines also the normalization and the physical dimensions of the field. In this case the field has dimension of inverse length.

We do not need to solve the field equations for $\hat{\phi}(x^\rho)$ in curved spacetime. Instead, the (weak) interaction with gravity will occur in flat spacetime as explained below. Therefore, it is natural to choose a decomposition of the field $\hat{\phi}(x^\rho)$ that is obtained as a solution of the Klein-Gordon field equations $\square\hat{\phi}(x^\rho)=(\partial_\mu \partial^\mu-m^2)\hat{\phi}(x^\rho)=0$ in flat spacetime, which reads
\begin{align}\label{field:expansion:appendix}
\hat{\phi}(x^\rho)  =\int \textrm{d}^3k\,\left[\hat{a}_{\boldsymbol{k}} u_{\boldsymbol{k}} + \hat{a}^{\dagger}_{\boldsymbol{k}} u^*_{\boldsymbol{k}}\right],
\end{align}
where the $u_{\boldsymbol{k}} (x^{\mu})=(2\sqrt{(2\pi)^3\omega_{\boldsymbol{k}}})^{-1/2}\,e^{i\,k_{\mu}\,x^{\mu}}$ are plane waves in Minkowski spacetime, we have introdued $k_{\mu}\,x^{\mu}=-\omega_{\boldsymbol{k}}\,t+\boldsymbol{k}\cdot\boldsymbol{x}$, and the frequency reads $\omega_{\boldsymbol{k}}:=\sqrt{|\boldsymbol{k}|^2+m^2}$. Crucially, the annihilation and creation operators $\hat{a}_{\boldsymbol{k}}$, $\hat{a}^{\dagger}_{\boldsymbol{k}}$ satisfy the canonical commutation relations $[\hat{a}_{\boldsymbol{k}},\hat{a}^{\dagger}_{\boldsymbol{k}'}]=\delta^3({\boldsymbol{k}}-{\boldsymbol{k}}')$ with our choice of field normalization, while all other vanish \cite{Srednicki:2007}.

This procedure allows us to define the vacuum state $|0_\textrm{S}\rangle$ of the non interacting matter sector (or system S) by the constraint $\hat{a}_{\boldsymbol{k}}|0_\textrm{S}\rangle=0$ for all $\boldsymbol{k}$, while particle states are obtained in the standard fashion by acting on the vacuum state with appropriate powers of creation operators. One note of care is that Fock states are not normalized in the usual sense. That is, if $\ket{1_{\boldsymbol{k}}}:=\hat{a}_{\boldsymbol{k}}^\dag|0_\textrm{S}\rangle$ is a one particle state with sharp momentum $\boldsymbol{k}$, we then have $\langle1_{\boldsymbol{k}}|1_{\boldsymbol{k}'}\rangle=\delta^3({\boldsymbol{k}}-{\boldsymbol{k}}')$, and thus $\ket{1_{\boldsymbol{k}}}$ has infinite norm. We will deal with this problem below by introducing suitable wave-packets.

Finally, the normal-ordered free Hamiltonian $:\hat{H}_{0,\phi}:$ for the field reads
\begin{align}\label{free:hamiltonian:field:appendix}
:\hat{H}_{0,\phi}:=\int \textrm{d}^3k\,\omega_{\boldsymbol{k}}\,\hat{a}^{\dagger}_{\boldsymbol{k}}\hat{a}_{\boldsymbol{k}}.
\end{align}

%---------------------------%
\subsection{Linearized quantum gravity}\label{gravity:quantization:appendix}
%---------------------------%
We assume that all dynamics take place in regions of spacetime where the spacetime is essentially flat. For example, if a massive body (such as a planet) is present, we assume that all processes of interest occur far enough from it. If no massive body is present, the particles of interest have masses that are low enough to avoid inducing regions of high curvature. This allows us to assume that any gravitational contribution can be modelled as a small metric perturbation, which gives us the following expression for the full metric:
\begin{align} \label{perturbed:metric:appendix}
g_{\mu\nu}=\eta_{\mu\nu}+\epsilon\,h_{\mu\nu},
\end{align}
where $\epsilon\ll1$ is the perturbative parameter that governs the weak-gravity regime of interest, and $\eta_{\mu\nu}=\textrm{diag}(-1,1,1,1)$ is the Minkowski metric. Notice that, in the literature \cite{Flanagan:Hughes:2005}, it is customary to write \eqref{perturbed:metric:appendix} as $g_{\mu\nu}=\eta_{\mu\nu}+h_{\mu\nu}$ and impose that $|h_{\mu\nu}|\ll1$. Here we choose to have the ``smallness'' of the metric perturbation $h_{\mu\nu}$ controlled by $\epsilon$, which is a parameter that will be determined later in terms of the relevant energy scales of the system.

The next ingredient that is required is the quantization of the metric perturbation $h_{\mu\nu}$. The logic of this approach is that we do not study the dynamics of fields in a classical fixed curved background but we assume that gravity is itself a dynamical quantized field that weakly interacts with matter in flat spacetime. We follow the canonical quantization procedure of the graviton in a weak field regime that has already been developed in the literature \cite{Gupta:1952,Bose:Mazumdar:2022}. We promote $h_{\mu\nu}$ to an operator $\hat{h}_{\mu\nu}$ and write $\hat{h}_{\mu\nu}=\hat{\gamma}_{\mu\nu}-\frac{1}{2}\eta_{\mu\nu}\hat{\gamma}$, where $\hat{\gamma}_{\mu\nu}$ and $\hat{\gamma}$ are distinct modes that can be identified as the transversal spin-2 and longitudinal spin-0 ones. In the following they are treated as independent variables for the purpose of the computations. Note that the dependence of these quantities on the coordinates $x^\mu$ is suppressed for the sake of brevity when it is not explicitly required.

In this perturbative regime we can expand the fields $\hat{\gamma}_{\mu\nu}$ and $\hat{\gamma}$ in terms of Fourier modes, and obtain
\begin{subequations}\label{metric:components:perturbation:quantization:appendix}
    \begin{align}
\hat{\gamma}_{\mu\nu}=&\mathcal{A}\int \textrm{d}^3k\frac{\sqrt{\hbar}}{\sqrt{2(2\pi)^3\omega^{\textrm{G}}_{\boldsymbol{k}}}}\left[\hat{P}_{\mu\nu}(\boldsymbol{k})e^{i\boldsymbol{k}\cdot\boldsymbol{x}}+\hat{P}_{\mu\nu}^\dag(\boldsymbol{k}) e^{-i\boldsymbol{k}\cdot\boldsymbol{x}}\right],\\
\hat{\gamma}=&2\mathcal{A}\int \textrm{d}^3k\frac{\sqrt{\hbar}}{\sqrt{2(2\pi)^3\omega^{\textrm{G}}_{\boldsymbol{k}}}}\left[\hat{P}(\boldsymbol{k}) e^{i\boldsymbol{k}\cdot\boldsymbol{x}}+\hat{P}^\dag(\boldsymbol{k}) e^{-i\boldsymbol{k}\cdot\boldsymbol{x}}\right].
\end{align}
\end{subequations}
Here we have introduced the coupling constant $\mathcal{A}:=\sqrt{16\pi G}=\sqrt{16\pi}/m_\textrm{P}$, the graviton frequency $\omega^{\textrm{G}}_{\boldsymbol{k}}:=|\boldsymbol{k}|$, while $\hat{P}_{\mu\nu}(\boldsymbol{k})$ and $\hat{P}_{\mu\nu}^\dag(\boldsymbol{k})$ denote the spin-$2$ graviton  annihilation and the creation operator respectively, and analogously for the spin-$0$ operators $\hat{P}(\boldsymbol{k})$ and $\hat{P}^\dagger(\boldsymbol{k})$. The constant $\hbar$ has been restored for completeness of presentation.

These expressions must be supplemented by the only nonvanishing commutation relations 
\begin{subequations}
    \begin{align}
[\hat{P}_{\mu\nu}(\boldsymbol{k}),\hat{P}_{\mu'\nu'}^\dag(\boldsymbol{k}')]=&(\eta_{\mu\mu'}\eta_{\nu\nu'}+\eta_{\mu\nu'}\eta_{\mu'\nu})\delta^3(\boldsymbol{k}-\boldsymbol{k}'),\\
[\hat{P}(\boldsymbol{k}),\hat{P}^\dag(\boldsymbol{k}')]=&-\delta^3(\boldsymbol{k}-\boldsymbol{k}'),
\end{align}\label{commutation:relations:graviton:appendix}
\end{subequations}
and we provide also the following expression for later convenience:
{\small
\begin{subequations}
\begin{align}
    \left[\hat{h}_{\mu\nu}(t,\boldsymbol{x}),\hat{h}_{\mu'\nu'}(t',\boldsymbol{x}')\right]=&8\pi G i\,\frac{\Gamma_{\mu\nu\mu'\nu'}}{|\boldsymbol{x}-\boldsymbol{x}'|}\,\left[\delta((t-t')+|\boldsymbol{x}-\boldsymbol{x}'|)-\delta((t-t')-|\boldsymbol{x}-\boldsymbol{x}'|)\right],\label{h:field:commutator}\\
    \left\{\hat{h}_{\mu\nu}(t,\boldsymbol{x}),\hat{h}_{\mu'\nu'}(t',\boldsymbol{x}')\right\}=&16 G \frac{\Gamma_{\mu\nu\mu'\nu'}}{-(t-t')^2+|\boldsymbol{x}-\boldsymbol{x}'|^2}\equiv16 G D_{\mu\nu,\mu'\nu'}(x-y),\label{h:field:anticommutator}\\
    \bra{0_\textrm{G}}\hat{h}_{\mu\nu}(t,\boldsymbol{x})\hat{h}_{\mu'\nu'}(t',\boldsymbol{x}')\ket{0_\textrm{G}}=&8 G D_{\mu\nu,\mu'\nu'}(x-y)+4\pi G i\,\frac{\Gamma_{\mu\nu\mu'\nu'}}{|\boldsymbol{x}-\boldsymbol{x}'|}\,\left[\delta((t-t')+|\boldsymbol{x}-\boldsymbol{x}'|)-\delta((t-t')-|\boldsymbol{x}-\boldsymbol{x}'|)\right]\label{h:expectation:value:vacuum:appendix},
\end{align}
\end{subequations}
}

\noindent where here we have defined $\Gamma_{\mu\nu\mu'\nu'}:=\eta_{\mu\mu'}\eta_{\nu\nu'}+\eta_{\mu\nu'}\eta_{\mu'\nu}-\eta_{\mu\nu}\eta_{\mu'\nu'}$ for notational purposes only and $\ket{0_\textrm{G}}$ is the graviton vacuum state. The set of constants $\Gamma_{\mu\nu\mu'\nu'}$ satisfies $\Gamma_{\mu\nu\mu'\nu'}\eta^{\mu\nu}=-2\eta_{\mu'\nu'}$ and that $\Gamma_{\mu\nu\mu'\nu'}\eta^{\mu'\nu'}=-2\eta_{\mu\nu}$. We have also introduced the \emph{graviton propagator}-like notation $D_{\mu\nu,\mu'\nu'}(x-x')$ for convenience \cite{Donoghue:Ivanov:2017,Biswas:Bose:2023}. Note that the graviton propagator proper is defined as the time ordered object $\bra{0_\textrm{G}}\overset{\leftarrow}{\mathcal{T}}\hat{h}_{\mu\nu}(t,\boldsymbol{x})\hat{h}_{\mu'\nu'}(t',\boldsymbol{x}')\ket{0_\textrm{G}}$. 
For later reference we will use the notation 
\begin{equation}\label{tilde:D:appendix}
    \tilde{D}_{\mu\nu,\mu'\nu'}(x-y):=\frac{1}{8G}\bra{0_\textrm{G}}\hat{h}_{\mu\nu}(t,\boldsymbol{x})\hat{h}_{\mu'\nu'}(t',\boldsymbol{x}')\ket{0_\textrm{G}}.
\end{equation}
The commutator \eqref{h:field:commutator} has been obtained by employing the identity
\begin{equation*}
    \int \frac{\textrm{d}^3k}{(2\pi)^3}\frac{\sin\bigl(|\boldsymbol{k}|t+\boldsymbol{x}\cdot\boldsymbol{k}\bigr)}{|\boldsymbol{k}|}=\frac{1}{4\pi|\boldsymbol{x}|}\left(\delta(|\boldsymbol{x}|-t)-\delta(|\boldsymbol{x}|+t)\right).
\end{equation*}
We also note that
\begin{align*}
    &\int\textrm{d}t\textrm{d}t'\left[\delta((t-t')+|\boldsymbol{x}-\boldsymbol{x}'|)-\delta((t-t')-|\boldsymbol{x}-\boldsymbol{x}'|)\right]\\
    &= \int\textrm{d}t\textrm{d}t'(\vartheta(t-t')+\vartheta(t'-t))\left[\delta((t-t')+|\boldsymbol{x}-\boldsymbol{x}'|)-\delta((t-t')-|\boldsymbol{x}-\boldsymbol{x}'|)\right]\\
    &= \int\textrm{d}t\textrm{d}t'\left[\vartheta(t'-t)\delta((t-t')+|\boldsymbol{x}-\boldsymbol{x}'|)-\vartheta(t-t')\delta((t-t')-|\boldsymbol{x}-\boldsymbol{x}'|)\right]\\
    &= \int\textrm{d}t\textrm{d}t'\left[\vartheta(t-t')\delta((t-t')-|\boldsymbol{x}-\boldsymbol{x}'|)-\vartheta(t-t')\delta((t-t')-|\boldsymbol{x}-\boldsymbol{x}'|)\right]=0,
\end{align*}
where to achieve the last line we have swapped the integration variables and used the fact that the Dirac-delta is symmetric. Thus
\begin{align}
    \int\textrm{d}t\textrm{d}t'\tilde{D}_{\mu\nu,\mu'\nu'}(x-y)= \int\textrm{d}t\textrm{d}t'D_{\mu\nu,\mu'\nu'}(x-y),
\end{align}
which will be of use later.

The free normal-ordered Hamiltonian $\hat{H}_{0,\textrm{G}}$ for the graviton therefore reads
\begin{align}\label{free:hamiltonian:gravity:sector:appendix}
\hat{H}_{0,\textrm{G}}=\frac{1}{2}\int \textrm{d}^3k\,\omega^{\textrm{G}}_{\boldsymbol{k}}\,\left[\hat{P}_{\mu\nu}^\dag(\boldsymbol{k})\hat{P}^{\mu\nu}(\boldsymbol{k})-2\hat{P}^\dag(\boldsymbol{k})\hat{P}(\boldsymbol{k})\right],
\end{align}
and it is easy to show that it induces the canonical time evolution for the operators $\hat{P}_{\mu\nu}$ and $\hat{P}$. Recall that $\omega^{\textrm{G}}_{\boldsymbol{k}}=|\boldsymbol{k}|$ for gravitons.

Finally, the number operator reads
\begin{align}\label{number:operator:gravity:sector:appendix}
\hat{N}_\textrm{G}=\frac{1}{2}\int \textrm{d}^3k\left[\hat{P}_{\mu\nu}^\dag(\boldsymbol{k})\hat{P}^{\mu\nu}(\boldsymbol{k})-2\hat{P}^\dag(\boldsymbol{k})\hat{P}(\boldsymbol{k})\right].
\end{align}

%---------------------------%
\subsection{Interaction within linearized gravity}\label{linearized:gravity:appendix}
%---------------------------%
Since we work in the Schr\"odinger picture, we need to find the time evolution operator $\hat{U}(t)$ that implements the dynamics. This requires us to compute the Hamiltonian $\hat{H}$ of the system, which in turn depends on the stress energy tensor (density) $\hat{T}_{\mu\nu}$ that encodes the properties of the matter content of the background. In the lack of other sources of gravity the stress-energy tensor will be given by
\begin{align}
    \hat{T}_{\mu\nu}=&\partial_{\mu}\hat{\phi}\partial_{\nu}\hat{\phi}-\frac{1}{2}\eta_{\mu\nu}\left[\partial^{\rho}\hat{\phi}\partial_{\rho}\hat{\phi}+m^2\hat{\phi}^2\right].
\end{align}\label{stress:energy:tensor:appendix}
Notice that the stress-energy tensor correctly has the dimension of an energy density and we employ the flat-spacetime expression since interaction with gravity is given by the dynamics below.

It is convenient to give a explicit expression of $\hat{T}_{\mu\nu}$ in terms of field operators. It reads
\begin{align}\label{stress:energy:tensor:explicit:appendix}
\hat{T}_{\mu\nu}(t,\boldsymbol{x})=&-\frac{1}{2}\int\frac{\textrm{d}^3k\textrm{d}^3k'}{2(2\pi)^3}\frac{2k_\mu k'_\nu-k^\rho k'_\rho\eta_{\mu\nu}+m^2\eta_{\mu\nu}}{\sqrt{\omega_{\boldsymbol{k}}\omega_{\boldsymbol{k}'}}}(\hat{a}_{\boldsymbol{k}}\hat{a}_{\boldsymbol{k}'}e^{i k_\mu x^\mu}e^{i k'_\mu x^\mu}+\hat{a}_{\boldsymbol{k}}^\dag\hat{a}_{\boldsymbol{k}'}^\dag e^{-i k_\mu x^\mu}e^{-i k'_\mu x^\mu})\nonumber\\
&+\int\frac{\textrm{d}^3k\textrm{d}^3k'}{2(2\pi)^3}\frac{k_\mu k'_\nu+k_\nu k'_\mu-k^\rho k'_\rho\eta_{\mu\nu}-m^2\eta_{\mu\nu}}{\sqrt{\omega_{\boldsymbol{k}}\omega_{\boldsymbol{k}'}}}\hat{a}_{\boldsymbol{k}}^\dag\hat{a}_{\boldsymbol{k}'}e^{-i k_\mu x^\mu}e^{i k'_\mu x^\mu}.
\end{align}

The full Hamiltonian $\hat{H}(t)$ is thus given by 
\begin{align}\label{hamiltonian:perturbative:form:appendix}
\hat{H}:=\hat{H}_0+\epsilon\hat{H}_\textrm{I},
\end{align}
where $\hat{H}_\textrm{I}$ is the interaction Hamiltonian with the $\epsilon$ removed.
Such term has been obtained in the literature \cite{Gupta:1952} and, modulo the $\epsilon$ parameter, it has the expression
\begin{align}\label{interaction:hamiltonian:appendix}
\hat{H}_\textrm{I}=-\frac{1}{2}\int \textrm{d}^3x\, \hat{h}^{\mu\nu}(\boldsymbol{x}):\hat{T}_{\mu\nu}(\boldsymbol{x}):.
\end{align}
This operator satisfies all of the invariance properties required by the theory \cite{Gupta:1952}, and the key aspect here is that this is consistent within perturbation theory only. We take the normal ordered expression $:\hat{T}_{\mu\nu}(x^\sigma):$ for the stress energy tensor to remove some of the spurious divergence associated with the vacuum energy that appear in field theory.
Here we will be using the notation $\hat{H}_\textrm{I}(t):=\hat{U}_0^\dag(t)\hat{H}_\textrm{I}\hat{U}_0(t)\equiv-\frac{1}{2}\int \textrm{d}^3x\, \hat{h}^{\mu\nu}(t,\boldsymbol{x}):\hat{T}_{\mu\nu}(t,\boldsymbol{x}):$

We now seek an expression for the Hamiltonian $\hat{H}_\textrm{I}$ that can be used effectively in our work. It is immediate to use the expressions\eqref{metric:components:perturbation:quantization:appendix} and \eqref{stress:energy:tensor:explicit:appendix} of the metric perturbation and the stress-energy tensor in momentum space that, in the interaction picture, the interaction Hamiltonian $\hat{H}_\textrm{I}(t)$ will be composed of products of operators of the form $\hat{P}_{\mu\nu}(\boldsymbol{k})-\hat{P}(\boldsymbol{k})\eta_{\mu\nu}$, $\hat{a}_{\boldsymbol{k}}\hat{a}_{\boldsymbol{k}'}$, $\hat{a}_{\boldsymbol{k}}^\dag\hat{a}_{\boldsymbol{k}'}$, as well as their Hermitian conjugates. The crucial aspect is that each creation operator in the product will be accompanied by a phase factor of the form $e^{-i k_\mu x^\mu}$, while any annihilation operator will be accompanied by a phase factor of the form $e^{i k_\mu x^\mu}$. We are working in perturbation theory  and therefore we invoke the rotating wave approximation (RWA) \cite{Burgarth:Facchi:2024,Heib:Lageyre:2025}, which allows us to drop the term with the form $(\hat{P}_{\mu\nu}(\boldsymbol{k})-\hat{P}(\boldsymbol{k})\eta_{\mu\nu})\hat{a}_{\boldsymbol{k}}\hat{a}_{\boldsymbol{k}'}$ together with their Hermitian conjugates when considering long enough times.

Therefore, we are left with the Hamiltonian $\hat{H}=\hat{H}_0+\hat{H}^\Phi_{\textrm{I,RWA}}$, where 
\begin{align}\label{rotating:wave:Hamiltonian}
\hat{H}^\Phi_{\textrm{I,RWA}}=&\frac{\sqrt{\pi\hbar G}}{2}\int \frac{\textrm{d}^3k\textrm{d}^3k'}{\sqrt{2(2\pi)^3\omega_{\boldsymbol{k}}\omega_{\boldsymbol{k}'}|\boldsymbol{k}+\boldsymbol{k}'|}}(2k_\mu k'_\nu-k^\rho k'_\rho\eta_{\mu\nu}+m^2\eta_{\mu\nu})(\hat{P}^{\mu\nu}(\boldsymbol{k}+\boldsymbol{k}')-\hat{P}(\boldsymbol{k}+\boldsymbol{k}')\eta^{\mu\nu})\hat{a}_{\boldsymbol{k}}^\dag\hat{a}_{\boldsymbol{k}'}^\dag \nonumber\\
&+\frac{\sqrt{\pi\hbar G}}{2}\int \frac{\textrm{d}^3k\textrm{d}^3k'}{\sqrt{2(2\pi)^3\omega_{\boldsymbol{k}}\omega_{\boldsymbol{k}'}|\boldsymbol{k}+\boldsymbol{k}'|}}(2k_\mu k'_\nu-k^\rho k'_\rho\eta_{\mu\nu}+m^2\eta_{\mu\nu})(\hat{P}^{\mu\nu}{}^\dag(\boldsymbol{k}+\boldsymbol{k}')-\hat{P}^\dag(\boldsymbol{k}+\boldsymbol{k}')\eta^{\mu\nu})\hat{a}_{\boldsymbol{k}}\hat{a}_{\boldsymbol{k}'} \nonumber\\
&-\sqrt{\pi\hbar G}\int \frac{\textrm{d}^3k\textrm{d}^3k'}{\sqrt{2(2\pi)^3\omega_{\boldsymbol{k}}\omega_{\boldsymbol{k}'}|\boldsymbol{k}-\boldsymbol{k}'|}}(k_\mu k'_\nu+k_\nu k'_\mu-k^\rho k'_\rho\eta_{\mu\nu}-m^2\eta_{\mu\nu})\nonumber\\
&\times\left[\hat{P}^{\mu\nu}(\boldsymbol{k}-\boldsymbol{k}')-\hat{P}(\boldsymbol{k}-\boldsymbol{k}')\,\eta^{\mu\nu}+\hat{P}^{\mu\nu}{}^\dag(\boldsymbol{k}'-\boldsymbol{k})-\hat{P}^\dag(\boldsymbol{k}'-\boldsymbol{k})\,\eta^{\mu\nu}\right]\hat{a}_{\boldsymbol{k}}^\dag\hat{a}_{\boldsymbol{k}'}.
\end{align}
Here $\hbar$ has been restored for the sake of completeness.

From now on, when we mention the interaction Hamiltonian we implicitly assume that it is written in the rotating wave approximation, and thus the subscript is dropped for convenience.

%---------------------------%
\section{CHARACTERIZING THE INITIAL STATE OF THE SYSTEM}\label{Supp:Mat:Initial:State}
%---------------------------%
The next step is to characterize the particle and its state. This requires us to provide a prescription to localize field excitations, together with a way to define states of such localized objects. The theory of localized field excitations can be formalized in a more abstract and elegant fashion than the one employed here, and we leave the interested reader to the relevant literature \cite{Hollands:Wald:2015,Mieling:2026}.

%---------------------------%
\subsection{Localized particles and localized particle states}
%---------------------------%
We start by introducing the localized objects together with the initial states of interest.
To do so we first note that we wish to consider a quantum state of a single localized particle excitation that can be found in one of two positions called left (L) and right (R). Such localized states can be constructed as follows. First, we introduce the \emph{smeared} (or \emph{extended}) operators
\begin{subequations}\label{extended:operators:appendix}
\begin{align}
\hat{a}_\textrm{L}:=&\int \textrm{d}^3k\, F_{\boldsymbol{k}_0}(\boldsymbol{k})\,e^{i\,\boldsymbol{x}_\textrm{L}\cdot\boldsymbol{k}}\,\hat{a}_{\boldsymbol{k}}\\
\hat{a}_\textrm{R}:=&\int \textrm{d}^3k\, F_{\boldsymbol{k}_0}(\boldsymbol{k})\,e^{i\,\boldsymbol{x}_\textrm{R}\cdot\boldsymbol{k}}\,\hat{a}_{\boldsymbol{k}},
\end{align}
\end{subequations}
that can be used to create a \textit{localized} single particle with momentum distribution $F_K(\boldsymbol{k}):=F_{\boldsymbol{k}_0}(\boldsymbol{k})e^{i\,\boldsymbol{x}_K\cdot\boldsymbol{k}}$. 

We then require that
\begin{equation*}
    [\hat{a}_K,\hat{a}_{K'}^\dag]=\delta_{KK'},
\end{equation*}
which implies that the functions $F_K(\boldsymbol{k})$ are (ortho-)normalized in the sense that $\int \mathrm{d}^3k\,|F_{\boldsymbol{k}_0}(\boldsymbol{k})|^2e^{i\,(\boldsymbol{x}_K-\boldsymbol{x}_{K'})\cdot\boldsymbol{k}}=\delta_{KK'}$. Here, as well as in the remainder of this work, we use the subscript $K$ that can take values $K=\textrm{L,R}$. 

We can now introduce the states $|1_K\rangle$ as
\begin{subequations}
\begin{align}
|1_\textrm{L}\rangle:=\hat{a}^\dag_\textrm{L}|0\rangle=&\int \textrm{d}^3k\, F_{\boldsymbol{k}_0}^*(\boldsymbol{k})\,e^{-i\,\boldsymbol{x}_\textrm{L}\cdot\boldsymbol{k}}\,\hat{a}^{\dag}_{\boldsymbol{k}}|0\rangle,\\
|1_\textrm{R}\rangle:=\hat{a}^\dag_\textrm{R}|0\rangle=&\int \textrm{d}^3k\, F_{\boldsymbol{k}_0}^*(\boldsymbol{k})\,e^{-i\,\boldsymbol{x}_\textrm{R}\cdot\boldsymbol{k}}\,\hat{a}^{\dag}_{\boldsymbol{k}}|0\rangle.
\end{align}
\end{subequations}
The commutation properties of the extended operators imply that the states $|1_\textrm{L}\rangle$ and $|1_\textrm{R}\rangle$ form and orthonormal set, i.e., $\langle1_K|1_{K'}\rangle=\delta_{KK'}$.

We note that the original basis $\mathcal{B}=\{u_{\boldsymbol{k}}\}$ of plane-waves $\boldsymbol{k}$ that was used to write the expansion of the field \eqref{field:expansion:appendix} is infinite dimensional. Thus, we will not be able to decompose the full space of modes of matter using $F_\textrm{L}$ and $F_\textrm{R}$ alone. We therefore need to introduce the functions $F_{\underline{\lambda}}$ who, together with $\{F_K\}|_{K=\textrm{L},\textrm{R}}$, form the orthonormal basis $\mathcal{B}_\textrm{ext}=\{F_\textrm{L},F_\textrm{R},\{F_{\underline{\lambda}}\}\}$ of extended modes. Here, $\underline{\lambda}$ is an appropriate label for the remaining modes, which we also call \emph{system environment} modes, and we have $\int\textrm{d}^3k F_{\underline{\lambda}}^*(\boldsymbol{k})  F_K(\boldsymbol{k}) =0$ while $\int\textrm{d}^3k F_{\underline{\lambda}}^*(\boldsymbol{k}) F_{\underline{\lambda}'}(\boldsymbol{k}) =\delta_{\underline{\lambda}\underline{\lambda}'}$.
We can then also define the extended operators $\hat{a}_{\underline{\lambda}}$ of the system environment in the analogous fashion to \eqref{extended:operators:appendix}, which in turn allows us to obtain the states $\ket{1_{\underline{\lambda}}}:=\int \textrm{d}^3k\, F_{\underline{\lambda}}^*(\boldsymbol{k})\,\hat{a}^{\dag}_{\boldsymbol{k}}|0\rangle$ that satisfy $\langle1_{\underline{\lambda}}|1_{\underline{\lambda}}'\rangle=\delta_{\underline{\lambda}\underline{\lambda}'}$, as well as $\langle1_{\underline{\lambda}}|1_K\rangle=0$.

Given all of the considerations above, we say that a particle state with $m_{\textrm{L}}$ excitations in the left mode, $m_{\textrm{R}}$ excitations in the right mode, and $m_{\underline{\lambda}}$ excitations in each environment mode $\underline{\lambda}$, in this notation reads $|m_{\textrm{L}}m_{\textrm{R}}\{m_{\underline{\lambda}}\}\rangle$. This notation will become convenient below. Conveniently, $|m_{\textrm{L}}m_{\textrm{R}}0_\textrm{E}\rangle\equiv|m_{\textrm{L}}m_{\textrm{R}}\rangle$ are states that contain solely particles in the system left and right modes.

\vspace{0.4cm}

\noindent \textit{Characterization of the particle in momentum space}---Given the distribution $F_{\boldsymbol{k}_0}(\boldsymbol{k})$ we first define the function $\rho_{\boldsymbol{k}_0}(\boldsymbol{k})\equiv\rho(\boldsymbol{k}-\boldsymbol{k}_0):=|F_{\boldsymbol{k}_0}(\boldsymbol{k})|^2=|F(\boldsymbol{k}-\boldsymbol{k}_0)|^2$, where the lack of subscript in any of the smeared functions implies a vanishing first moment $\bar{\boldsymbol{k}}$, defined as
\begin{align}
    \bar{\boldsymbol{k}}:=\int \textrm{d}^3k\, \boldsymbol{k}\, \rho_{\boldsymbol{k}_0}(\boldsymbol{k}).
\end{align}
The momentum distributions considered here will have a characteristic size $\sigma$ and a peak value $\boldsymbol{k}_0$, and thus we find
\begin{align}
    \bar{\boldsymbol{k}}
    =\int \textrm{d}^3k\, \boldsymbol{k}\, \rho(\boldsymbol{k}-\boldsymbol{k}_0)
    =\int \textrm{d}^3k\, (\boldsymbol{k}+\boldsymbol{k}_0)\, \rho(\boldsymbol{k})=\boldsymbol{k}_0.
\end{align}
Thus, $\boldsymbol{k}_0$ will be referred to as the \emph{average momentum}.

The variance $\Delta k$ of the distribution is defined by
\begin{align}
    (\Delta k)^2:=\int \textrm{d}^3k\, |\boldsymbol{k}|^2\, \rho_{\boldsymbol{k}_0}(\boldsymbol{k}/\sigma)-|\bar{\boldsymbol{k}}|^2,
\end{align}
where $\sigma$ is the scale of the momentum distribution.

In our case we have $\boldsymbol{k}_0=0$ and thus find $(\Delta k)^2=\sigma^2\int \textrm{d}^3u\, |\boldsymbol{u}|^2\, \rho'(\boldsymbol{u})=\sigma^2\Gamma^2_\rho$, which implies
\begin{align}
    \Delta k=\sigma\Gamma_\rho,
\end{align}
where we have introduced the form-factor $\Gamma_\rho=(\int \textrm{d}^3u\, |\boldsymbol{u}|^2\, \rho'(\boldsymbol{u}))^{1/2}$, where $\boldsymbol{u}:=\boldsymbol{k}/\sigma$ is a dimensionless variable, and we note that $\rho'(\boldsymbol{u})=\sigma^3\rho(\boldsymbol{u})$ is a distribution without physical dimension. The form factor $\Gamma_\rho$ is solely determined by the exact functional form of the momentum distribution, and is therefore a numerical constant.

\vspace{0.4cm}

\noindent \textit{Characterization of the particle in configuration space}---We now characterize the particle in configuration space. Given the momentum amplitude function $F_K(\boldsymbol{k})=F_{\boldsymbol{k}_0}(\boldsymbol{k})e^{i\,\boldsymbol{x}_K\cdot\boldsymbol{k}}$ of the particle excitations, we can seek their spatial amplitude by means of Fourier transform. We have $\tilde{F}_{\boldsymbol{x}_K}(\boldsymbol{x})\equiv\tilde{F}(\boldsymbol{x}-\boldsymbol{x}_K):=\int \frac{\textrm{d}^3k}{(2\pi)^3} e^{-i\,\boldsymbol{k}\cdot(\boldsymbol{x}-\boldsymbol{x}_K)}F_{\boldsymbol{k}_0}(\boldsymbol{k})=e^{-i\,\boldsymbol{k}_0\cdot(\boldsymbol{x}-\boldsymbol{x}_K)}\int \frac{\textrm{d}^3k}{(2\pi)^3} e^{-i\,\boldsymbol{k}\cdot(\boldsymbol{x}-\boldsymbol{x}_K)}F(\boldsymbol{k})$, which is a function in position space of characteristic and average position that we now proceed to determine. For later convenience, we define $\tilde{\rho}^\textrm{s}_{\boldsymbol{x}_K}(\boldsymbol{x})\equiv\tilde{\rho}^\textrm{s}(\boldsymbol{x}-\boldsymbol{x}_K):=(2\pi)^3|\tilde{F}(\boldsymbol{x}-\boldsymbol{x}_K)|^2$ as the probability density of finding the particle at position $\boldsymbol{x}_K$, which is normalized by $\int \textrm{d}^3x\,\tilde{\rho}^\textrm{s}_{\boldsymbol{x}_K}(\boldsymbol{x})=1$. Crucially, note that the spatial probability density $\tilde{\rho}^\textrm{s}(\boldsymbol{x})$ does \emph{not} coincide with the Fourier transform of the momentum density, defined as $\tilde{\rho}(\boldsymbol{x}):=\int\frac{\textrm{d}^3k}{(2\pi)^3}\rho(\boldsymbol{k})e^{i \boldsymbol{k}\cdot\boldsymbol{x}}$.

We define the average position $\bar{\boldsymbol{x}}$ as
\begin{align}
    \bar{\boldsymbol{x}}:=\int \textrm{d}^3x\, \boldsymbol{x}\, \tilde{\rho}^\textrm{s}_{\boldsymbol{x}_K}(\boldsymbol{x}).
\end{align}
This allows us to compute
\begin{align*}
\bar{\boldsymbol{x}}=\int \textrm{d}^3x\, \boldsymbol{x}\, \tilde{\rho}^\textrm{s}(\boldsymbol{x}-\boldsymbol{x}_K)
=\int \textrm{d}^3x\, (\boldsymbol{x}+\boldsymbol{x}_K)\, \tilde{\rho}^\textrm{s}(\boldsymbol{x})
=\boldsymbol{x}_K\int \textrm{d}^3x\, \tilde{\rho}^\textrm{s}(\boldsymbol{x})+\int \textrm{d}^3x\, \boldsymbol{x}\, \tilde{\rho}^\textrm{s}(\boldsymbol{x})
=\boldsymbol{x}_K,
\end{align*}
where the last step is obtained by realizing that $\int \textrm{d}^3x\, \boldsymbol{x}\, \tilde{\rho}^\textrm{s}(\boldsymbol{x})=0$ by definition, since $\tilde{\rho}^\textrm{s}(\boldsymbol{x})$ is a distribution with vanishing first moment.
Thus, we have that $\bar{\boldsymbol{x}}=\boldsymbol{x}_K$ and in the following we will refer to $\boldsymbol{x}_K$ as the average position of the particle.

We can also compute the characteristic size $l_0$ of the particle as the variance of the distribution $\tilde{\rho}^\textrm{s}(\boldsymbol{x}-\boldsymbol{x}_K)$, as done above for the momentum distribution. We introduce
\begin{align}
    l_0^2:=\int \textrm{d}^3x\, |\boldsymbol{x}|^2\, \tilde{\rho}^\textrm{s}(\boldsymbol{x}-\boldsymbol{x}_K)-|\bar{\boldsymbol{x}}|^2.
\end{align}
Simple algebra allows us to obtain
\begin{align*}
    l_0^2
    =&\int \textrm{d}^3x\, |\boldsymbol{x}|^2\, \tilde{\rho}^\textrm{s}(\boldsymbol{x}-\boldsymbol{x}_K)-|\boldsymbol{x}_K|^2\\
    =&\int \textrm{d}^3x\, (|\boldsymbol{x}|^2+|\boldsymbol{x}_K|^2+2\boldsymbol{x}\cdot\boldsymbol{x}_K)\, \tilde{\rho}^\textrm{s}(\boldsymbol{x})-|\boldsymbol{x}_K|^2\\
    =&\int \textrm{d}^3x\, |\boldsymbol{x}|^2\, \tilde{\rho}^\textrm{s}(\boldsymbol{x})+|\boldsymbol{x}_K|^2\,\int \textrm{d}^3x\,  \tilde{\rho}^\textrm{s}(\boldsymbol{x})+2\boldsymbol{x}_K\cdot\int \textrm{d}^3x\, \boldsymbol{x}\, \tilde{\rho}^\textrm{s}(\boldsymbol{x})-|\boldsymbol{x}_K|^2\\
    =&\int \textrm{d}^3x\, |\boldsymbol{x}|^2\, \tilde{\rho}^\textrm{s}(\boldsymbol{x})
     =\frac{1}{\sigma^2}\tilde{\Gamma}_{\tilde{\rho}^\textrm{s}}^2,
\end{align*}
where to obtain the last line we have normalized all variables in the integrals appropriately by $\sigma$. Here, the new form factor reads $\tilde{\Gamma}_{\tilde{\rho}^\textrm{s}}:=(\int \textrm{d}^3u\, |\boldsymbol{u}|^2\, \tilde{\rho}^{\textrm{s}\prime}(\boldsymbol{u}))^{1/2}$, where $\boldsymbol{u}$ is a dimensionless variable and $\tilde{\rho}^{\textrm{s}\prime}(\boldsymbol{u})=\sigma^{-3}\tilde{\rho}^\textrm{s}(\boldsymbol{u})$ is a dimensionless function. The form factor $\tilde{\Gamma}_{\tilde{\rho}^\textrm{s}}$ is solely determined by the exact functional form of the spatial distribution, and is therefore a numerical constant.

Thus, we have found that
\begin{align}
l_0=\sigma^{-1}\tilde{\Gamma}_{\tilde{\rho}^\textrm{s}}.
\end{align}
We will then refer to $l_0$ as the \emph{effective size of the particle}. Using the expression for $\Delta k$ we find
\begin{align}
l_0=\mu\frac{\hbar}{\Delta k}
\end{align}
which, modulo the numerical ratio $\mu:=\frac{\tilde{\Gamma}_{\tilde{\rho}^\textrm{s}}}{\Gamma_\rho}$, gives the expected inverse relation between the effective size $l_0$ of the particle in configuration space and the spread $\Delta k$ of the momentum distribution in momentum space.

\begin{figure}[ht!]
\includegraphics[width=0.5\linewidth]{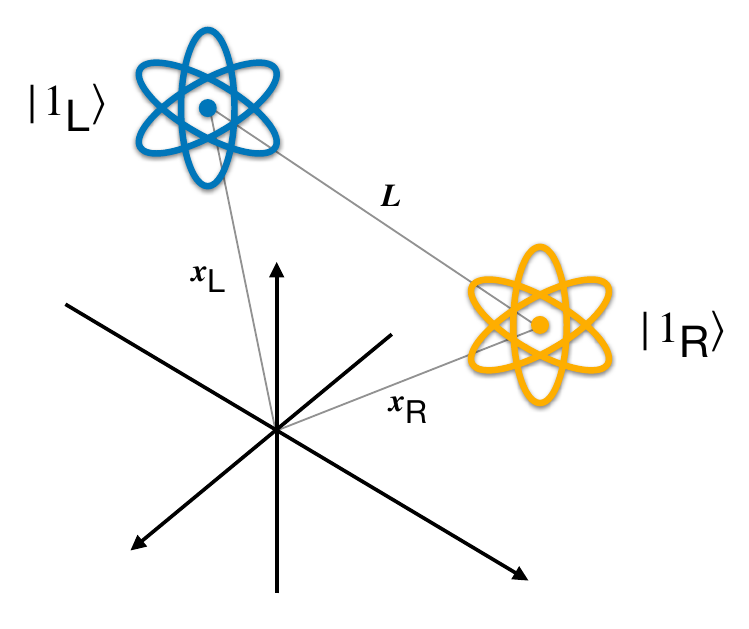}
\caption{Pictorial representation of the extended particle states. The particle can be localized on the Left `L' or Right `R', which label two arbitrary positions for convenience and without loss of generality. Each configuration is assigned a vector $|1_\textrm{L}\rangle$ or $|1_\textrm{R}\rangle$ in the Hilbert space. The distance from the origin of both initial locations is $L=|\boldsymbol{L}|\equiv|\boldsymbol{x}_\textrm{R}-\boldsymbol{x}_\textrm{L}|$.}\label{Fig:LR}
\end{figure}

Finally, we also compute the distance between the two spatial configurations, denoted as $L=|\boldsymbol{L}|$, where $\boldsymbol{L}$ is obtained from 
\begin{equation}
    \boldsymbol{L}:=\int \textrm{d}^3x\, \boldsymbol{x}\,\tilde{\rho}^\textrm{s}_{\boldsymbol{x}_
    \textrm{R}}(\boldsymbol{x})-\int \textrm{d}^3x\, \boldsymbol{x}\,\tilde{\rho}^\textrm{s}_{\boldsymbol{x}_
    \textrm{L}}(\boldsymbol{x})\equiv\boldsymbol{x}_\textrm{R}-\boldsymbol{x}_\textrm{L}
\end{equation}
as expected.
A depiction of the possible position configurations can be found in Figure~\ref{Fig:LR}.

\vspace{0.4cm}

\noindent \textit{Gaussian momentum distribution}---In order to obtain analytical results for a better understanding of the system, we assume that the momentum distribution $F$ is a Gaussian as a function of the modulus of the momentum. The normalization condition $\int \textrm{d}^3k |F(\boldsymbol{k})|^2=1$ implies that
\begin{align}
    F(\boldsymbol{k})=\frac{1}{\sqrt{2\pi\sqrt{2\pi}\sigma^3}}e^{-\frac{|\boldsymbol{k}|^2}{4\sigma^2}}
\end{align}
is the sought-after normalized function in $3+1$ dimensions.
Note that, crucially, $F(\boldsymbol{k}+\boldsymbol{k}')F(\boldsymbol{k}-\boldsymbol{k}')=F(\sqrt{2}\boldsymbol{k})F(\sqrt{2}\boldsymbol{k}')$ for this choice of function and this property will allow for a great simplification of the expressions as we will see below.

%---------------------------%
\subsection{Initial state of a single localized excitation}
%---------------------------%
We are now in the position of introducing the initial state of the whole system.
We consider the following (family of) initial states
\begin{align} \label{global:initial:state:appendix}
\hat{\rho}(0)=&\hat{\rho}_\textrm{S}(0)\otimes\left(\bigotimes_{\underline{\lambda}}|0_{\underline{\lambda}}\rangle\langle 0_{\underline{\lambda}}|\right)\otimes \hat{\rho}_\textrm{G}(0)\equiv\hat{\rho}_{\textrm{SE}}(0)\otimes \hat{\rho}_\textrm{G}(0),
\end{align}
where the initial reduced state of the particle system S reads
\begin{align} \label{system:initial:states:appendix}
\hat{\rho}_\textrm{S}(0)=&\alpha\ket{1_\textrm{L}}\bra{1_\textrm{L}}+(1-\alpha)\ket{1_\textrm{R}}\bra{1_\textrm{R}}+\beta\ket{1_\textrm{L}}\bra{1_\textrm{R}}+\beta^*\ket{1_\textrm{R}}\bra{1_\textrm{L}}\,,
\end{align}
while the reduced state of the gravity sector reads $\hat{\rho}_\textrm{G}(0):=|0_\textrm{G}\rangle\langle0_\textrm{G}|$. This latter choice allows us to associate all effects solely to self gravity of the particle.

We note that the defining parameters of \eqref{system:initial:states:appendix} cannot be chosen completely arbitrarily. In particular, in order to have a valid physical state we require $0\leq\alpha\leq1$, $-1/2\leq|\beta|\leq1/2$, and $(\alpha-1/2)^2+|\beta|^2\leq1/4$. It is useful to recall that in the case $\alpha=1/2$ and $\beta=0$ one has a maximally mixed state $\hat{\rho}_\textrm{S}(0)=\frac{1}{2}\ket{1_\textrm{L}}\bra{1_\textrm{L}}+\frac{1}{2}\ket{1_\textrm{R}}\bra{1_\textrm{R}}$, while for $\alpha=|\beta|=1/2$ one has a pure state $\hat{\rho}_\textrm{S}(0)=\ket{\theta}\bra{\theta}$ with $\ket{\theta}=\frac{1}{\sqrt{2}}(\ket{1_\textrm{L}}+e^{-i\theta}\ket{1_\textrm{R}})$ and $\theta=\arg(\beta)$. 

The state $\hat{\rho}_{\textrm{SE}}(0)$ is constructed starting from the reduced state $\hat{\rho}_\textrm{S}(0)$ of the system S and adding the tensor product the vacuum for \textit{all} other modes of excitation labeled by $\underline{\lambda}$.
In particular, we note that the vacuum state $|0_\textrm{SE}\rangle$ of the matter sector reads
\begin{equation}
   |0_\textrm{SE}\rangle=|0_\textrm{S}\rangle\otimes |0_\textrm{E}\rangle=|0_\textrm{S}\rangle \otimes\left(\underset{\underline{\lambda}}{\otimes}|0_{\underline{\lambda}}\rangle\right),
\end{equation}
where we have conveniently introduced $|0_\textrm{E}\rangle:= \underset{\underline{\lambda}}{\otimes}|0_{\underline{\lambda}}\rangle$.

%---------------------------%
\subsection{Massive and static particle regime}\label{ssec: massive static particles}
%---------------------------%
We continue by introducing here the \textit{massive} and \textit{static} regime for the states \eqref{system:initial:states:appendix} of interest, a regime that has also been considered in the literature \cite{Joos:Zeh:1985,Blencowe:2013,Anastopoulos:Hu:2013,Bose:Mazumdar:2022}.
It is defined as follows:

\vspace{0.4cm}

\noindent \textit{Static particles}---We say that a particle is static if its average momentum vanishes, and the variance of the momentum distribution is small with respect to an appropriate energy scale. 

As demonstrated above, the average momentum is $\boldsymbol{k}_0$, and thus in the remainder of this work we will assume that $\boldsymbol{k}_0=0$. Determining that the variance $\Delta k=\sigma\Gamma_f$ of the momentum distribution is small in a meaningful way cannot be done using the parameters of the distribution alone, since the average momentum vanishes and it cannot provide a momentum scale. Thus, we need to meaningfully compare $\Delta k$ with another relevant quantity, which we do below.

\vspace{0.4cm}

\noindent \textit{Massive particles}---We say that a particle is massive if its rest energy is much larger than its kinetic energy. Concretely, since the average momentum in our case vanishes, we define the massive regime as that where
\begin{align}
    mc^2\gg\hbar\,c\,\Delta k,
\end{align}
where $\Delta k=\sigma\Gamma_\rho$ is the variance of the momentum distribution. Here the dimensions have been restored for clarity.
In this regime we also have that
\begin{align*}
\hbar\,\omega_{\boldsymbol{k}}=\sqrt{ m^2c^4+\hbar^2\,c^2|\boldsymbol{k}|^2}\approx mc^2\left(1+\frac{1}{2}\frac{(\Delta k)^2}{m^2c^2}\right)
\end{align*}
to lowest order for particles whose momentum is well localized around zero with variance $\Delta k$.

We have thus found the desired condition: we require $\frac{\Delta k}{m}\ll1$, where we recall that $\Delta k\propto\sigma$ is the effective size of the momentum distribution. Thus, we equivalently require that $\frac{\sigma}{m}\ll1$ when the form factor is a number that does not invalidate this equivalent statement. The static regime is equivalent to the non-relativistic regime \cite{Anastopoulos:Hu:2014,Bose:Mazumdar:2022}. 

It is not difficult to see that, in this regime, we have $\frac{1}{m}\hat{H}_{0,\phi}\ket{1_K}=\ket{1_K}+\mathcal{O}((\Delta k/m)^2)$, which in turn implies that $E_0:=\textrm{Tr}(\hat{H}_{0,\phi}\hat{\rho}(0))\approx m$ to lowest order. Once more, this is consistent with the starting point, namely that, in the massive static regime, the particle has energy $E_0=mc^2$. Thus, we can conclude that
\begin{align}\label{massive:approximation:main:property}
\hat{U}_0(t)|n_\textrm{L}n_\textrm{R}\rangle\approx&e^{-i\,(n_\textrm{L}+n_\textrm{R})mt}|n_\textrm{L}n_\textrm{R}\rangle
\end{align}
for a time $t$ much smaller than $t_{\textrm{diff}}$, where $|n_\textrm{L}n_\textrm{R}\rangle\equiv|n_\textrm{L}n_\textrm{R}\{0_{\underline{\lambda}}\}\rangle$ and
\begin{align}\label{diffusion:timescale:appendix}
t_{\textrm{diff}}:=\frac{\ell_0^2m}{\hbar}
\end{align}
is the natural time-scale that determines the validity of the massive static approximation, which we call the \emph{diffusion time}. 
Note that a more precise definition of the diffusion timescale would be 
\begin{align*}
t_{\textrm{diff}}:=\frac{2\hbar m}{(\Delta k)^2}=\frac{2}{\mu^2}\frac{\ell_0^2m}{\hbar}    
\end{align*}
Above we have dropped the numerical constant $2/\mu^2$ since it is highly dependent on the shape of the particle and does not add information on the fundamentals of the theory. Equivalently, we are assuming that $\mu\approx\mathcal{O}(1)$ and that \eqref{diffusion:timescale:appendix} gives the correct qualitative understanding of the phenomena of interest. Of course, the more correct expression can be used when necessary, for example if $\mu$ is a very large or small number.

The eigenvalue equation \eqref{massive:approximation:main:property} implies that the wave packets that define the particle excitations do not spread in time at least for times (significantly) smaller than $t_{\textrm{G}}$, a feature expected for a generic wave-packet in quantum field theory \cite{Ford:Oconnell:2002,Srednicki:2007,Edery:2021}. Most importantly, the fact that particle states $|pn\{0_{\underline{\lambda}}\}\rangle$ are eigenstates of the operator $\hat{U}^{(0)}(t)$ implies that the initial reduced state $\hat{\rho}_{\textrm{SE}}(0)$ of the System and Environment is invariant under the free evolution $\hat{U}^{(0)}(t)$, that is, it evolves to lowest order as
\begin{align}\label{useful:lowest:order:evolution:appendix}
\hat{U}^{(0)}(t)\,\hat{\rho}_{\textrm{SE}}(0)\,\hat{U}^{(0)\dag}(t)\approx\hat{\rho}_{\textrm{SE}}(0)\,.
\end{align}
We interpret this as the fact that massive static particles do not move in time.

%---------------------------%
\section{TIME EVOLUTION}\label{suppl_sec_3}
%---------------------------%
In this section we build up the technology necessary to compute the time evolution of the system.
The time evolution of a quantum system is given by the von Neumann equation $\hat{\rho}(t)=\hat{U}(t)\,\hat{\rho}(0)\,\hat{U}^\dag(t)$, where the time evolution operator $\hat{U}(t)$ reads
\begin{align} \label{time:evolution:operator:appendix}
\hat{U}(t)=\overset{\leftarrow}{\mathcal{T}}\,\exp\left[-i\,\int_0^t\,dt'\,\hat{H}(t')\right],
\end{align}
and $\hat{H}(t)=\hat{H}_0+\hat{H}_\textrm{I}$ is the Hamiltonian of the system. In the present case, the starting full Hamiltonian $\hat{H}$ is time independent and therefore $\hat{U}(t)=e^{-i\hat{H}t}$. Here, $\overset{\leftarrow}{\mathcal{T}}$ is the time-ordering operator.

In the following we will need to trace over the graviton subsector in order to obtain the particle's reduced state. To do this, we first need introduce tools to perform this operation.

\vspace{0.4cm}

 %---------------------------%
\subsection{Identity operators and trace in the one-graviton and one-particle subspace}
%---------------------------%

\noindent \textit{Identity operator and trace in the one-graviton}---A first technical issue that needs to be resolved is that of the identity operator in the one-graviton subspace. The idea is that if we have a state $\ket{\Psi}$ that contains one graviton, i.e., $\hat{N}_\textrm{G}\ket{\Psi}=\ket{\Psi}$, we would like to have an expression for the operator $\mathds{1}_{1\textrm{p,G}}$ that acts according to $\mathds{1}_{1\textrm{p,G}}\ket{\Psi}=\ket{\Psi}$ for any single-graviton state $\ket{\Psi}$.
To achieve our goal, let us introduce 
\begin{align}
    |1_{F,F'}\rangle:=\int\textrm{d}^3k \left[F^{\mu\nu}(\boldsymbol{k})\hat{P}_{\mu\nu}^\dag(\boldsymbol{k})+F'(\boldsymbol{k})\hat{P}^\dag(\boldsymbol{k})\right]\ket{0_\textrm{G}},
\end{align}
where $F_{\mu\nu}(\boldsymbol{k})=F_{\nu\mu}(\boldsymbol{k})$ and $F'(\boldsymbol{k})$ are functions normalized by the constraint $\langle1_{F,F'}|1_{F,F'}\rangle=\int\textrm{d}^3k [2F^{\mu\nu*}(\boldsymbol{k})F_{\mu\nu}(\boldsymbol{k})-|F'(\boldsymbol{k})|^2]=1$.
It is immediate to employ the number operator \eqref{number:operator:gravity:sector:appendix} and the commutation relation \eqref{commutation:relations:graviton:appendix} to see that
\begin{align*}
\hat{N}_\textrm{G}\hat{P}^\dag_{\mu\nu}\ket{0_\textrm{G}}=\hat{P}^\dag_{\mu\nu}\ket{0_\textrm{G}},
\qquad
\textrm{and}
\qquad
\hat{N}_\textrm{G}\hat{P}^\dag\ket{0_\textrm{G}}=\hat{P}^\dag\ket{0_\textrm{G}},
\end{align*}
and therefore
\begin{align*}
    \hat{N}_\textrm{G}|1_{F,F'}\rangle=|1_{F,F'}\rangle.
\end{align*}
This means that $|1_{F,F'}\rangle$ is a good candidate for the generic one-graviton state.

We therefore introduce the operator 
\begin{align}\label{identity:operator:one:graviton:subspace:appendix}
    \mathds{1}_{1\textrm{p,G}}:=\frac{1}{2}\int\textrm{d}^3k\hat{P}^\dag_{\mu\nu}(\boldsymbol{k})\ket{0_\textrm{G}}\bra{0_\textrm{G}}\hat{P}^{\mu\nu}(\boldsymbol{k})-\int\textrm{d}^3k\hat{P}^\dag(\boldsymbol{k})\ket{0_\textrm{G}}\bra{0_\textrm{G}}\hat{P}(\boldsymbol{k}).
\end{align}
It is immediate to prove that $ \mathds{1}_{1\textrm{p,G}}^2= \mathds{1}_{1\textrm{p,G}}$, which makes $ \mathds{1}_{1\textrm{p,G}}$ a projector, and that $ \mathds{1}_{1\textrm{p,G}}|1_{F,F'}\rangle=|1_{F,F'}\rangle$. Thus, we conclude that $\mathds{1}_{1\textrm{p,G}}$ is a good definition for the identity operator restricted to the single-graviton subspace.

An additional tool of interest is that of the trace $\textrm{Tr}_{1\textrm{ps}}(\hat{A})$ in the single-particle subspace of an operator $\hat{A}$. Given single-particle states $|1_{F,F'}\rangle$, and the mixed states that can be constructed from the projectors $\ket{1_{F,F'}}\bra{1_{F,F'}}$, it is easy to see that
\begin{align}\label{trace:one:graviton:subspace:appendix}
    \textrm{Tr}_{1\textrm{ps}}(\hat{A}):=\frac{1}{2}\int\textrm{d}^3k\bra{0_\textrm{G}}\hat{P}^{\mu\nu}(\boldsymbol{k})\hat{A}\hat{P}^\dag_{\mu\nu}(\boldsymbol{k})\ket{0_\textrm{G}}-\int\textrm{d}^3k\bra{0_\textrm{G}}\hat{P}(\boldsymbol{k})\hat{A}\hat{P}^\dag(\boldsymbol{k})\ket{0_\textrm{G}}.
\end{align}
It is immediate to use the state $|1_{F,F'}\rangle$ introduced above to see that $\textrm{Tr}_{1\textrm{ps}}(\ket{1_{F,F'}}\bra{1_{F,F'}})=\int\textrm{d}^3k [2F^{\mu\nu*}(\boldsymbol{k})F_{\mu\nu}(\boldsymbol{k})-|F'(\boldsymbol{k})|^2]=1$ as expected. Extension to mixed states is immediate but not contemplated in this work.

\vspace{0.4cm}

\noindent \textit{Identity operator in the one-particle subspace}---Here we follow the logic above to introduce the identity operator $\mathds{1}_{1\textrm{p,SE}}$ in the one-particle System-Environment subspace for the field, which will become useful later. 
It is immediate to see that in the basis $\mathcal{B}:=\{\ket{\textrm{L}},\ket{\textrm{R}},\{\ket{1_{\underline{\lambda}}}\}\}$ we can write
\begin{align}\label{identity:operator:one:particle:SE:subspaace:appendix}
    \mathds{1}_{1\textrm{p,SE}}:=\ket{1_\textrm{L}0_\textrm{E}}\bra{1_\textrm{L}0_\textrm{E}}+\ket{1_\textrm{R}0_\textrm{E}}\bra{1_\textrm{R}0_\textrm{E}}+\int \textrm{d}\underline{\lambda}\ket{0_\textrm{S}1_{\underline{\lambda}}}\bra{0_\textrm{S}1_{\underline{\lambda}}},
\end{align}
which is the operator we are looking for since it is a projector  $\mathds{1}_{1\textrm{p,SE}}^2=\mathds{1}_{1\textrm{p,SE}}$, and it enjoys the action $\mathds{1}_{1\textrm{p,SE}}\ket{\Psi}=\ket{\Psi}$ on any one-particle state.

In the expressions above we have specified the fact that, as is always the case in quantum field theory \cite{Srednicki:2007}, one excitation in a mode means that all other modes are left in the vacuum state. The notation is meant to convey the hidden subtlety just mentioned.

%---------------------------%
\subsection{Time evolution in the perturbative regime}
%---------------------------%
Before proceeding with the specific case at hand, we briefly provide general expressions for the time evolution in the perturbative regime, i.e., when the coupling is small.

We start by considering the time-ordered operator $\hat{E}(t)\equiv\overset{\leftarrow}{\mathcal{T}}e^{-i\int_0^t dt'\hat{A}(t')}$, where $\hat{A}=\hat{A}^\dag$. Concretely, we have that $\hat{E}(t)$ is defined by
\begin{subequations}
    \begin{align}
        \hat{E}(t)\equiv\overset{\leftarrow}{\mathcal{T}}e^{-i\int_0^t dt'\hat{A}(t')}=&\mathds{1}-i\int_0^t dt'\hat{A}(t')-\int_0^t dt'\int_0^{t'} dt''\hat{A}(t')\hat{A}(t'')+\ldots,\\
        \hat{E}^\dag(t)\equiv\overset{\rightarrow}{\mathcal{T}}e^{i\int_0^t dt'\hat{A}(t')}=&\mathds{1}+i\int_0^t dt'\hat{A}(t')-\int_0^t dt'\int_0^{t'} dt''\hat{A}(t'')\hat{A}(t')+\ldots.
    \end{align}
\end{subequations}
Note the change in the ordering of the operators in the hermitian conjugate, which is crucial since in general $[\hat{A}(t),\hat{A}(t')]\neq0$. It is also important to note that $\hat{E}(t)\hat{E}^\dag(t)=\mathds{1}$ by construction.

For the sake of simplicity, in the following we use the compact notation 
\begin{subequations}
    \begin{align}
        \hat{E}(t)\equiv\overset{\leftarrow}{\mathcal{T}}e^{-i\int_0^t dt'\hat{A}(t')}=&\mathds{1}-i\hat{A}_1(t)-\hat{A}_{2,\leftarrow}(t)+\ldots,\\
        \hat{E}^\dag(t)\equiv\overset{\rightarrow}{\mathcal{T}}e^{i\int_0^t dt'\hat{A}(t')}=&\mathds{1}-\hat{A}_1(t)-\hat{A}_{2,\rightarrow}(t)+\ldots,
    \end{align}
\end{subequations}
where we have introduced the quantities
\begin{subequations}
    \begin{align}
    \hat{A}_1(t):=&\int_0^t dt'\hat{A}(t'),\\
    \hat{A}_{2,\leftarrow}(t):=&\int_0^t dt'\int_0^{t'} dt''\hat{A}(t')\hat{A}(t''),\\
    \hat{A}_{2,\rightarrow}(t):=&\int_0^t dt'\int_0^{t'} dt''\hat{A}(t'')\hat{A}(t'),
\end{align}
\end{subequations}
and all further contributions $\hat{A}_{n,\leftrightarrow}(t)$ can be defined in a similar fashion if necessary.
Note that $\hat{A}_{n,\rightarrow}(t)=\hat{A}_{n,\leftarrow}^\dag(t)$.

The expressions above are completely general. 
Let us now assume that the operator $\hat{A}(t)$ is governed by a small parameter $\epsilon\ll1$. We can then treat the formal expansions of $\hat{E}(t)$ and $\hat{E}^\dag(t)$ as perturbative expansions. In this case, the unitarity condition $\hat{E}(t)\hat{E}^\dag(t)=\mathds{1}$ imposes relations on the the terms $\hat{A}_{n,\leftrightarrow}(t)$ in the expansion. Focusing on the condition up to second order it is immediate to see that the zeroth and first order constraints are automatically satisfied, while the second order imposes the non-trivial constraint
    \begin{align}\label{perturbative:constraint:appendix}
        \hat{A}_1^2(t)=\hat{A}_{2,\leftarrow}(t)+\hat{A}_{2,\rightarrow}(t).
    \end{align}
It is important to remark that this constraint is true only in the perturbative regime.

We now wish to compute the evolution $\hat{\rho}(t)=\hat{U}(t)\hat{\rho}(0)\hat{U}^\dag(t)$ of an initial quantum state $\hat{\rho}(0)$ in the perturbative regime. We consider a generic interaction-picture approach and write
\begin{align*} \label{time:evolution:operator:perturbative:expansion:appendix}
\hat{U}(t)=&\hat{U}_0(t)\hat{U}_\textrm{I}(t),
\end{align*}
where we have introduced
\begin{align}
    \hat{U}_\textrm{I}(t):=&\overset{\leftarrow}{\mathcal{T}}e^{-i\int_0^t dt'\hat{U}_0^\dagger(t')\hat{H}_\textrm{I}\hat{U}_0(t')},
\end{align} 
and it is the interaction Hamiltonian $\hat{H}_\textrm{I}$ that induces a perturbative correction to the dynamics.
This motivates us to set $\hat{E}(t)\equiv\hat{U}_\textrm{I}(t)$ as well as $\hat{A}(t)=\hat{U}_0^\dagger(t)\hat{H}_\textrm{I}\hat{U}_0(t)$ in order to use the perturbative notation developed above.
We adapt the expression derived above and have  
\begin{align}
    \hat{\rho}(t)=&\hat{U}_0(t)\left(\mathds{1}-i\hat{H}_1(t)-\hat{H}^\leftarrow_2(t)\right)\,\hat{\rho}(0)\left(\mathds{1}+i\hat{H}_1(t)-\hat{H}^\rightarrow_2(t)\right)\hat{U}_0^\dag(t)\nonumber\\
=&\hat{\rho}_0(t)+i\left[\hat{\rho}_0(t),\hat{H}_{1*}(t)\right]+\hat{H}_{1*}(t)\hat{\rho}_0(t)\hat{H}_{1*}(t)-\hat{H}^\leftarrow_{2*}(t)\hat{\rho}_0(t)-\hat{\rho}_0(t)\hat{H}^\rightarrow_{2*}(t),
\end{align}
where we have conveniently introduced
\begin{subequations}
    \begin{align}
        \hat{\rho}_0(t):=&\hat{U}_0(t)\hat{\rho}(0)\hat{U}_0^\dag(t),\\
        \hat{H}_{1*}(t):=&\hat{U}_0(t)\hat{H}_1(t)\hat{U}_0^\dag(t),\\
        \hat{H}_{2*}^{\leftrightarrow}(t):=&\hat{U}_0(t)\hat{H}^{\leftrightarrow}_2(t)\hat{U}_0^\dag(t).
    \end{align}
\end{subequations}
We now use the constraint \eqref{perturbative:constraint:appendix} to write
{\small
\begin{align*}
    \hat{H}^{\leftarrow}_2(t)\hat{\rho}(0)+\hat{\rho}(0)\hat{H}^{\rightarrow}_2(t)
    =&\frac{1}{2}\left[\hat{H}^{\leftarrow}_2(t),\hat{\rho}(0)\right]+\frac{1}{2}\hat{H}^{\leftarrow}_2(t)\hat{\rho}(0)+\frac{1}{2}\hat{\rho}(0)\hat{H}^{\leftarrow}_2(t)+\frac{1}{2}\left[\hat{\rho}(0),\hat{H}^{\rightarrow}_2(t)\right]+\frac{1}{2}\hat{\rho}(0)\hat{H}^{\rightarrow}_2(t)+\frac{1}{2}\hat{H}^{\rightarrow}_2(t)\hat{\rho}(0)\\
    =&\frac{1}{2}\left[\hat{H}^{\leftarrow}_2(t),\hat{\rho}(0)\right]-\frac{1}{2}\left[\hat{H}^{\rightarrow}_2(t),\hat{\rho}(0)\right]+\frac{1}{2}\left\{\hat{H}^{\rightarrow}_2(t)+\hat{H}^{\leftarrow}_2(t),\hat{\rho}(0)\right\}\\
    =&\frac{1}{2}\left[\hat{H}^{\leftarrow}_2(t)-\hat{H}^{\rightarrow}_2(t),\hat{\rho}(0)\right]+\frac{1}{2}\left\{\hat{H}_1^2(t),\hat{\rho}(0)\right\},
\end{align*}
}
which we distill into the final relation
\begin{align}
    \hat{H}^{\leftarrow}_2(t)\hat{\rho}(0)+\hat{\rho}(0)\hat{H}^{\rightarrow}_2(t)
    =&\frac{1}{2}\left[\hat{H}^{\leftarrow}_2(t)-\hat{H}^{\rightarrow}_2(t),\hat{\rho}(0)\right]+\frac{1}{2}\left\{\hat{H}_1^2(t),\hat{\rho}(0)\right\}.
\end{align}
We can immediately adapt this expression to the full case of interest and therefore obtain
\begin{align}\label{generic:perturbative:state:time:evoilution:best:form:appendix}
    \hat{\rho}(t)=&\hat{\rho}_0(t)-i\left[\hat{H}_{1*}(t),\hat{\rho}_0(t)\right]+\hat{H}_{1*}(t)\hat{\rho}_0(t)\hat{H}_{1*}(t)-\frac{1}{2}\left\{\hat{H}_{1*}^2(t),\hat{\rho}_0(t)\right\}-\frac{1}{2}\left[\hat{H}^{\leftarrow}_{2*}(t)-\hat{H}^{\rightarrow}_{2*}(t),\hat{\rho}_0(t)\right].
\end{align}
This is the starting point for the task of interest to our work.

 %---------------------------%
\subsection{Renormalization of the time evolution operator}
%---------------------------%
We now turn to an issue of crucial importance that must be addressed before continuing.
We set out to compute the simple quantity $p_0(t):=|\bra{0}\hat{U}(t)\ket{0}|^2$ using the perturbative method described above applied to the case of interest in this work. To second order we have:
\begin{align*}
    p_0(t)=|\bra{0}\hat{U}_0(t)\hat{U}_\mathrm{I}(t)\ket{0}|^2=&|\bra{0}(\mathds{1}-i\hat{H}_1(t)-\hat{H}^{\leftarrow}_2(t))\ket{0}|^2=1+(\bra{0}\hat{H}_1(t)\ket{0})^2-\bra{0}\hat{H}^{\leftarrow}_2(t)\ket{0}-\bra{0}\hat{H}^{\rightarrow}_2(t)\ket{0}\\
    =&1-\left[\bra{0}\hat{H}^2_1(t)\ket{0}-(\bra{0}\hat{H}_1(t)\ket{0})^2\right]=1-\bra{0}\hat{H}^2_1(t)\ket{0},
\end{align*}
where we have used the fact that the vacuum state is assumed to have vanishing energy and thus $\bra{0}\hat{U}_0(t)=\bra{0}$, the perturbative constraint \eqref{perturbative:constraint:appendix}, and the fact that $\bra{0_\textrm{G}}\hat{H}_1(t)\ket{0_\textrm{G}}=0$, as can be immediately verified from the form of the interaction Hamiltonian.

Let us now defined $p^{(2)}_0(t):=\bra{0}\hat{H}^2_1(t)\ket{0}\geq0$ for convenience, which gives us $p_0(t)=1-p^{(2)}_0(t)$. It is not difficult to use all expressions given above, as well as a lengthy computation that is not illuminating and is therefore left as an exercise to the reader, to show that
\begin{align*}
p^{(2)}_0(t)=G\int_0^t\mathrm{d}t'\mathrm{d}t''\int\mathrm{d}^3x\mathrm{d}^3y F(t'-t'',\boldsymbol{x}-\boldsymbol{y})=G\int_0^t\mathrm{d}t'\mathrm{d}t''\int\mathrm{d}^3x\mathrm{d}^3z F(t'-t'',\boldsymbol{z}),
\end{align*}
which diverges since there is no functional dependence on $\boldsymbol{x}$ inside the integrand. The expression above has been obtained by the change of variables $\boldsymbol{y}\rightarrow\boldsymbol{z}=\boldsymbol{x}-\boldsymbol{y}$.

We can see that such divergence occurs also by employing \eqref{stress:energy:tensor:explicit:appendix} and computing the following expectation value that would appear in $p^{(2)}_0(t)$. By appropriately translating the spatial integration variables we have: 
\begin{align*}
p^{(2)}_0(t)\propto\int_0^t\mathrm{d}t'\mathrm{d}t''\int\textrm{d}^3k\textrm{d}^3k' \mathcal{F}(\boldsymbol{k},\boldsymbol{k}')\int\mathrm{d}^3x\mathrm{d}^3y e^{-i (k_\mu+k_\mu')x^{\mu}}=V \int_0^t\mathrm{d}t'\mathrm{d}t''\int\textrm{d}^3k\textrm{d}^3k' \mathcal{F}(\boldsymbol{k},\boldsymbol{k}')\int\mathrm{d}^3x e^{-i (k_\mu+k_\mu')x^{\mu}},
\end{align*}
where $V:=\int\mathrm{d}^3y$ is the divergent spatial volume. 

One way to cure this problem would be to introduce a volume cutoff, such as replacing each spatial integral by $\int\mathrm{d}^3y\rightarrow\int\mathrm{d}^3ye^{-\sigma_0^2|\boldsymbol{y}|^2}$, for finite cutoff parameter $\sigma_0$. Such finite-volume ad-hoc methods are used in quantum field theory in order to cure divergences \cite{Srednicki:2007}.
Instead of proceeding this way, we note that the rotating wave interaction Hamiltonian does not suffer this problem. In fact, it is easy to verify that
\begin{align*}
    \bra{0}\hat{H}^\Phi{}^2_{1,\textrm{RWA}}(t)\ket{0}=0.
\end{align*}
This further supports the use of $\hat{H}^\Phi{}_{\textrm{I,RWA}}(t)$ for the dynamics of the system.

%---------------------------%
\subsection{Time evolution due to self-gravity}
%---------------------------%
We now proceed with computing the reduced state $\hat{\rho}_\mathrm{S}(t):=\textrm{Tr}_\textrm{E,G}(\hat{\rho}(t))$ of the system as a function of time. The main expression to be used is \eqref{generic:perturbative:state:time:evoilution:best:form:appendix}, which we further refine below employing the explicit form
\begin{align}
    \hat{\rho}(0)=\hat{\rho}_\textrm{SE}(0)\otimes\ket{0_\textrm{G}}\bra{0_\textrm{G}}
\end{align}
 of the initial state.
In the following we will use the expression $\hat{U}_0(t)=e^{-i\hat{H}_0t}$. 

We now can move on to time evolution in the perturbative regime. Let us consider a Hamiltonian $\hat{H}=\hat{H}_0+\hat{H}_\textrm{I}$, where $\hat{H}_\textrm{I}$ will be the perturbative interaction part. From now on it is understood that $\hat{H}_\textrm{I}\equiv\hat{H}^\Phi{}_{\textrm{I,RWA}}(t)$, as mentioned before, in order to reduce the burden due to the notation. 

The expression for the state at time $t$ has already been obtained to second order in \eqref{generic:perturbative:state:time:evoilution:best:form:appendix}, which we re-print here for convenience:
\begin{align*}
    \hat{\rho}(t)=&\hat{\rho}_0(t)-i\left[\hat{H}_{1*}(t),\hat{\rho}_0(t)\right]+\hat{H}_{1*}(t)\hat{\rho}_0(t)\hat{H}_{1*}(t)-\frac{1}{2}\left\{\hat{H}_{1*}^2(t),\hat{\rho}_0(t)\right\}-\frac{1}{2}\left[\hat{H}^{\leftarrow}_{2*}(t)-\hat{H}^{\rightarrow}_{2*}(t),\hat{\rho}_0(t)\right].
\end{align*}
Defining $\hat{\rho}_\textrm{SE}(t):=\textrm{Tr}_\textrm{G}(\hat{\rho}(t))$ and introducing $\hat{\rho}_{0,\textrm{SE}}(t):=\hat{U}_0(t)\hat{\rho}_{\textrm{SE}}(0)\hat{U}_0^\dag(t)$ for convenience of presentation, we can then use the cyclic properties of the trace in the graviton subsector to obtain
\begin{align}\label{generic:perturbative:reduced:state:time:evoilution:general:form:appendix}
    \hat{\rho}_\textrm{SE}(t)=&\hat{\rho}_{0,\textrm{SE}}(t)-i\left[\bra{0_\textrm{G}}\hat{H}_{1*}(t)\ket{0_\textrm{G}},\hat{\rho}_{0,\textrm{SE}}(t)\right]+\textrm{Tr}_\textrm{G}\left(\hat{H}_{1*}(t)\hat{\rho}_0(t)\hat{H}_{1*}(t)\right)-\frac{1}{2}\left\{\bra{0_\textrm{G}}\hat{H}_{1*}^2(t)\ket{0_\textrm{G}},\hat{\rho}_{0,\textrm{SE}}(t)\right\}\nonumber\\
    &-\frac{1}{2}\left[\bra{0_\textrm{G}}\hat{H}^{\leftarrow}_{2*}(t)\ket{0_\textrm{G}}-\bra{0_\textrm{G}}\hat{H}^{\rightarrow}_{2*}(t)\ket{0_\textrm{G}},\hat{\rho}_{0,\textrm{SE}}(t)\right].
\end{align} 
Crucially, note that $\hat{\rho}_{0,\textrm{SE}}(t)\approx\hat{\rho}_{\textrm{SE}}(0)$ in the static massive regime, as seen in \eqref{useful:lowest:order:evolution:appendix}. We also note that it is immediate to verify that $\bra{0_\textrm{G}}\hat{h}_{\mu\nu}(t,\boldsymbol{x})\ket{0_\textrm{G}}=0$ implies $\bra{0_\textrm{G}}\hat{H}_{1*}(t)\ket{0_\textrm{G}}=0$, as argued in the subsection on renormalization of the time evolution operator. Furthermore, we have that $\hat{H}_{2*,\textrm{SE}}(t):=-\frac{i}{2}(\bra{0_\textrm{G}}\hat{H}^{\leftarrow}_{2*}(t)\ket{0_\textrm{G}}-\bra{0_\textrm{G}}\hat{H}^{\rightarrow}_{2*}(t)\ket{0_\textrm{G}})$ is a Hermitian operator defined in the reduced field-space only. 
Thus, we have 
\begin{align*}
    \hat{\rho}_\textrm{SE}(t)=&\hat{\rho}_{0,\textrm{SE}}(t)-i\left[\hat{H}_{2*,\textrm{SE}}(t),\hat{\rho}_{0,\textrm{SE}}(t)\right]+\textrm{Tr}_\textrm{G}\left(\hat{H}_{1*}(t)\ket{0_\textrm{G}}\hat{\rho}_{0,\textrm{SE}}(t)\bra{0_\textrm{G}}\hat{H}_{1*}(t)\right)-\frac{1}{2}\left\{\bra{0_\textrm{G}}\hat{H}_{1*}^2(t)\ket{0_\textrm{G}},\hat{\rho}_{0,\textrm{SE}}(t)\right\}.
\end{align*}
The expression above can be written conveniently as
\begin{align}\label{generic:perturbative:reduced:state:time:evoilution:bettering:form:appendix}
    \hat{\rho}_\textrm{SE}(t)=&\hat{U}_*(t)\hat{\rho}_{0,\textrm{SE}}(t)\hat{U}_*^\dag(t)+\textrm{Tr}_\textrm{G}\left(\hat{H}_{1*}(t)\ket{0_\textrm{G}}\hat{\rho}_{0,\textrm{SE}}(t)\bra{0_\textrm{G}}\hat{H}_{1*}(t)\right)-\frac{1}{2}\left\{\bra{0_\textrm{G}}\hat{H}_{1*}^2(t)\ket{0_\textrm{G}},\hat{\rho}_{0,\textrm{SE}}(t)\right\},
\end{align}
where $\hat{U}_*(t):=\exp[-i\hat{H}_{2*,\textrm{SE}}(t)]$, which better highlights the separate contributions of an effective unitary evolution component against the contributions of the non-unitary component.

We now use the definition \eqref{trace:one:graviton:subspace:appendix} of the trace on the gravitational sector to compute $\textrm{Tr}_\textrm{G}\left(\hat{H}_{1*}(t)\ket{0_\textrm{G}}\hat{\rho}_{0,\textrm{SE}}(t)\bra{0_\textrm{G}}\hat{H}_{1*}(t)\right)$, and find
{\small
\begin{align*}
    \textrm{Tr}_\textrm{G}\left(...\right)=&\frac{1}{2}\int\textrm{d}^3k\bra{0_\textrm{G}}\hat{P}^{\mu\nu}(\boldsymbol{k})\hat{H}_{1*}(t)\ket{0_\textrm{G}}\hat{\rho}_{0,\textrm{SE}}(t)\bra{0_\textrm{G}}\hat{H}_{1*}(t)\hat{P}^\dag_{\mu\nu}(\boldsymbol{k})\ket{0_\textrm{G}}\\
    &-\int\textrm{d}^3k\bra{0_\textrm{G}}\hat{P}(\boldsymbol{k})\hat{H}_{1*}(t)\ket{0_\textrm{G}}\hat{\rho}_{0,\textrm{SE}}(t)\bra{0_\textrm{G}}\hat{H}_{1*}(t)\hat{P}^\dag(\boldsymbol{k})\ket{0_\textrm{G}}\\
    =&\frac{1}{8}\int_0^t\textrm{d}t'\textrm{d}t''\int\textrm{d}^3x\textrm{d}^3y\int\textrm{d}^3k\bra{0_\textrm{G}}\hat{P}^{\mu\nu}(\boldsymbol{k})\hat{h}^{\mu'\nu'}(t'-t,\boldsymbol{x})\ket{0_\textrm{G}}\bra{0_\textrm{G}}\hat{h}^{\mu''\nu''}(t''-t,\boldsymbol{y})\hat{P}^\dag_{\mu\nu}(\boldsymbol{k})\ket{0_\textrm{G}}\\
    &\times:\hat{T}_{\mu'\nu'}(t'-t,\boldsymbol{x}):\hat{\rho}_{0,\textrm{SE}}(t):\hat{T}_{\mu''\nu''}(t''-t,\boldsymbol{y}):\\
    &-\frac{1}{4}\int\textrm{d}^3k\bra{0_\textrm{G}}\hat{P}(\boldsymbol{k})\hat{h}^{\mu'\nu'}(t'-t,\boldsymbol{x})\ket{0_\textrm{G}}\bra{0_\textrm{G}}\hat{h}^{\mu''\nu''}(t''-t,\boldsymbol{y})\hat{P}^\dag(\boldsymbol{k})\ket{0_\textrm{G}}\\
    &\times:\hat{T}_{\mu'\nu'}(t'-t,\boldsymbol{x}):\hat{\rho}_{0,\textrm{SE}}(t):\hat{T}_{\mu''\nu''}(t''-t,\boldsymbol{y}):\\
    =&\frac{1}{8}\int_0^t\textrm{d}t'\textrm{d}t''\int\textrm{d}^3x\textrm{d}^3y\int\textrm{d}^3k\bra{0_\textrm{G}}\hat{h}^{\mu''\nu''}(t''-t,\boldsymbol{y})\hat{P}^\dag_{\mu\nu}(\boldsymbol{k})\ket{0_\textrm{G}}\bra{0_\textrm{G}}\hat{P}^{\mu\nu}(\boldsymbol{k})\hat{h}^{\mu'\nu'}(t'-t,\boldsymbol{x})\ket{0_\textrm{G}}\\
    &\times:\hat{T}_{\mu'\nu'}(t'-t,\boldsymbol{x}):\hat{\rho}_{0,\textrm{SE}}(t):\hat{T}_{\mu''\nu''}(t''-t,\boldsymbol{y}):\\
    &-\frac{1}{4}\int\textrm{d}^3k\bra{0_\textrm{G}}\hat{h}^{\mu''\nu''}(t''-t,\boldsymbol{y})\hat{P}^\dag(\boldsymbol{k})\ket{0_\textrm{G}}\bra{0_\textrm{G}}\hat{P}(\boldsymbol{k})\hat{h}^{\mu'\nu'}(t'-t,\boldsymbol{x})\ket{0_\textrm{G}}\\
    &\times:\hat{T}_{\mu'\nu'}(t'-t,\boldsymbol{x}):\hat{\rho}_{0,\textrm{SE}}(t):\hat{T}_{\mu''\nu''}(t''-t,\boldsymbol{y}):\\
    =&\frac{1}{4}\int_0^t\textrm{d}t'\textrm{d}t''\int\textrm{d}^3x\textrm{d}^3y\left[\frac{1}{2}\int\textrm{d}^3k\bra{0_\textrm{G}}\hat{h}^{\mu''\nu''}(t''-t,\boldsymbol{y})\hat{P}^\dag_{\mu\nu}(\boldsymbol{k})\ket{0_\textrm{G}}\bra{0_\textrm{G}}\hat{P}^{\mu\nu}(\boldsymbol{k})\hat{h}^{\mu'\nu'}(t'-t,\boldsymbol{x})\ket{0_\textrm{G}}\right.\\
    &\left.-\int\textrm{d}^3k\bra{0_\textrm{G}}\hat{h}^{\mu''\nu''}(t''-t,\boldsymbol{y})\hat{P}^\dag(\boldsymbol{k})\ket{0_\textrm{G}}\bra{0_\textrm{G}}\hat{P}(\boldsymbol{k})\hat{h}^{\mu'\nu'}(t'-t,\boldsymbol{x})\ket{0_\textrm{G}}\right]\\
    &\times:\hat{T}_{\mu'\nu'}(t'-t,\boldsymbol{x}):\hat{\rho}_{0,\textrm{SE}}(t):\hat{T}_{\mu''\nu''}(t''-t,\boldsymbol{y}):\\
    =&\frac{1}{4}\int_0^t\textrm{d}t'\textrm{d}t''\int\textrm{d}^3x\textrm{d}^3y\bra{0_\textrm{G}}\hat{h}^{\mu''\nu''}(t''-t,\boldsymbol{y})\hat{h}^{\mu'\nu'}(t'-t,\boldsymbol{x})\ket{0_\textrm{G}}:\hat{T}_{\mu'\nu'}(t'-t,\boldsymbol{x}):\hat{\rho}_{0,\textrm{SE}}(t):\hat{T}_{\mu''\nu''}(t''-t,\boldsymbol{y}):
\end{align*}
}

\noindent where we have used the identity \eqref{identity:operator:one:graviton:subspace:appendix} in the last step.

It is then immediate to see that
\begin{align*}
    \bra{0_\textrm{G}}\hat{H}_{1*}^2(t)\ket{0_\textrm{G}}=\frac{1}{4}\int_0^t\textrm{d}t'\textrm{d}t''\int\textrm{d}^3x\textrm{d}^3y\bra{0_\textrm{G}}\hat{h}^{\mu''\nu''}(t''-t,\boldsymbol{y})\hat{h}^{\mu'\nu'}(t'-t,\boldsymbol{x})\ket{0_\textrm{G}}:\hat{T}_{\mu'\nu'}(t'-t,\boldsymbol{x}):\,:\hat{T}_{\mu''\nu''}(t''-t,\boldsymbol{y}):
\end{align*}
Thus, we employ the graviton propagator-like quantity $D_{\mu\nu,\mu'\nu'}(x-y)$ introduced in \eqref{h:expectation:value:vacuum:appendix} to obtain 
{\small
\begin{align}\label{time:evolution:matrix:representation:appendix}
    \hat{\rho}_\textrm{SE}(t)
    =&\hat{U}_*(t)\hat{\rho}_\textrm{SE}(0)\hat{U}_*^\dag(t)+2\frac{m^2}{m_\textrm{P}^2}\int_0^t\textrm{d}t'\textrm{d}t''\int\frac{\textrm{d}^3x\textrm{d}^3y}{m^2}\tilde{D}_{\mu\nu,\mu'\nu'}(x-y)\,\left[\hat{T}^{\mu\nu}_x\hat{\rho}_\textrm{SE}(0)\hat{T}^{\mu'\nu'}_y-\frac{1}{2}\left\{\hat{T}^{\mu'\nu'}_y\hat{T}^{\mu\nu}_x,\hat{\rho}_\textrm{SE}(0)\right\}\right],
\end{align}
}
\noindent where $\hat{T}^{\mu\nu}_x\equiv:\hat{T}^{\mu\nu}(t',\boldsymbol{x}):$ and $\hat{T}^{\mu\nu}_y\equiv:\hat{T}^{\mu\nu}(t'',\boldsymbol{y}):$ for easiness of presentation. Equation \eqref{time:evolution:matrix:representation:appendix} is our main expression. 

These expressions allow us to identify the key perturbative control parameter $\epsilon$ mentioned above, which reads
\begin{align}
    \epsilon:=\frac{m}{m_\textrm{P}}.
\end{align}
That this is the relevant parameter is in line with directly related works in the literature \cite{Marletto:Vedral:2017,Aziz:Howl:2025}.

%---------------------------%
\section{EVALUATION OF THE TIME DEPENDENT DENSITY MATRIX ELEMENTS}\label{sec_state_elements}
%---------------------------%
We proceed with using the general form of the time-evolved state above to obtain explicit expressions in the reduced single-particle subsector.

%---------------------------%
\subsection{Diagrammatic-like approach to dynamics}
%---------------------------%
In order to proceed further we find it useful to better understand the implications of our main expression \eqref{time:evolution:matrix:representation:appendix}. To do so, we would like to provide an intuition using an approach \textit{à la Feynman}. 
We apply this understanding to the main expression \eqref{time:evolution:matrix:representation:appendix}, which we re-print here for convenience
\begin{align*}
    \hat{\rho}_\textrm{SE}(t)
    =&\hat{U}_*(t)\hat{\rho}_\textrm{SE}(0)\hat{U}_*^\dag(t)+2\frac{m^2}{m_\textrm{P}^2}\int_0^t\textrm{d}t'\textrm{d}t''\int\frac{\textrm{d}^3x\textrm{d}^3y}{m^2}\tilde{D}_{\mu\nu,\mu'\nu'}(x-y)\,\left[\hat{T}^{\mu\nu}_x\hat{\rho}_\textrm{SE}(0)\hat{T}^{\mu'\nu'}_y-\frac{1}{2}\left\{\hat{T}^{\mu'\nu'}_y\hat{T}^{\mu\nu}_x,\hat{\rho}_\textrm{SE}(0)\right\}\right].
\end{align*}
For the sake of the discussion, we here ignore the unitary part and focus on the non-unitary one.

As argued above, we can project the state $\hat{\rho}_\textrm{SE}(t)$ onto the one-particle states $\ket{1_K}$ and $\ket{1_{\underline{\lambda}}}$ and then trace over the system-environment subsector. This means that the coefficients of interest will be determined by terms of the form
$\bra{1_K}\hat{T}^{\mu\nu}_x\hat{\rho}_\textrm{SE}(0)\hat{T}^{\mu'\nu'}_y\ket{1_{K'}}$, $\bra{1_{\underline{\lambda}}}\hat{T}^{\mu\nu}_x\hat{\rho}_\textrm{SE}(0)\hat{T}^{\mu'\nu'}_y\ket{1_{\underline{\lambda}}}$, as well as $\bra{1_K}\hat{T}^{\mu'\nu'}_y\hat{T}^{\mu\nu}_x\ket{1_{K'}}$. The second term will contribute as a consequence of tracing over $\hat{T}^{\mu\nu}_x\hat{\rho}_\textrm{SE}(0)\hat{T}^{\mu'\nu'}_y$ in the one-particle system-environment subsector.

It is convenient to use the one-particle sector identity and write
\begin{align}
    \bra{1_K}\hat{T}^{\mu'\nu'}_y\hat{T}^{\mu\nu}_x\ket{1_{K'}}
    =\sum_{K''=\textrm{L,R}}\bra{1_K}\hat{T}^{\mu'\nu'}_y\ket{1_{K''}}\bra{1_{K''}}\hat{T}^{\mu\nu}_x\ket{1_{K'}}
    +\sum_{\underline{\lambda}}\bra{1_K}\hat{T}^{\mu'\nu'}_y\ket{1_{\underline{\lambda}}}\bra{1_{\underline{\lambda}}}\hat{T}^{\mu\nu}_x\ket{1_{K'}},
\end{align}
which motivates us to introduce the quantities:
\begin{subequations}\label{key:q:quantities:appendix}
 \begin{align}
Q_{KK'K''K'''}(t):=&2\int_0^t\textrm{d}t'\textrm{d}t''\int\frac{\textrm{d}^3x\textrm{d}^3y}{m^2}\tilde{D}_{\mu\nu,\mu'\nu'}(x-y)\,\bra{1_K}\hat{T}^{\mu'\nu'}_y\ket{1_{K'}}\bra{1_{K''}}\hat{T}^{\mu\nu}_x\ket{1_{K'''}},\\
\tilde{Q}_{KK'K''K'''}(t):=&2\int_0^t\textrm{d}t'\int_0^{t'}\textrm{d}t''\int\frac{\textrm{d}^3x\textrm{d}^3y}{m^2}\tilde{D}_{\mu\nu,\mu'\nu'}(x-y)\,\bra{1_K}\hat{T}^{\mu'\nu'}_y\ket{1_{K'}}\bra{1_{K''}}\hat{T}^{\mu\nu}_x\ket{1_{K'''}}.
\end{align}   
\end{subequations}
The quantities above are defined through the key object $\bra{1_K}\hat{T}_{\mu\nu}(t,\boldsymbol{x})\ket{1_{K'}}$. We here study some of its properties for later use.
The expression \eqref{stress:energy:tensor:explicit:appendix} of the stress-energy tensor allows us to conclude that $\bra{1_K}\hat{T}_{\mu\nu}(t,\boldsymbol{x})\ket{1_{K'}}$ is computed using the effective expression
\begin{align}
\hat{T}_{\mu\nu}(t,\boldsymbol{x})=&\int\frac{\textrm{d}^3k\textrm{d}^3k'}{2(2\pi)^3}t_{\mu\nu}(k,k')e^{i (k'_\mu-k_\mu) x^\mu}\hat{a}_{\boldsymbol{k}}^\dag\hat{a}_{\boldsymbol{k}'},
\end{align}
where we have defined $t_{\mu\nu}(k,k'):=\frac{k_\mu k'_\nu+k_\nu k'_\mu-k^\rho k'_\rho\eta_{\mu\nu}-m^2\eta_{\mu\nu}}{\sqrt{\omega_{\boldsymbol{k}}\omega_{\boldsymbol{k}'}}}$ for simplicity of presentation.

We then have that
 \begin{align}
    \bra{1_K}\hat{T}_{\mu\nu}(t,\boldsymbol{x})\ket{1_{K'}}=&\bra{0_\textrm{S}}\left[\hat{a}_K,\left[\hat{T}_{\mu\nu}(t,\boldsymbol{x}),\hat{a}_{K'}^\dag\right]\right]\ket{0_\textrm{S}}\nonumber\\
    =&\int\frac{\textrm{d}^3k\textrm{d}^3k'}{2(2\pi)^3}t_{\mu\nu}(k,k')e^{i(\omega_{\boldsymbol{k}}-\omega_{\boldsymbol{k}'})t}e^{-i (\boldsymbol{k}-\boldsymbol{k}')\cdot\boldsymbol{x}}e^{i \boldsymbol{k}\cdot\boldsymbol{x}_K}e^{-i \boldsymbol{k}'\cdot\boldsymbol{x}_{K'}}F(\boldsymbol{k})F^*(\boldsymbol{k}')\\
    =&m\,\eta_{\mu0}\eta_{\nu0}\int\frac{\textrm{d}^3k\textrm{d}^3k'}{(2\pi)^3}e^{-i (\boldsymbol{k}-\boldsymbol{k}')\cdot\boldsymbol{x}}e^{i \boldsymbol{k}\cdot\boldsymbol{x}_K}e^{-i \boldsymbol{k}'\cdot\boldsymbol{x}_{K'}}F(\boldsymbol{k})F^*(\boldsymbol{k}'),
\end{align}   
where the last expression has been obtained by invoking the static massive regime. Crucially, $\bra{1_K}\hat{T}_{\mu\nu}(t,\boldsymbol{x})\ket{1_{K'}}$ is time independent in this regime. 

We now write $\boldsymbol{x}_\textrm{R}=\boldsymbol{x}_\textrm{L}+\boldsymbol{L}$. Thus, it is immediate to use the simultaneous $t'\leftrightarrow t''$, $\boldsymbol{x}\leftrightarrow\boldsymbol{y}$, and $(\mu\nu
)\leftrightarrow(\mu'\nu')$ exchange properties of the integrals within the quantities above to see that
\begin{subequations}\label{Q:quantities:symmetries:appendix}
    \begin{align}
        Q_{KK'K''K'''}(t)=&Q_{K''K'''KK'}(t)\\
        Q_\textrm{RRRR}(t)=&Q_\textrm{LLLL}(t).
    \end{align}
\end{subequations}
We proceed by computing $Q_{KK'K''K'''}(t)$ explicitly. We have
{\small
\begin{align*}
   &Q_{KK'
    K''K'''}(t)=\frac{2}{m^2}\int_0^t\textrm{d}t'\textrm{d}t''\int\textrm{d}^3x\textrm{d}^3y
    \tilde{D}^{\mu\nu,\mu'\nu'}(x-y)\bra{1_K}\hat{T}_{\mu\nu}(t'-t,\boldsymbol{x})\ket{1_{K'}}\bra{1_{K''}}\hat{T}_{\mu'\nu'}(t''-t,\boldsymbol{y})\ket{1_{K'''}}\\
    &=\frac{2}{m^2}\int_0^t\textrm{d}t'\textrm{d}t''\int\textrm{d}^3x\textrm{d}^3y
    D^{\mu\nu,\mu'\nu'}(x-y)\bra{1_K}\hat{T}_{\mu\nu}(t'-t,\boldsymbol{x})\ket{1_{K'}}\bra{1_{K''}}\hat{T}_{\mu'\nu'}(t''-t,\boldsymbol{y})\ket{1_{K'''}}\\
    &=2\int_0^t\textrm{d}t'\textrm{d}t''\int\textrm{d}^3x\textrm{d}^3y\int\frac{\textrm{d}^3k\textrm{d}^3k'\textrm{d}^3p\textrm{d}^3p'}{(2\pi)^6}\frac{e^{-i (\boldsymbol{k}-\boldsymbol{k}')\cdot\boldsymbol{x}}e^{i \boldsymbol{k}\cdot\boldsymbol{x}_K}e^{-i \boldsymbol{k}'\cdot\boldsymbol{x}_{K'}}e^{-i (\boldsymbol{p}-\boldsymbol{p}')\cdot\boldsymbol{y}}e^{i \boldsymbol{p}\cdot\boldsymbol{x}_{K''}}e^{-i \boldsymbol{p}'\cdot\boldsymbol{x}_{K'''}}}{-(t'-t'')^2+|\boldsymbol{x}-\boldsymbol{y}|^2}
    F(\boldsymbol{k})F^*(\boldsymbol{k}')
    F(\boldsymbol{p})F^*(\boldsymbol{p}')\\
    &=\int_0^t\textrm{d}t'\textrm{d}t''\int\textrm{d}^3x\int\frac{\textrm{d}^3k\textrm{d}^3k'\textrm{d}^3p\textrm{d}^3p'}{4\pi^3}\frac{e^{-i (\boldsymbol{k}-\boldsymbol{k}')\cdot\boldsymbol{x}}e^{i \boldsymbol{k}\cdot\boldsymbol{x}_K}e^{-i \boldsymbol{k}'\cdot\boldsymbol{x}_{K'}}e^{i \boldsymbol{p}\cdot\boldsymbol{x}_{K''}}e^{-i \boldsymbol{p}'\cdot\boldsymbol{x}_{K'''}}}{-(t'-t'')^2+|\boldsymbol{x}|^2}\delta^3(\boldsymbol{k}-\boldsymbol{k}'+\boldsymbol{p}-\boldsymbol{p}')F(\boldsymbol{k})F^*(\boldsymbol{k}')
    F(\boldsymbol{p})F^*(\boldsymbol{p}')    
\end{align*}   
}

\noindent The second line was obtained by using the property (proven above) that states that $\int_0^t\textrm{d}t'\textrm{d}t''\tilde{D}^{\mu\nu,\mu'\nu'}(x-y)=\int_0^t\textrm{d}t'\textrm{d}t''D^{\mu\nu,\mu'\nu'}(x-y)$.
We then perform a change of variables $\boldsymbol{u}:=(\boldsymbol{k}+\boldsymbol{k}')/2$, $\boldsymbol{v}:=(\boldsymbol{k}-\boldsymbol{k}')/2$, $\boldsymbol{u}':=(\boldsymbol{p}+\boldsymbol{p}')/2$, and $\boldsymbol{v}':=(\boldsymbol{p}-\boldsymbol{p}')/2$, and have

{\small
\begin{align*}
    Q_{KK'
    K''K'''}(t)
    =&4\pi^2\int\frac{\textrm{d}^3u\textrm{d}^3v\textrm{d}^3u'\textrm{d}^3v'}{(2\pi)^3}
    e^{i \boldsymbol{u}\cdot(\boldsymbol{x}_K-\boldsymbol{x}_{K'})}
    e^{i \boldsymbol{v}\cdot(\boldsymbol{x}_K+\boldsymbol{x}_{K'})}e^{i \boldsymbol{u}'\cdot(\boldsymbol{x}_{K''}-\boldsymbol{x}_{K'''})}e^{i \boldsymbol{v}'\cdot(\boldsymbol{x}_{K''}+\boldsymbol{x}_{K'''})}\\
&\times\frac{\sin^2(t|\boldsymbol{v}|)}{|\boldsymbol{v}|^3}
\delta^3(\boldsymbol{v}+\boldsymbol{v}')F(\sqrt{2}\boldsymbol{u})F(\sqrt{2}\boldsymbol{v})
    F(\sqrt{2}\boldsymbol{u}')F(\sqrt{2}\boldsymbol{v}')\\
    =&\int\frac{\textrm{d}^3u\textrm{d}^3v\textrm{d}^3u'}{16\pi}
    e^{i \boldsymbol{u}\cdot(\boldsymbol{x}_K-\boldsymbol{x}_{K'})}
    e^{i \boldsymbol{v}\cdot(\boldsymbol{x}_K+\boldsymbol{x}_{K'}-\boldsymbol{x}_{K''}-\boldsymbol{x}_{K'''})}e^{i \boldsymbol{u}'\cdot(\boldsymbol{x}_{K''}-\boldsymbol{x}_{K'''})}\frac{\sin^2(t|\boldsymbol{v}|)}{|\boldsymbol{v}|^3}
F(\sqrt{2}\boldsymbol{u})F^2(\sqrt{2}\boldsymbol{v})
    F(\sqrt{2}\boldsymbol{u}')\\
    =&\frac{\sqrt{\pi}}{16\sqrt{2}}\sigma^3e^{-\frac{\sigma^2}{2}|\boldsymbol{x}_K-\boldsymbol{x}_{K'}|^2}
    e^{-\frac{\sigma^2}{2}|\boldsymbol{x}_{K''}-\boldsymbol{x}_{K'''}|^2}\int\textrm{d}^3v
    e^{i \boldsymbol{v}\cdot(\boldsymbol{x}_K+\boldsymbol{x}_{K'}-\boldsymbol{x}_{K''}-\boldsymbol{x}_{K'''})}\frac{\sin^2(t|\boldsymbol{v}|)}{|\boldsymbol{v}|^3}
F^2(\sqrt{2}\boldsymbol{v})\\
    =&\frac{1}{64\pi}e^{-\frac{\sigma^2}{2}|\boldsymbol{x}_K-\boldsymbol{x}_{K'}|^2}
    e^{-\frac{\sigma^2}{2}|\boldsymbol{x}_{K''}-\boldsymbol{x}_{K'''}|^2}\int\textrm{d}^3v
    e^{i \boldsymbol{v}\cdot(\boldsymbol{x}_K+\boldsymbol{x}_{K'}-\boldsymbol{x}_{K''}-\boldsymbol{x}_{K'''})}\frac{\sin^2(t|\boldsymbol{v}|)}{|\boldsymbol{v}|^3}
e^{-\frac{1}{\sigma^2}|\boldsymbol{v}|^2}\\
=&\frac{1}{16}e^{-\frac{\sigma^2}{2}|\boldsymbol{x}_K-\boldsymbol{x}_{K'}|^2}
    e^{-\frac{\sigma^2}{2}|\boldsymbol{x}_{K''}-\boldsymbol{x}_{K'''}|^2}
    \int_0^{+\infty}\textrm{d}v
    \frac{\sin(\sigma|\boldsymbol{x}_K+\boldsymbol{x}_{K'}-\boldsymbol{x}_{K''}-\boldsymbol{x}_{K'''}|v)}{\sigma|\boldsymbol{x}_K+\boldsymbol{x}_{K'}-\boldsymbol{x}_{K''}-\boldsymbol{x}_{K'''}|}\frac{\sin^2(\sigma\,t\,v)}{v^2}
e^{-v^2}.
\end{align*}
We are motivated to introduce 
    \begin{align}
    Q_x(t):=&\frac{1}{32}\int_0^{+\infty}\textrm{d}v
    \frac{\sin(\sigma\,x\,v)}{\sigma\,x}\frac{\sin^2(\sigma\,t\,v)}{v^2}e^{-v^2},
    \end{align}
where $x\in\{0,L,2L\}$. In particular, we have $Q(t):=Q_{KKKK}(t)=\frac{1}{32}\int_0^{+\infty}\textrm{d}v
    \frac{\sin^2(\sigma\,t\,v)}{v}e^{-v^2}$. 
Note that
\begin{align}
        Q(t)=&\frac{1}{32}\int_0^{+\infty}\textrm{d}v
    \frac{\sin^2(\sigma\,t\,v)}{v}e^{-v^2}=\frac{1}{64}\int_0^{+\infty}\textrm{d}v
    \frac{1-\cos(2\sigma\,t\,v)}{v}e^{-v^2}=\frac{1}{64}\sum_{n=1}(-1)^{n+1}\frac{(2\sigma t)^{2n}}{(2n)!}\int_0^{+\infty}\textrm{d}v\,v^{2n-1}
    e^{-v^2}\nonumber\\
    =&\frac{(\sigma t)^2}{32}\sum_{n=0}(-1)^n\frac{(2\sigma t)^{2n+2}}{(2n+2)!}\Gamma(n+1)=\frac{(\sigma\,t)^2}{64}{}_2 F_2(1,1;3/2,2;-(\sigma\,t)^2),
\end{align}
where ${}_2 F_2(a,b;c,d;z)$ is a generalized hypergeometric function. Note that $x^2{}_2 F_2(1,1;3/2,2,-x^2)\approx \ln x$ for $x\gg1$.

\vspace{0.4cm}

\noindent We now compute the quantity $\tilde{Q}_{KK'K''K'''}(t)$, which comes into play only in the expression $\hat{H}_{2*,\textrm{SE}}(t)=-\frac{i}{2}(\bra{0_\textrm{G}}\hat{H}^{\leftarrow}_{2*}(t)\ket{0_\textrm{G}}-\bra{0_\textrm{G}}\hat{H}^{\rightarrow}_{2*}(t)\ket{0_\textrm{G}})$ that drives the unitary part of the effective time evolution. It reads
\begin{align}
\tilde{Q}^{(2)}_{KK''K''K'}(t):=
    -\frac{i}{2m^2}\int_0^t\textrm{d}t'\int_0^{t'}\textrm{d}t''\int\textrm{d}^3x\textrm{d}^3y
    \tilde{D}^{\mu\nu,\mu'\nu'}(x-y)\bra{1_K}\hat{T}_{\mu\nu}(t'-t,\boldsymbol{x})\ket{1_{K''}}\bra{1_{K''}}\hat{T}_{\mu'\nu'}(t''-t,\boldsymbol{y})\ket{1_{K'}}\nonumber\\
    +\frac{i}{2m^2}\int_0^t\textrm{d}t'\int_0^{t'}\textrm{d}t''\int\textrm{d}^3x\textrm{d}^3y
    \tilde{D}^{\mu\nu,\mu'\nu'}(y-x)\bra{1_K}\hat{T}_{\mu'\nu'}(t''-t,\boldsymbol{y})\ket{1_{K''}}\bra{1_{K''}}\hat{T}_{\mu\nu}(t'-t,\boldsymbol{x})\ket{1_{K'}}
\end{align}
We have already noted that elements of the form $\bra{1_K}\hat{T}_{\mu\nu}(t'-t,\boldsymbol{x})\ket{1_{K''}}$ are time independent. We therefore drop the time dependence for such elements and use the symmetric property $\tilde{D}^{\mu\nu,\mu'\nu'}(x-y)=\tilde{D}^{\mu'\nu',\mu\nu}$ to find
\begin{align*}
\tilde{Q}^{(2)}_{KK''K''K'}(t):=
    -\frac{i}{2m^2}\int_0^t\textrm{d}t'\int_0^{t'}\textrm{d}t''\int\textrm{d}^3x\textrm{d}^3y
    \tilde{D}^{\mu\nu,\mu'\nu'}(x-y)\bra{1_K}\hat{T}_{\mu'\nu'}(\boldsymbol{x})\ket{1_{K''}}\bra{1_{K''}}\hat{T}_{\mu\nu}(\boldsymbol{y})\ket{1_{K'}}\nonumber\\
    +\frac{i}{2m^2}\int_0^t\textrm{d}t'\int_0^{t'}\textrm{d}t''\int\textrm{d}^3x\textrm{d}^3y
    \tilde{D}^{\mu\nu,\mu'\nu'}(y-x)\bra{1_K}\hat{T}_{\mu'\nu'}(\boldsymbol{y})\ket{1_{K''}}\bra{1_{K''}}\hat{T}_{\mu\nu}(\boldsymbol{x})\ket{1_{K'}}
\end{align*}
We then note that $\tilde{D}^{\mu\nu,\mu'\nu'}(x-y)$ is given by two contributions: one symmetric in the exchange $t\leftrightarrow t'$ and proportional to $D^{\mu\nu,\mu'\nu'}(x-y)$, while the second is antisymmetric in the exchange $t\leftrightarrow t'$, as can be seen in the implicit definition \eqref{tilde:D:appendix} defined via \eqref{h:expectation:value:vacuum:appendix}.

Thus, we have
{\small
\begin{align*}
\tilde{Q}^{(2)}_{KK''K''K'}(t)=&-\frac{\pi}{2}\int_0^t\textrm{d}t'\int_0^{t'}\textrm{d}t''\int\textrm{d}^3x\textrm{d}^3y\int\frac{\textrm{d}^3k\textrm{d}^3k'\textrm{d}^3p\textrm{d}^3p'}{(2\pi)^6}\frac{e^{-i (\boldsymbol{k}-\boldsymbol{k}')\cdot\boldsymbol{x}}e^{i \boldsymbol{k}\cdot\boldsymbol{x}_K}e^{-i \boldsymbol{k}'\cdot\boldsymbol{x}_{K''}}e^{-i (\boldsymbol{p}-\boldsymbol{p}')\cdot\boldsymbol{y}}e^{i \boldsymbol{p}\cdot\boldsymbol{x}_{K''}}e^{-i \boldsymbol{p}'\cdot\boldsymbol{x}_{K'}}}{|\boldsymbol{x}-\boldsymbol{y}|}\\
    &\times
    F(\boldsymbol{k})F^*(\boldsymbol{k}')
    F(\boldsymbol{p})F^*(\boldsymbol{p}')\delta((t'-t'')-|\boldsymbol{x}-\boldsymbol{y}|)\vartheta(t'-|\boldsymbol{x}-\boldsymbol{y}|)\\
    =&-\frac{\pi}{2}\int_0^t\textrm{d}t'\int\textrm{d}^3x\textrm{d}^3y\int\frac{\textrm{d}^3k\textrm{d}^3k'\textrm{d}^3p\textrm{d}^3p'}{(2\pi)^6}\frac{e^{-i (\boldsymbol{k}-\boldsymbol{k}')\cdot\boldsymbol{x}}e^{i \boldsymbol{k}\cdot\boldsymbol{x}_K}e^{-i \boldsymbol{k}'\cdot\boldsymbol{x}_{K''}}e^{-i (\boldsymbol{k}-\boldsymbol{k}'+\boldsymbol{p}-\boldsymbol{p}')\cdot\boldsymbol{y}}e^{i \boldsymbol{p}\cdot\boldsymbol{x}_{K''}}e^{-i \boldsymbol{p}'\cdot\boldsymbol{x}_{K'}}}{|\boldsymbol{x}|}\\
    &\times
    F(\boldsymbol{k})F^*(\boldsymbol{k}')
    F(\boldsymbol{p})F^*(\boldsymbol{p}')\vartheta(t'-|\boldsymbol{x}|)\\
    =&-\frac{\pi}{2}\int_0^t\textrm{d}t'\int\textrm{d}^3x\int\frac{\textrm{d}^3k\textrm{d}^3k'\textrm{d}^3p\textrm{d}^3p'}{(2\pi)^3}\vartheta(t'-|\boldsymbol{x}|)\frac{e^{-i (\boldsymbol{k}-\boldsymbol{k}')\cdot\boldsymbol{x}}e^{i \boldsymbol{k}\cdot\boldsymbol{x}_K}e^{-i \boldsymbol{k}'\cdot\boldsymbol{x}_{K''}}e^{i \boldsymbol{p}\cdot\boldsymbol{x}_{K''}}e^{-i \boldsymbol{p}'\cdot\boldsymbol{x}_{K'}}}{|\boldsymbol{x}|}\\
    &\times
    F(\boldsymbol{k})F^*(\boldsymbol{k}')
    F(\boldsymbol{p})F^*(\boldsymbol{p}')\delta^3(\boldsymbol{k}-\boldsymbol{k}'+\boldsymbol{p}-\boldsymbol{p}')\\
    =&-\pi\int_0^t\textrm{d}t'\int\frac{\textrm{d}^3k\textrm{d}^3k'\textrm{d}^3p\textrm{d}^3p'}{(2\pi)^2}e^{i \boldsymbol{k}\cdot\boldsymbol{x}_K}e^{-i \boldsymbol{k}'\cdot\boldsymbol{x}_{K''}}e^{i \boldsymbol{p}\cdot\boldsymbol{x}_{K''}}e^{-i \boldsymbol{p}'\cdot\boldsymbol{x}_{K'}}\frac{1-\cos(|\boldsymbol{k}-\boldsymbol{k}'|t')}{|\boldsymbol{k}-\boldsymbol{k}'|^2}
    \delta^3(\boldsymbol{k}-\boldsymbol{k}'+\boldsymbol{p}-\boldsymbol{p}')\\
    &\times
    F(\boldsymbol{k})F^*(\boldsymbol{k}')
    F(\boldsymbol{p})F^*(\boldsymbol{p}')\\
    =&-\pi\,t\,\int\frac{\textrm{d}^3k\textrm{d}^3k'\textrm{d}^3p\textrm{d}^3p'}{(2\pi)^2}e^{i \boldsymbol{k}\cdot\boldsymbol{x}_K}e^{-i \boldsymbol{k}'\cdot\boldsymbol{x}_{K''}}e^{i \boldsymbol{p}\cdot\boldsymbol{x}_{K''}}e^{-i \boldsymbol{p}'\cdot\boldsymbol{x}_{K'}}\frac{1-\textrm{sinc}(|\boldsymbol{k}-\boldsymbol{k}'|t)}{|\boldsymbol{k}-\boldsymbol{k}'|^2}
    \delta^3(\boldsymbol{k}-\boldsymbol{k}'+\boldsymbol{p}-\boldsymbol{p}')\\
    &\times
    F(\boldsymbol{k})F^*(\boldsymbol{k}')
    F(\boldsymbol{p})F^*(\boldsymbol{p}')
\end{align*}
}

\noindent We now change variables as done before and obtain
{\small
\begin{align*}
\tilde{Q}^{(2)}_{KK''K''K'}(t)=&\frac{\pi}{64}\,t\,\int\frac{\textrm{d}^3u\textrm{d}^3u'\textrm{d}^3v}{(2\pi)^2}e^{i \boldsymbol{u}\cdot(\boldsymbol{x}_K-\boldsymbol{x}_{K''})}e^{i \boldsymbol{u}'\cdot(\boldsymbol{x}_{K''}-\boldsymbol{x}_{K'})}e^{i \boldsymbol{v}\cdot(\boldsymbol{x}_K-\boldsymbol{x}_{K'})}\frac{1-\textrm{sinc}(2|\boldsymbol{v}|t)}{2|\boldsymbol{v}|^2}
    F(\sqrt{2}\boldsymbol{u})
    F(\sqrt{2}\boldsymbol{u}')|F(\sqrt{2}\boldsymbol{v})|^2\\
    =&\frac{1}{512\pi^2\sigma^6}\,t\,\int\frac{\textrm{d}^3u\textrm{d}^3u'\textrm{d}^3v}{(2\pi)^2}e^{i \boldsymbol{u}\cdot(\boldsymbol{x}_K-\boldsymbol{x}_{K''})}e^{i \boldsymbol{u}'\cdot(\boldsymbol{x}_{K''}-\boldsymbol{x}_{K'})}e^{i \boldsymbol{v}\cdot(\boldsymbol{x}_K-\boldsymbol{x}_{K'})}\frac{1-\textrm{sinc}(2|\boldsymbol{v}|t)}{2|\boldsymbol{v}|^2}e^{-\frac{|\boldsymbol{u}|^2}{2\sigma^2}}e^{-\frac{|\boldsymbol{u}'|^2}{2\sigma^2}}e^{-\frac{|\boldsymbol{v}|^2}{\sigma^2}}\\
    =&\frac{1}{256\pi}\,\sigma\,t\,e^{-\sigma^2\frac{|\boldsymbol{x}_K-\boldsymbol{x}_{K''}|^2}{2}}e^{-\sigma^2\frac{|\boldsymbol{x}_{K''}-\boldsymbol{x}_{K'}|^2}{2}}\int\textrm{d}^3v\,e^{\sigma\,i \boldsymbol{v}\cdot(\boldsymbol{x}_K-\boldsymbol{x}_{K'})}\frac{1-\textrm{sinc}(2|\boldsymbol{v}|\sigma\,t)}{2|\boldsymbol{v}|^2}e^{-|\boldsymbol{v}|^2}\\
    =&\frac{1}{256}\,\sigma\,t\,e^{-\sigma^2\frac{|\boldsymbol{x}_K-\boldsymbol{x}_{K''}|^2}{2}}e^{-\sigma^2\frac{|\boldsymbol{x}_{K''}-\boldsymbol{x}_{K'}|^2}{2}}\int_{-\infty}^{+\infty}\textrm{d}v\,\textrm{sinc}(\sigma\, v|\boldsymbol{x}_K-\boldsymbol{x}_{K'}|)(1-\textrm{sinc}(2\,v\,\sigma\,t))e^{-v^2}.
\end{align*}
}
The expression above informs us that $\tilde{Q}^{(2)}_{KK''K''K'}(t)$ has the following properties:
\begin{subequations}
    \begin{align}
        \tilde{Q}^{(2)}_{KK''K''K'}(t)=&\tilde{Q}^{(2)*}_{KK''K''K'}(t)\\
        \tilde{Q}^{(2)}_{KK''K''K'}(t)=&\tilde{Q}^{(2)}_{K'K''K''K}(t)\\
        \tilde{Q}^{(2)}_{KK''K''K}(t)=&\tilde{Q}^{(2)}_{K''KKK''}(t)\\
         \tilde{Q}^{(2)}(t):=\tilde{Q}^{(2)}_{KKKK}(t)=&\frac{\sqrt{\pi}}{256}\sigma\,t\left(1+\frac{\sqrt{\pi}}{2}\frac{\textrm{erf}(\sigma\,t)}
         {\sigma\,t}\right)\\
          \tilde{Q}^{(2)}_{KK''K''K}(t)=&e^{-\sigma^2|\boldsymbol{x}_K-\boldsymbol{x}_{K''}|^2}\tilde{Q}^{(2)}(t).
    \end{align}
\end{subequations}
Note that $ \tilde{Q}^{(2)}(t)=\frac{\sqrt{\pi}}{256}\sigma\,t$ for $\sigma\,t\gg1$.
We are motivated to introduce 
    \begin{align}
    \tilde{Q}_x(t):=&\frac{1}{256}\int_{-\infty}^{+\infty}\textrm{d}v\,\textrm{sinc}(\sigma\, v\,x)(1-\textrm{sinc}(2\,v\,\sigma\,t))e^{-v^2},
    \end{align}
where $x\in\{0,L\}$.

We the  conclude that the renormalized version (see below) of $\bra{1_K}\hat{H}_{2*,\textrm{SE}}(t)\ket{1_{K'}}$ reads
\begin{align}
     \bra{1_K}\hat{H}_{2*,\textrm{SE}}(t)\ket{1_{K'}}=\tilde{Q}^{(2)}_{K\textrm{LL}K'}(t)+\tilde{Q}^{(2)}_{K\textrm{RR}K'}(t).
\end{align}

\vspace{0.4cm}

\noindent \textbf{Fyenman-like diagrammtic approach: basic ingredients}---We now introduce a diagrammatic approach \textit{á la Feynman} to the evolution of the system. Instead of looking at the standard picture in sharp momentum space \cite{Srednicki:2007,Biswas:Bose:2023}, which for our purposes would not help since we are not using any sharp momentum approximations \cite{Biswas:Bose:2023}, we use an effective approach where we map the sharp momentum picture is mapped to a smeared particle picture, which effectively replaces integrals with sums and allows for quantum coherence to be included. We stress that, while this new approach is technically different, the results it gives should be independent of this ``choice of basis''. Below we will also discuss ``renormalization'' issues, but note that in general the full theory is not renormalizable \cite{Bergshoeff:Hohm:2009}.

The new picture is composed of extended particles and interaction vertices between such particles. To achieve our goal we list the main modified ingredients:
\begin{enumerate}
    \item \textit{Propagators:} we introduce lines for each ``type of particle''. Since we are interested in differentiating between the excitations in the system S and those in the environment E, we will have the following:
    \begin{itemize}
        \item \textit{System particle}: we will use a solid line for the excitations in the system. Such lines are labeled by the generic system-label $K$.
        \item \textit{Environment particle}: we will use a fuzzy solid line for the excitations in the system-environment. Such lines are labelled by $\underline{\lambda}$
        \item \textit{Gravitons}: we will use a dashed line for the excitations in the system-environment. Such lines are labelled generally by $h$.
        \item \textit{Mixed propagator:} the presence of coherence requires us to introduce a new type of line, which cannot exist in standard approaches to quantum field theory: this is a line composed by one system S and one environment E propagator joining at a point, which we highlight with a dot and to which we give the weight $\beta$.
    \end{itemize}
    \item \textit{Vertex:} the vertex consists of three lines joining at a point, one of which is always a graviton propagator, while the other two can be of any particle type (i.e., system or environment). The weight of the vertex is $\sqrt{G}$.
    \item \textit{Fusion rules:} two lines of the same type can be joined, if the are ``loose'' at least at one end. Any such fused lines that end up connecting two vertices represent a sum over the degrees of freedom of the type of particle, in analogy to the fact that internal lines in standard quantum field theory indicate integration over sharp momenta \cite{Srednicki:2007}. In the case of internal graviton lines, this provides a graviton-like propagator $\tilde{D}_{\mu\nu,\mu'\nu'}(x-y)$.
\end{enumerate}
An illustration of the basic ingredients is given in Figure~\ref{Fig:FDB}.

\begin{figure}[ht!]
\includegraphics[width=\linewidth]{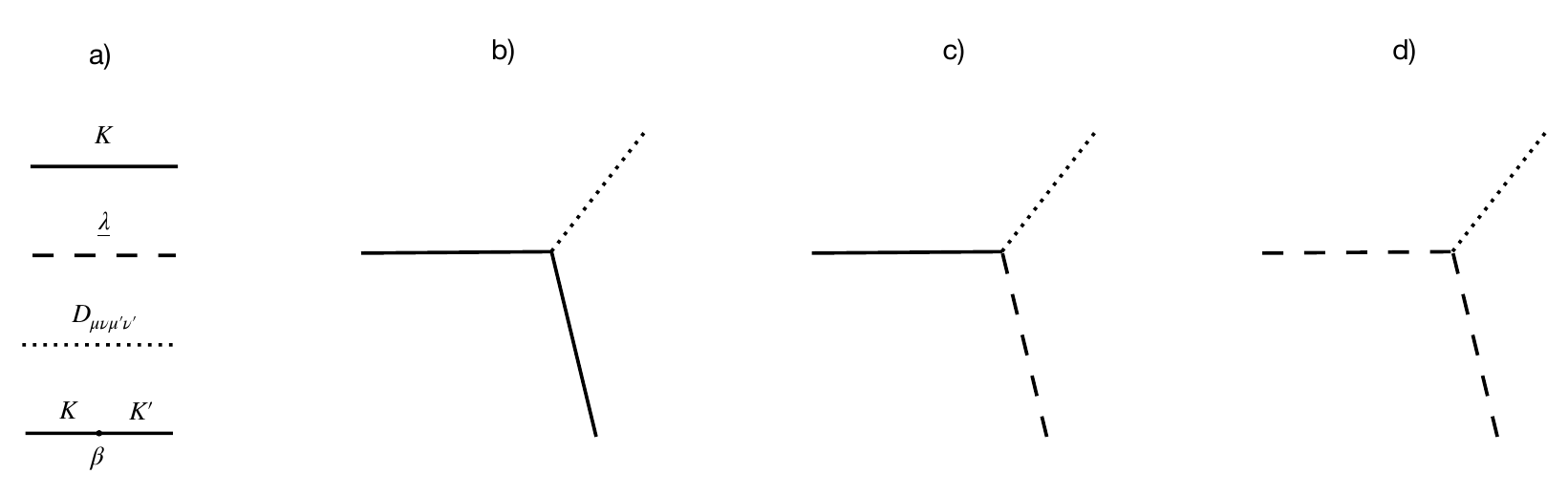}
\caption{Pictorial representation of the basic tools for the diagrammatic approach presented here. In panel a) the lines associated to the ``free propagation'' are given, including the one with weight $\beta$ which allows for quantum coherence, while in the remaining b)-d) panels Feynman vertices between the particles (represented by the solid and dashed lines) and the graviton (represented by the dotted line) are given.}\label{Fig:FDB}
\end{figure}

In order to determine which types of diagrams will represent our interactions, we need to find a relation between the number of external sources, propagators and vertices. In the standard $\phi^3$ one has $E=2P-3V$, where $E$ stands for the number of external sources, $P$ stands for number of propagators, and $V$ stands for the number of vertices \cite{Srednicki:2007}. In our case, we need to account for the different types of particles. It is not difficult to extend the logic of the simple $\phi^3$ theory to our case by the following procedure: we introduce the vectors $\vec{E}:=(E_\textrm{S},E_\textrm{E},E_\textrm{G})^\textrm{Tp}$, $\vec{P}:=(P_\textrm{S},P_\textrm{E},P_\textrm{G})^\textrm{Tp}$, $\vec{V}_{20}:=(2V,0,V)^\textrm{Tp}$, $\vec{V}_{11}:=(V,V,V)^\textrm{Tp}$, and $\vec{V}_{02}:=(0,2V,V)^\textrm{Tp}$, where it is clear that $E_j$ is the number of external sources of type $j=\textrm{S},\textrm{E},\textrm{G}$, $P_j$ s the number of external propagators of type $j$, and $(V_{nm})_j$ is the number of vertices of the type listed above.\footnote{Here the superscript Tp stands for transposition.} Then, since each propagator adds two sources of its type, while each vertex removes an appropriate number sources per type, we have the relation
\begin{align}
    \vec{E}=2\vec{P}-\vec{V}_{20}-\vec{V}_{11}-\vec{V}_{02},
\end{align}
with the constraint that $E_j\geq0$ for $j=1,2,3$ since there cannot be negative numbers of sources.
This relation can be conveniently written in terms of its components as
\begin{align}\label{explicit:diagram:constraint:appendix}
    \begin{pmatrix}
        E_\textrm{S}\\
        E_\textrm{E}\\
        E_\textrm{G}
    \end{pmatrix}
    =
    \begin{pmatrix}
        2P_\textrm{S}-2V_{20}-V_{11}\\
        2P_\textrm{E}-2V_{02}-V_{11}\\
        2P_\textrm{G}-V_{20}-V_{02}-V_{11}
    \end{pmatrix}.
\end{align}
We define $E:=\sum_j E_j$, $P:=\sum_j P_j$, and $V:=V_{20}+V_{02}+V_{11}$, and we find that these quantities satisfy the constraint
\begin{align}\label{general:diagram:constraint:appendix}
    E=2P-3V
\end{align}
which is the general constraint for a theory where three fields meet at a point, i.e., a cubic theory, as discussed before.

\vspace{0.4cm}

\noindent \textbf{Fyenman-like diagrammtic approach: $E=1,2,3$ diagrams up to second order}---The properties listed above constrain the types of diagrams that can appear. We here compute to lowest nontrivial order (where by lowest order we mean that there is at least one vertex of any type) all diagrams that have no external line, one external line of any type, and two external line of any type. 

\begin{itemize}
    \item \textit{Vacuum diagrams}: We start by seeking all vacuum diagrams to lowest order. They are defined by $\vec{E}=0$ and thus $E=0$. The constraint \eqref{general:diagram:constraint:appendix} informs us that we will have $P=3$ and $V=2$ to lowest order, and therefore the diagrams of interest will be of second order. It is immediate to see that $0=E_j=2P_j-(V_{20})_j-(V_{11})_j-(V_{02})_j$ to lowest order has as only unique solution $P_j=1$ for all $j$ and $\vec{V}_{20}=\vec{V}_{02}=0$, and thus there are exactly two vertices of type $\vec{V}_{11}$. This diagram is given in Figure~\ref{Fig:FD1}a.
    \item \textit{Tadpole-like diagrams}: We continue by seeking all vacuum diagrams that have one external line of any type (also known as tadpole diagrams \cite{Srednicki:2007}). This means that $1=E=E_\textrm{S}+E_\textrm{E}+E_\textrm{G}$. The constraint \eqref{general:diagram:constraint:appendix} informs us that we will have $P=2$ and $V=1$ to lowest order, and therefore the diagrams of interest will be of first order. This sets the constraint $1=V=V_{20}+V_{02}+V_{11}$. Inserting this latter constraint into the expression for $E_\textrm{G}$ gives us $E_\textrm{G}=2P_\textrm{G}-1$, which has as only solution $E_\textrm{G}=P_\textrm{G}=1$ given the constraints of the problem. In turn, this implies $E_\textrm{S}=E_\textrm{E}=0$ as well as $P_\textrm{S}+P_\textrm{E}=1$ since there are two propagators in total. Inserting this in the expressions \eqref{explicit:diagram:constraint:appendix} gives us $P_\textrm{S}-V_{20}=P_\textrm{E}-V_{02}$. Given that there is one propagator left to assign, and there is only one vertex in the system, this constraint has as only solutions $P_\textrm{S}=V_{20}=1$ and $P_\textrm{E}=V_{02}=0$, or viceversa. These two diagrams are given in Figure~\ref{Fig:FD1}b and Figure~\ref{Fig:FD1}c.
    \item \textit{Propagator-like diagrams with two system external lines}: We finally move on to our last case, that of diagrams with two external lines. We restrict ourselves to diagrams that have only system external lines, which are those of interest to this work. Thus, $E_\textrm{E}=E_\textrm{G}=0$. This case is defined by $2=E=E_\textrm{S}$ and $E_\textrm{E}=E_\textrm{G}=0$. The constraint \eqref{general:diagram:constraint:appendix} informs us that we will have $P=4$, $V=2$ , and no graviton external lines to lowest order, and therefore the diagrams of interest will be of second order.

    The expression for $E_\textrm{G}$ in \eqref{explicit:diagram:constraint:appendix} tells us that $P_\textrm{G}=1$ given our constraint here. We are thus left with $2=E_\textrm{S}+E_\textrm{E}$ and $3=P_\textrm{S}+P_\textrm{E}$. We then look at the expression for $E_\textrm{E}$ in \eqref{general:diagram:constraint:appendix} and see that it implies $2P_\textrm{E}=2V_{02}+V_{11}$. This, in turn, implies that either $V_{11}=0$ or $V_{11}=2$. We consider these cases separately:
    \begin{enumerate}
        \item \textit{Case} $V_{11}=0$. Here, we have $V_{20}+V_{02}=2$,  $1=P_\textrm{S}-V_{20}$, and $P_\textrm{E}=V_{02}$. The three possible cases are listed in the table below:

        \begin{table}[ht!]
    \begin{tabular}{|c|c|c|c|}
                  \hline
        $V_{20}$ & $V_{02}$ & $P_\textrm{S}$ & $P_\textrm{E}$ \\
       \hline
       \hline
       $2$  & $0$ & $3$ & $0$  \\
       \hline
       $1$  & $1$ & $2$ & $1$  \\
       \hline
       $0$  & $2$ & $1$ & $2$  \\
       \hline
    \end{tabular}
    \caption{Here we give the possible occurrences of Case $V_{11}=0$.}
    \label{tab:my_label2}
\end{table}
The last line is not connected. It is the product of two systems S propagators and a vacuum diagram composed of two $\vec{V}_{02}$ vertices. This case is discarded as it is not of interest.

        \item \textit{Case} $V_{11}=2$. Here, we have $V_{20}=V_{02}=0$, and thus $P_\textrm{S}=2$ and $P_\textrm{E}=1$.
    \end{enumerate}
    All diagrams of this case are given in Figure~\ref{Fig:FD1} and Figure~\ref{Fig:FD2}.
\end{itemize}

\begin{figure}[ht!]
\includegraphics[width=0.8\linewidth]{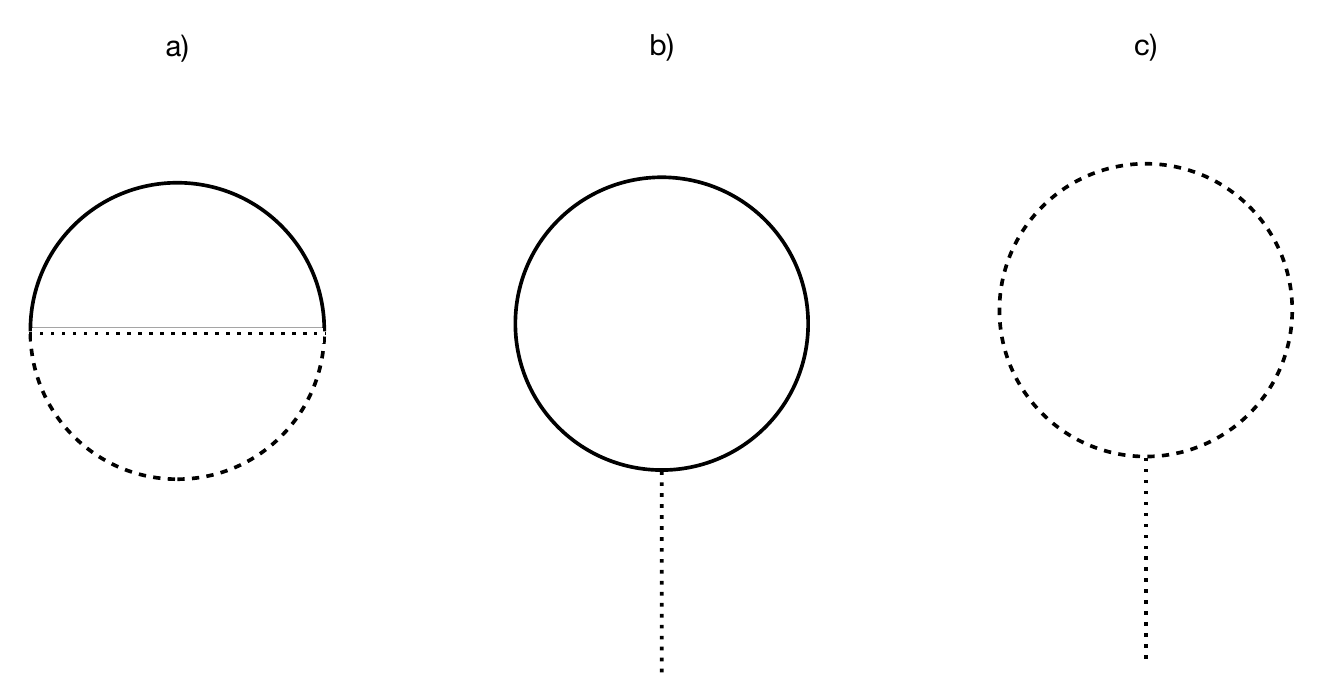}
\caption{Pictorial representations of all vacuum diagrams (panel a)), as well as all diagrams with one external line, or “tadpoles”,
(panels b) and c)).}\label{Fig:FD2}
\label{Fig:FD1}
\end{figure}
\begin{figure}[ht!]
\includegraphics[width=0.8\linewidth]{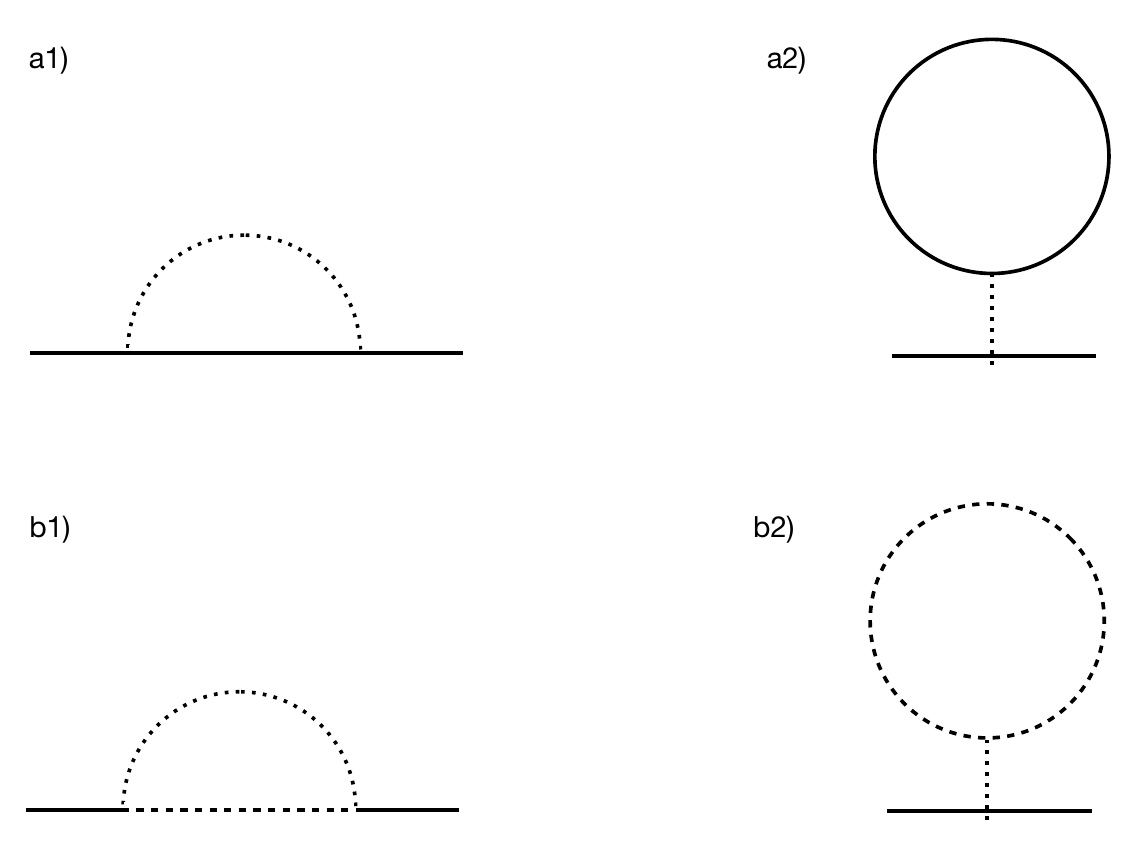}
\caption{Pictorial representation of diagrams with two external lines. Diagrams in panel b1) and b2) involve sums over all degrees of freedom $\underline{\lambda}$ of the system environment E (represented by the internal dahsed line).}\label{Fig:FD2}
\end{figure}

\noindent \textbf{Renormalization of tadpole terms}---We now note a problem that is common to interacting cubic field theories \cite{Srednicki:2007}. In particular, we require that
\begin{align}
    A^{\mu\nu}\bra{0}\hat{h}_{\mu\nu}(t,\boldsymbol{x})\ket{0}=0
\end{align}
for any function $A^{\mu\nu}$, where $\ket{0}$ is the full vacuum of the theory. This, however, cannot occur if tadpole-like diagrams exist. Thus, we employ the standard approach and add to the full Hamiltonian a counter-term 
\begin{align}\label{tadpole:counterterm:appendix}
    \hat{H}_\textrm{ct,1}:=\int\textrm{d}^3x A^{\mu\nu}(\boldsymbol{x})\hat{h}_{\mu\nu}(\boldsymbol{x}),
\end{align}
where $A^{\mu\nu}(\boldsymbol{x})$ is a function of $\epsilon=m/m_\textrm{P}$ such that $A^{\mu\nu}(\boldsymbol{x})=0$ for $\epsilon=0$. The functional expression of $A^{\mu\nu}(\boldsymbol{x})$ can be determined at each order in $\epsilon$ by requiring that it cancels tadpole diagrams. For example, the $\mathcal{O}(\epsilon)$ contribution should cancel the linear combination of the two diagrams given in Figure~\ref{Fig:FD1}a and Figure~\ref{Fig:FD1}b.
As is standard in quantum field theory, this contribution will compensate for the diagrams given in Figure~\ref{Fig:FD1}a2 and Figure~\ref{Fig:FD1}b2.

The counter-term \eqref{tadpole:counterterm:appendix} can be associated to the diagram given in Figure~\ref{Fig:FD3}a1, which is itself a counterterm and allows us to obtain all other counter-terms given, such the one given in Figure~\ref{Fig:FD3}b2, which compensates diagrams such as those in Figure~\ref{Fig:FD2}a2 and Figure~\ref{Fig:FD2}b2.

\begin{figure}[ht!]
\includegraphics[width=1\linewidth]{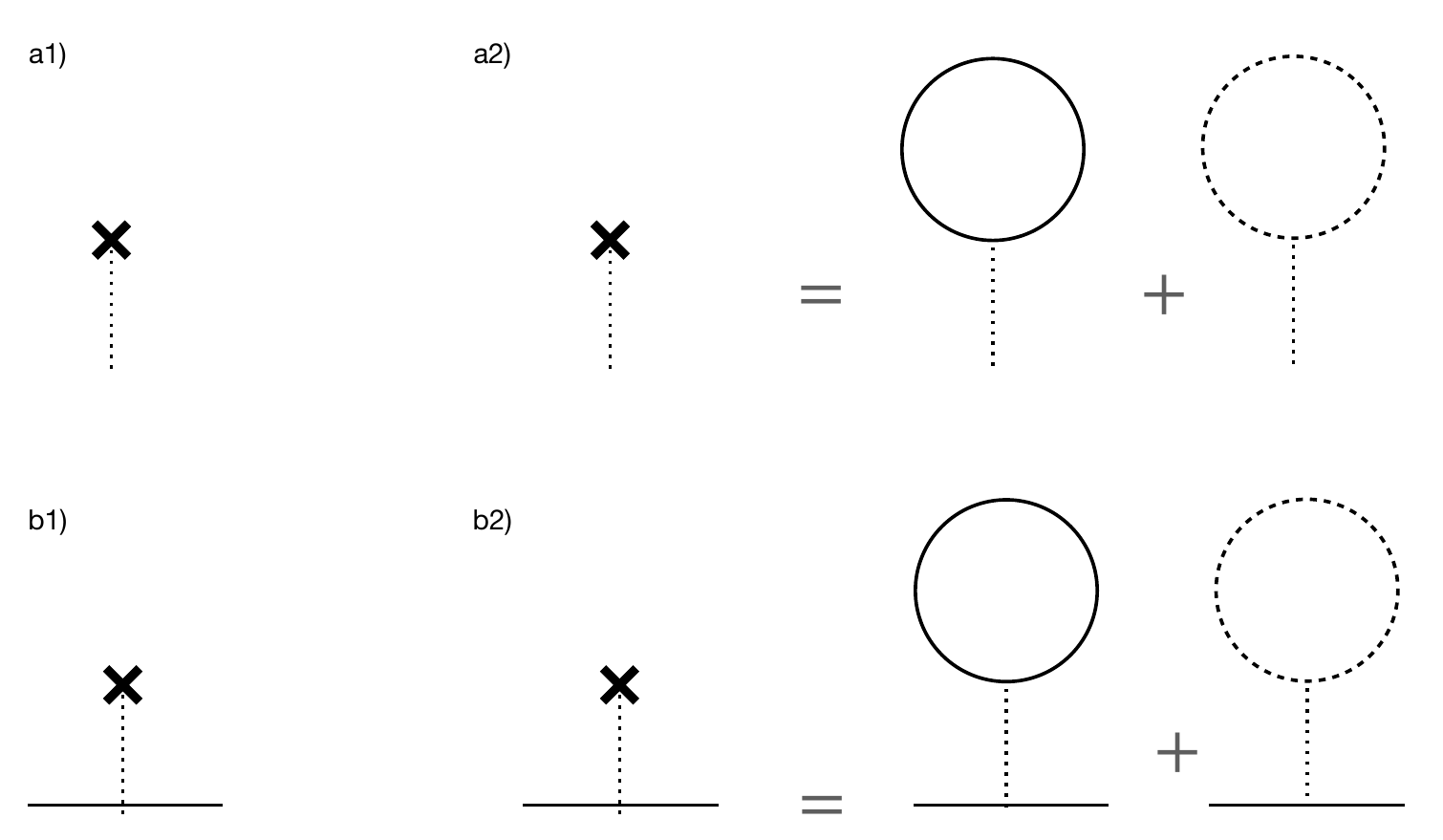}
\caption{Pictorial representation of the counterterms to divergent diagrams containing tadpoles (i.e., those that have one external line), which are divergent \cite{Srednicki:2007}.}\label{Fig:FD3}
\end{figure}

From now on, we assume that all \emph{one-particle irreducible} diagrams have been renormalized in this fashion \cite{Srednicki:2007}.

\vspace{0.4cm}

\noindent \textbf{Vacuum diagrams and their role in non-unitary dynamics}---In quantum field theory, vacuum diagrams do not contribute to physical processes \cite{Srednicki:2007}. The formal reason is that the normalization of the functional integral $Z(J)$ satisfies the normalization condition $Z(J=0)=1$. This, in turn, is implicitly derived as a consequence that the scattering processes of interest in the theory are unitary processes.

Here, we are adopting a picture of Feynman diagrams inspired by standard path-integral formulations, however, the evolution is not unitary. Interestingly, there has been little development of path-integral-like techniques that can be applied to states where quantum coherence is initially present \cite{Williams:Dougherty:2024}, and where the evolution is nonunitary \cite{Feynman:Vernon:1963,Caldeira:Leggett:1983,Cosacchi:Cygorek:2018}.

In our case, we see that the vacuum diagram is given by $\sum_K Q_{KK}(t)$, while the two diagrams $Q_{\textrm{LL}}(t)$ and $Q_{\textrm{RR}}(t)$ in Figure~\ref{Fig:FD2}b1 are obtained by selecting the same input and output type of excitation $K$, and correctly we have that the vacuum diagram is equivalent to the sum of these two diagrams. We can also have, due to the new mixed propagator with weight $\beta$, vacuum contributions of the type $\beta Q_{\textrm{LR}}(t)+\beta^*Q_{\textrm{RL}}(t)$ which, again, are compensated by appropriate diagrams of Figure~\ref{Fig:FD2}b1 when one of the external propagators is composed to a mixed propagator with weight $\beta$.

More concretely, we note that $\bra{1_K}\hat{T}^{\mu'\nu'}_y\ket{1_{\underline{\lambda}}}\bra{1_{\underline{\lambda}}}\hat{T}^{\mu\nu}_x\ket{1_{K'}}=\bra{1_{\underline{\lambda}}}\hat{T}^{\mu\nu}_x\ket{1_{K'}}\bra{1_K}\hat{T}^{\mu'\nu'}_y\ket{1_{\underline{\lambda}}}$, and therefore
\begin{align*}
    \int \textrm{d}^3x\textrm{d}^3y\tilde{D}_{\mu\nu\mu'\nu'}(\boldsymbol{x}-\boldsymbol{y})&\sum_{K=\textrm{L},\textrm{R}}\sum_{\underline{\lambda}}\bra{1_K}\hat{T}^{\mu'\nu'}_y\ket{1_{\underline{\lambda}}}\bra{1_{\underline{\lambda}}}\hat{T}^{\mu\nu}_x\ket{1_{K}}\\
    =&\int \textrm{d}^3x\textrm{d}^3y\tilde{D}_{\mu\nu\mu'\nu'}(\boldsymbol{x}-\boldsymbol{y})\sum_{\underline{\lambda}}\sum_{K=\textrm{L},\textrm{R}}\bra{1_{\underline{\lambda}}}\hat{T}^{\mu'\nu'}_y\ket{1_{K}}\bra{1_K}\hat{T}^{\mu\nu}_x\ket{1_{\underline{\lambda}}}
\end{align*}
due to the symmetric properties of $\tilde{D}_{\mu\nu\mu'\nu'}(\boldsymbol{x}-\boldsymbol{y})$.

We can then introduce a second counter term $\hat{H}_\textrm{ct,2}$ that reads
\begin{align}\label{propagator:counterterm:appendix}
\hat{H}_\textrm{ct,2}:=B[\ket{1_\textrm{L}}\bra{1_\textrm{L}}+\ket{1_\textrm{R}}\bra{1_\textrm{LR}}],
\end{align}
where $B$ vanishes for $\epsilon=0$ and to lowest order contributes to second order in $\epsilon$. Effectively, this is a renormalization of the propagator, as can be seen diagrammatically in Figure~\ref{Fig:FD4}, which also renormalizes corresponding vacuum diagrams as discussed here.

\begin{figure}[ht!]
\includegraphics[width=1\linewidth]{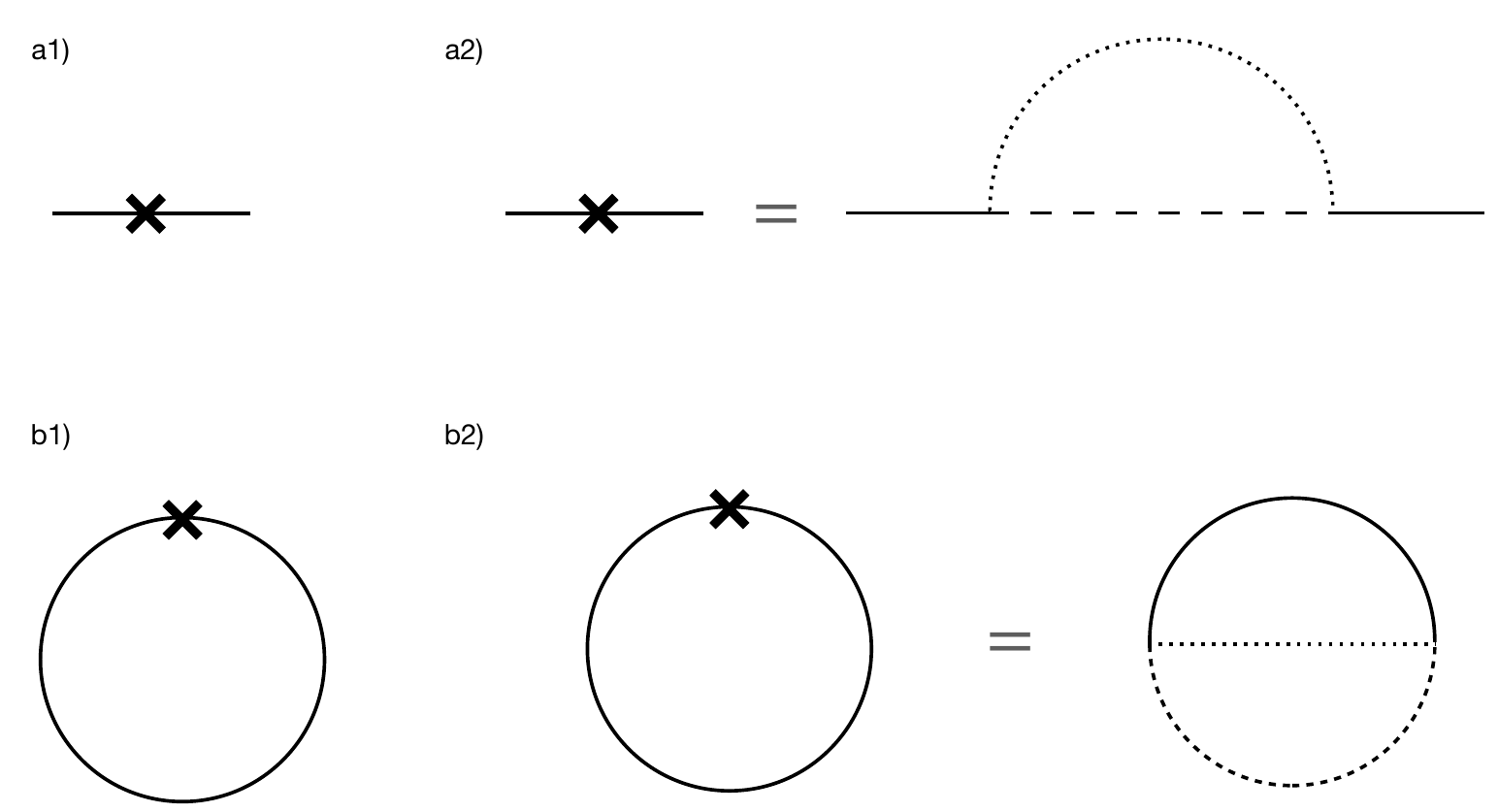}
\caption{Pictorial representation of the counterterms to divergent diagrams with two external lines (i.e., those that have an internal dahsed line) to order $\epsilon^2$.}\label{Fig:FD4}
\end{figure}

This discussion allows us to make the following important conclusion: when studying nonunitary dynamics, vacuum diagrams \emph{do not vanish} due to the normalization conditions of the functional integral $Z(J)$. Instead, they indicate the presence of ``leakage'' from the system of interest to other subsystems. Nevertheless, since the global dynamics including the environment must be unitary, the vacuum diagrams must collectively be compensated by other terms that appear in the system's state, and can therefore be discarded or ``renormalized'' if one wishes to condition the dynamics the those events where no leakage has occurred. This can be achieved by effectively setting to zero any diagram that involves internal propagators associated to the environment.

%---------------------------%
\subsection{Computation of the matrix elements}
%---------------------------%
We are now finally in the position of computing the elements $\hat{\rho}_{\textrm{S},KK'}(t)$ of the reduced state $\hat{\rho}_\textrm{S}(t)$ in the $\ket{1_\textrm{L}}$ and $\ket{1_\textrm{R}}$ basis, where $K,K'=\textrm{R,L}$. The most efficient way to achieve this goal is to compute
\begin{align}\label{system:coefficients:definition:appendix}
    \rho_{\textrm{S},KK'}(t):=\bra{1_K}\textrm{Tr}_\textrm{E}(\hat{\rho}_\textrm{SE}(t))\ket{1_{K'}}.
\end{align}
We now use the explicit form of the state $\hat{\rho}_\textrm{SE}(t)$ and the abstract basis $\mathcal{B}=\{\ket{\textrm{L}},\ket{\textrm{R}},\ket{1_{\underline{\lambda}}}\}$ to trace out the Environment E. We note that only one particle transitions are allowed in the rotating wave approximation. The trace can be achieved by using the general expression 
\begin{align}
    \textrm{Tr}_\textrm{E}(\rho_\textrm{SE})\equiv\bra{0_\textrm{E}}\hat{\rho}_\textrm{SE}\ket{0_\textrm{E}}+\sum_{\underline{\lambda}}\bra{1_{\underline{\lambda}}}\hat{\rho}_\textrm{SE}\ket{1_{\underline{\lambda}}},
\end{align}
where now we have used the notation $\ket{1_{\underline{\lambda}}0_{\underline{\lambda}'}0_{\underline{\lambda}''}...\}}\equiv\ket{1_{\underline{\lambda}}}:=\hat{a}^\dag_{\underline{\lambda}}\ket{0_\textrm{E}}$. In order to apply this trace we need to use the identity operator \eqref{identity:operator:one:particle:SE:subspaace:appendix} in the reduced one-particle subspace as follows $\hat{\rho}_\textrm{SE}(t)=\mathds{1}_{1\textrm{p,SE}}\hat{\rho}_\textrm{SE}(t)\mathds{1}_{1\textrm{p,SE}}$, and subsequently employ some straightforward algebra to find
\begin{align}\label{generic:perturbative:reduced:state:time:evoilution:final:general:approximated:form:appendix}
    \hat{\rho}_\textrm{S}(t)=&\rho_{\textrm{S},00}(t)\ket{0_\textrm{S}}\bra{0_\textrm{S}}+\sum_{K,K'=\textrm{L,R}}\rho_{\textrm{S},KK'}(t)\ket{K}\bra{K'},
\end{align}
which is precisely of the form of interest for the state.
Note that, since we work in the perturbative regime, we will have
\begin{align*}
    \rho_{\textrm{S},KK'}(t)=\rho_{\textrm{S},KK'}(0)+\epsilon^2\rho_{\textrm{S},KK'}^{(2)}(t),
\end{align*}
where $\epsilon=\frac{m}{m_\textrm{P}}$.

We now recall the results of the previous subsection and seek the contributing diagrams to the process of interest.  Below we discuss the different contributions. However, before proceeding, in Figure~\ref{Fig:FD5} we list the diagrams of the terms that are expected to effectively contribute. 

\begin{figure}[ht!]
\includegraphics[width=0.9\linewidth]{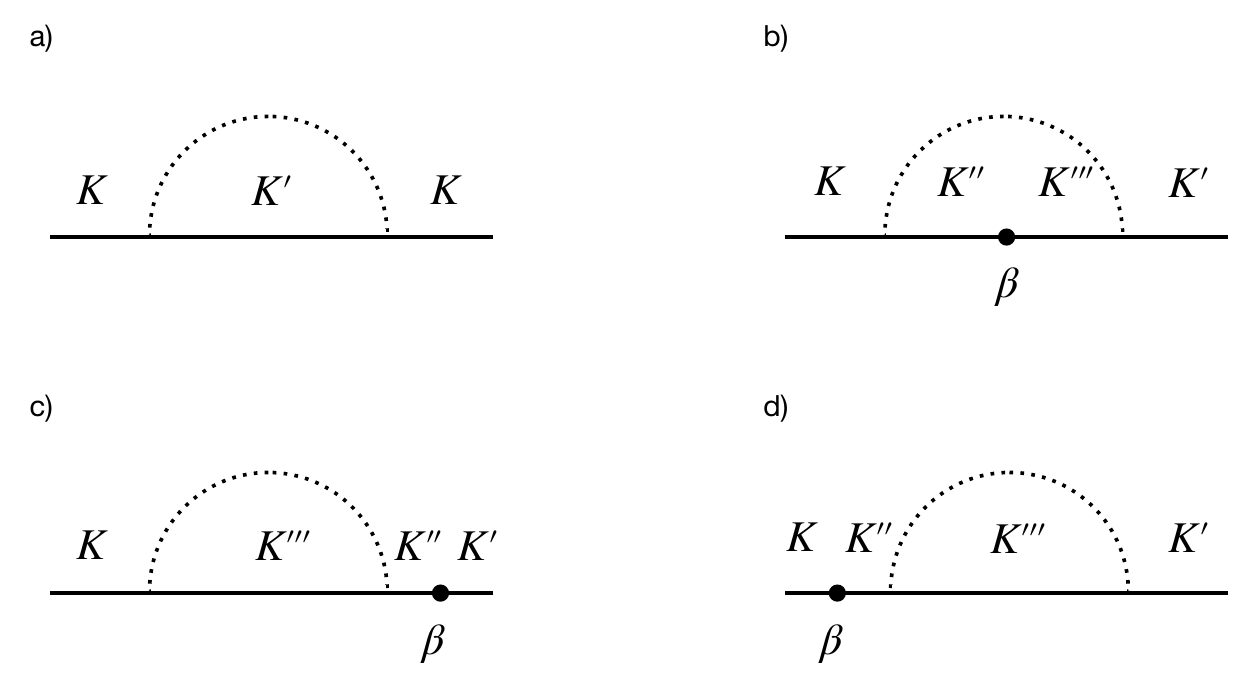}
\caption{Pictorial representation of all diagrams with two external lines that contribute, to second order in $\epsilon$, to the particle-in particle-out process of interest}\label{Fig:FD5}
\end{figure}

\vspace{0.4cm}

\noindent \textbf{Vacuum contribution}: The first term to be considered is $\rho_{\textrm{S},00}(t)\ket{0_\textrm{S}}\bra{0_\textrm{S}}$, which must occur at $\mathcal{O}(\epsilon^2)$ since there is no initial contribution to this term. The second order coefficient $\rho_{\textrm{S},00}^{(2)}(t)$ of this term is easily obtained by computing $\rho_{\textrm{S},00}^{(2)}(t)=\bra{0_\textrm{S}}\int \sum_{\underline{\lambda}}\bra{1_{\underline{\lambda}}}\hat{\rho}_\textrm{SE}\ket{1_{\underline{\lambda}}}\ket{0_\textrm{S}}$.

\vspace{0.4cm}

\noindent \textbf{Single-particle contributions}: The remaining set of terms to be considered is that of $\hat{\rho}_{\textrm{S},KK'}(t)$ defined above. It is immediate to see that they are obtained as linear contributions of elements $Q_{KK''K'''K'}$, $\tilde{Q}_{KK''K'''K'}$, $Q_{KK'}$, and $\tilde{Q}_{KK'}$.
Once the ``renormalization'' procedure has been applied, we know that the contributions $\tilde{Q}_{KK''K'''K'}$ and $\tilde{Q}_{KK'}$ will not be present.

\vspace{0.4cm}

\noindent \textbf{Full list of coefficients}: We now give a list of the explicit expressions of the second order coefficients $\rho^{(2)}_{\textrm{S},KK'}(t)$ for $K,K'\in\{\textrm{L,R}\}$, including $\rho^{(2)}_{\textrm{S},00}(t)$, which are the factors of the $\mathcal{O}(\epsilon^2=m^2/m_\textrm{P}^2)$ corrections to the state coefficients. 
We omit the explicit dependence on time in order to simplify the expressions, use the properties \eqref{Q:quantities:symmetries:appendix}, and find the full expressions
\begin{subequations}
    \begin{align}
    \rho^{(2)}_{\textrm{S},00}=&\alpha Q_\textrm{LL}+(1-\alpha) Q_\textrm{RR}+\beta Q_\textrm{RL}+\beta^*Q_\textrm{LR},\\
    \rho^{(2)}_{\textrm{S},\textrm{LL}}=&(1-2\alpha) Q_\textrm{LRRL}+ \frac{\beta}{2} Q_\textrm{RLLL}+\frac{\beta^*}{2} Q_\textrm{LRLL}-\frac{\beta^*}{2} Q_\textrm{LRRR}-\frac{\beta}{2} Q_\textrm{RLRR}-\alpha Q_\textrm{LL}-\frac{\beta^*}{2} Q_\textrm{LR}-\frac{\beta}{2} Q_\textrm{RL},\\
    \rho^{(2)}_{\textrm{S},\textrm{RR}}=&-(1-2\alpha) Q_\textrm{LRRL}+ \frac{\beta}{2} Q_\textrm{RLRR}+\frac{\beta^*}{2} Q_\textrm{LRRR}-\frac{\beta}{2} Q_\textrm{RLLL}-\frac{\beta^*}{2} Q_\textrm{LRLL}-(1-\alpha) Q_\textrm{RR}-\frac{\beta}{2} Q_\textrm{RL}-\frac{\beta^*}{2} Q_\textrm{LR},\\
    \rho^{(2)}_{\textrm{S},\textrm{LR}}=&\frac{(1-2\alpha)}{2} Q_\textrm{LRRR}-\frac{(1-2\alpha)}{2} Q_\textrm{LRLL}+ \beta Q_\textrm{LLRR}+\beta^* Q_\textrm{LRLR}-\beta Q_\textrm{LLLL}-\beta Q_\textrm{LRRL}
    \nonumber\\
    &
    -\frac{(1-\alpha)}{2} Q_\textrm{LR}-\frac{\alpha}{2} Q_\textrm{LR}-\frac{\beta}{2} Q_\textrm{LL}-\frac{\beta}{2} Q_\textrm{RR}-2i(1-2\alpha)\tilde{Q}_\textrm{LLLR}.
\end{align}
\end{subequations}

\vspace{0.4cm}

\noindent \textbf{List of renormalized coefficients}: Here we present the list of the coefficients given above where the renormalization scheme has been applied. We have
\begin{subequations}
    \begin{align}
    \rho^{(2)}_{\textrm{S},00}=&0,\\
    \rho^{(2)}_{\textrm{S},\textrm{LL}}=&(1-2\alpha) Q_\textrm{LRRL}+ \frac{\beta}{2} Q_\textrm{RLLL}+\frac{\beta^*}{2} Q_\textrm{LRLL}-\frac{\beta^*}{2} Q_\textrm{LRRR}-\frac{\beta}{2} Q_\textrm{RLRR},\\
    \rho^{(2)}_{\textrm{S},\textrm{RR}}=&-(1-2\alpha) Q_\textrm{LRRL}+ \frac{\beta}{2} Q_\textrm{RLRR}+\frac{\beta^*}{2} Q_\textrm{LRRR}-\frac{\beta}{2} Q_\textrm{RLLL}-\frac{\beta^*}{2} Q_\textrm{LRLL},\\
    \rho^{(2)}_{\textrm{S},\textrm{LR}}=&\frac{(1-2\alpha)}{2} Q_\textrm{LRRR}-\frac{(1-2\alpha)}{2} Q_\textrm{LRLL}+ \beta Q_\textrm{LLRR}+\beta^* Q_\textrm{LRLR}-\frac{\beta}{2} Q_\textrm{LLLL}-\beta Q_\textrm{LRRL}-\frac{\beta}{2} Q_\textrm{RRRR}-2i(1-2\alpha)\tilde{Q}_\textrm{LLLR}.
\end{align}
\end{subequations}
It is immediate to verify that $\rho^{(2)}_{\textrm{S},\textrm{LL}}+\rho^{(2)}_{\textrm{S},\textrm{RR}}=1$ now, which has compensated for the leaking of the modes out of the left-right system, as explained above.

\vspace{0.4cm}

\noindent \textbf{List of coefficients in the Gaussian momentum profile}: Here we present the list of the coefficients given above where the profile for the momentum distribution has been set to be Gaussian. In the case of real $\beta$ have
\begin{subequations}
    \begin{align}
    \rho^{(2)}_{\textrm{S},00}=&0,\\
    \rho^{(2)}_{\textrm{S},\textrm{LL}}=&(1-2\alpha) e^{-\sigma^2L^2}Q,\\
    \rho^{(2)}_{\textrm{S},\textrm{RR}}=&-(1-2\alpha) e^{-\sigma^2L^2}Q,\\
    \rho^{(2)}_{\textrm{S},\textrm{LR}}=& \beta\left[ Q_{2L}-Q\right]-2i(1-2\alpha)e^{-\frac{\sigma^2}{2}L^2}\tilde{Q}_L.
\end{align}
\end{subequations}

%---------------------------%
\section{TIME EVOLUTION AS A QUANTUM CHANNEL}\label{suppl_sec_channel}
%---------------------------%
The expression \eqref{generic:perturbative:reduced:state:time:evoilution:bettering:form:appendix} for the time evolution of the reduced System-Environment state can be therefore decomposed into two parts, which together form the \emph{channel}: the first, given by the first term, constitutes the unitary part of the evolution, while the second, given by the last two terms, constitutes the non-unitary part of the channel $\mathcal{N}$. Since both the unitary channel and the non-unitary channel occur at second order in the perturbative parameter $\epsilon$, they commute and can be treated separately. A pictorial representation of the process can be found in Figure~\ref{Fig:QC2}.

\begin{figure}[ht!]
\includegraphics[width=0.6\linewidth]{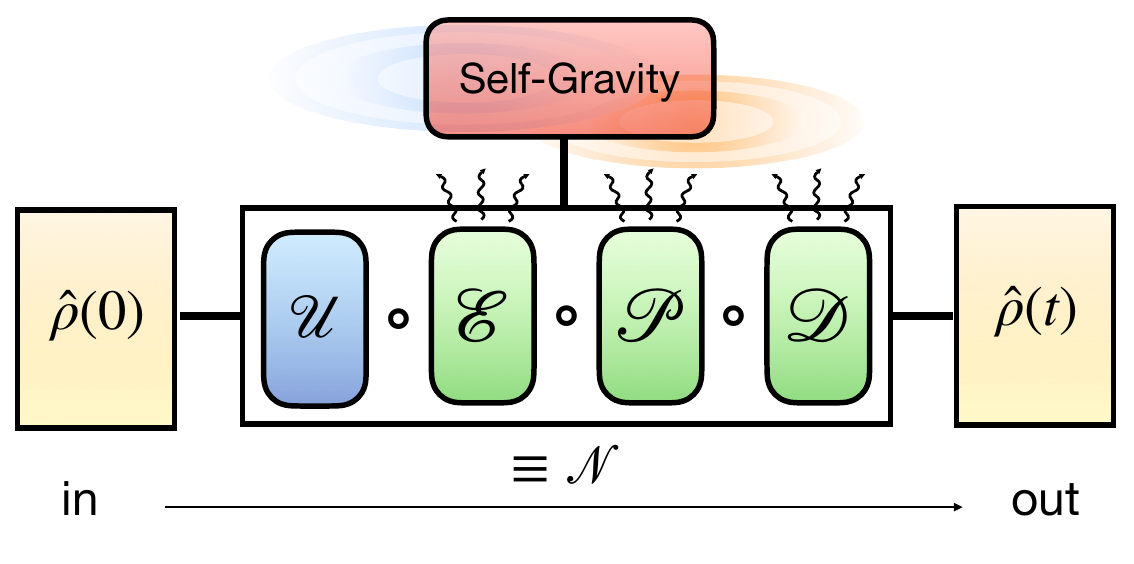}
\caption{\textbf{Self gravity as a quantum channel:} The figure gives a  pictorial representation of the time evolution of the reduced state $\hat{\rho}_\textrm{S}(t)$ of the system due to self gravity as quantum channel. The blue part is an effective unitary component, while the green part represent the genuine non-unitary component of the channel.}\label{Fig:QC2}
\end{figure}

In the following we will proceed to identify the components of the different channels. To do so, we will use a matrix notation that adapts to our setup. In particular, we will re-write \eqref{generic:perturbative:reduced:state:time:evoilution:final:general:approximated:form:appendix} conveniently
\begin{align}\label{useful:state:expression:for:channel:decompoisition:appendix}
    \hat{\rho}_\textrm{S}(t)=\rho_{\textrm{S},00}(t)\ket{0_\textrm{S}}\bra{0_\textrm{S}}\oplus
    \begin{pmatrix}
        \rho_{\textrm{S,LL}}(t) & \rho_{\textrm{S,LR}}(t)\\
        \rho_{\textrm{S,LR}}^*(t) & \rho_{\textrm{S,RR}}(t),
    \end{pmatrix}
\end{align}
where the matrix is defined in the $\textrm{L},\textrm{R}$ basis.

Since all channels occur at second order in $\epsilon=\frac{m}{m_\textrm{P}}$ and therefore commute, we work backwards within the chain of transformations
\begin{align}\label{channel:chain:decomposition:appendix}
    \hat{\rho}_\textrm{S}(0)
    \overset{\hat{U}}{\rightarrow}\hat{\rho}_{\textrm{S},1}(t)=\hat{U}\hat{\rho}_\textrm{S}(0)\hat{U}^\dag
    \overset{\mathcal{N}_2}{\rightarrow}\hat{\rho}_{\textrm{S},2}(t)=\mathcal{N}_2(\hat{\rho}_{\textrm{S},1}(t))
    \overset{\mathcal{N}_3}{\rightarrow}...
    \overset{\mathcal{N}_k}{\rightarrow}
    \hat{\rho}_\textrm{S}(t)=\mathcal{N}_k(\hat{\rho}_{\textrm{S},k-1}(t)),
\end{align}
which allows us to decompose the full channel in terms of elementary ones. This process (or, more precisely, its reverse) is achieved in full detail below. The crucial observation is that, at each step, we can consider the $\ell$-th channel to act on the initial state, since all channels commute and contribute to the same perturbation order.

%---------------------------%
\subsection{Basic channels}
%---------------------------%
Here we introduce the basic channels that will be useful in this work.
\begin{itemize}
    \item[*] \textbf{Unitary channel}---We start with considering the unitary channel. A unitary channel is the ``trivial'' channel implemented by a unitary operation $\hat{U}(\theta)$ via the transformation
\begin{align}
    \hat{\rho}(0)\rightarrow\hat{\rho}(\theta)=\hat{U}(\theta)\hat{\rho}(0)\hat{U}^\dag(\theta).
\end{align}
In the case of a small perturbation, we have $\hat{U}(\theta)=e^{-i \hat{H}_\textrm{I}(\theta)}\approx(\mathds{1}-i\hat{H}_\textrm{I}(\theta))$, and  we can then write the generic form of the channel as
\begin{align*}
    \hat{\rho}(0)\rightarrow\hat{\rho}(\theta)=\hat{\rho}(0)-i[\hat{H}_\textrm{I}(\theta),\hat{\rho}(0)].
\end{align*}
    \item[*] \textbf{Erasure channel}---We also consider an \emph{erasure channel} $\mathcal{E}$, which is implemented via the transformation
\begin{align}\label{generic:erasure:channel:appendix}
    \mathcal{E}(\hat{\rho}(0))\rightarrow\hat{\rho}(p)=p\ket{\textrm{e}}\bra{\textrm{e}}+(1-p)\hat{\rho}(0),
\end{align}
where $\ket{\textrm{e}}$ is a ``flag state that is orthogonal to the input Hilbert space'', while  $p$ is the probability of detecting the flag state \cite{Zhong:Oh:2023}. Note that $0\leq p\leq1$.
    \item[*] \textbf{Depolarization channel}---A depolarization channel $\mathcal{P}$ is defined by the transformation
\begin{align}\label{generic:depolarisation:channel:appendix}
    \mathcal{P}(\hat{\rho}(0))\rightarrow\hat{\rho}(\lambda)=\frac{\lambda}{d}\mathds{1}+(1-\lambda)\hat{\rho}(0),
\end{align}
where $d$ is the dimension of $\hat{\rho}(0)$ and $\lambda$ is the depolarization parameter. Note that the channel preserves the trace of the state. In our case, the state is effectively 2-dimensional after the erasure part has been taken care of, and therefore we set $d=2$.
    \item[*] \textbf{Dephasing channel}---A dephasing channel $\mathcal{D}$ is defined by the transformation
\begin{align}\label{generic:dephasing:channel:appendix}
    \mathcal{D}_\varphi(\hat{\rho}(0))\rightarrow\hat{\rho}(\varphi)
    =
    \begin{pmatrix}
        \rho_{00} & \rho_{01}(1-\varphi)\\
        \rho_{01}^*(1-\varphi) & \rho_{11}
    \end{pmatrix}
    ,
    \qquad
    \textrm{with}
    \qquad
    \hat{\rho}(0)\equiv
    \begin{pmatrix}
        \rho_{00} & \rho_{01}\\
        \rho_{01}^* & \rho_{11}
    \end{pmatrix}.
\end{align}
Here, $\varphi$ is the dephasing parameter. Note that the channel preserves the trace of the state.
\end{itemize}

%---------------------------%
\subsection{Full quantum channel due to self-gravity}
%---------------------------%
 We are now in a position to write explicitly the full channel $\mathcal{N}_\textrm{sg}$. We have that the evolution due to self gravity of the (reduced) initial state $\hat{\rho}_{\textrm{S,red}}(0)$ of the System S can be implemented by the quantum channel
 \begin{align}
     \mathcal{N}_\textrm{sg}\equiv \mathcal{E}_p\circ\mathcal{D}_\phi\circ\mathcal{P}_\lambda\circ \mathcal{U},
 \end{align}
 where we have denoted by $\mathcal{U}$ the unitary channel acting on the reduced state.
 
We apply the channels sequentially to our initial state $\hat{\rho}_\textrm{S}(0)$ and obtain the final state $\hat{\rho}_\textrm{S}(t)$ as $\hat{\rho}_\textrm{S}(0)\overset{\mathcal{N}_\textrm{sg}}{\rightarrow}\hat{\rho}_\textrm{S}(t)\equiv\mathcal{N}_\textrm{sg}(\hat{\rho}_\textrm{S}(0))$. In the perturbative regime that we are considering we find
\begin{align}
    \hat{\rho}_\textrm{S}(t)=p\ket{0_\textrm{S}}\bra{0_\textrm{S}}\oplus
    \begin{pmatrix}
        \alpha+\frac{\lambda}{d}-(\lambda+p)\alpha & \beta-(\lambda+\varphi+p)\beta-i\,h_2\\
        \beta^*-(\lambda+\varphi+p)\beta^*+i\,h_2 & (1-\alpha)+\frac{\lambda}{d}-(\lambda+p)(1-\alpha)
    \end{pmatrix},
\end{align}
where $h_2$ is the contribution due to the unitary part of the channel.

In the following we use the notation ``red'' for the reduced space of the left L and right R subsector only. Also, in the following we set $d=2$ since the reduced space just mentioned is effectively 2-dimensional.
Thus, we will use the notation $\hat{\rho}_\textrm{S}(t)=p\ket{0_\textrm{S}}\bra{0_\textrm{S}}\oplus\hat{\rho}_\textrm{S,red}(t)$, and focus on the effective state $\hat{\rho}_\textrm{S,red}(t)$. 

\vspace{0.2cm}
We now proceed to find the parameters $p,\lambda,\phi$ in terms of the quantities of the problem.
We have already argued that the unitary channel absorbs the free evolution and the interaction part  $\hat{H}_{\textrm{I}*,\textrm{red}}(t)$. It reads
\begin{align}
          \hat{U}_{*,\textrm{red}}(t)=&
            \mathds{1}
            -\sqrt{\pi}i(1-2\alpha)\tilde{Q}_2(t)\frac{e^{-\frac{1}{2}\sigma^2L^2}}{\sigma\,L}\,\epsilon^2\,\boldsymbol{\sigma}_\textrm{x},
\end{align}
where $\boldsymbol{\sigma}_\textrm{x}$ is the relevant Pauli matrix. 

The remaining parameters thus read
\begin{subequations}
    \begin{align}
        p=&\epsilon^2 (\alpha Q_\textrm{LL}+(1-\alpha) Q_\textrm{RR}+2\beta\Re( Q_\textrm{LR})),\\
        \lambda=&2\epsilon^2\,Q(t)\,e^{-\sigma^2L^2},\\
         \varphi=&\epsilon^2 \left((1-2e^{-\sigma^2L^2})\,Q(t)-Q_{2L}(t)\right).
    \end{align}
\end{subequations}
It is immediate to verify that the trace of the full state is preserved to second order, crucially by including the $\rho_{\textrm{S},00}(t)$ contribution. 

For completeness, we also write the renormalized parameters here. They read
\begin{subequations}\label{channel:parameters:appendix}
    \begin{align}
        p(t)=&0\\
        \lambda(t)=&2\epsilon^2\,Q(t)\,e^{-\sigma^2L^2},\\
         \varphi(t)=&\epsilon^2 \bigl((1-2e^{-\sigma^2L^2})\,Q(t)-Q_{2L}(t)\bigr).
    \end{align}
\end{subequations}

%---------------------------%
\subsection{Concrete effects}
%---------------------------%
We can now ask the question of the nature of the effects due to gravity. In order to achieve this goal, we note that the we need to ``renormalize'' the vacuum contribution, which is akin to conditioning all effects to the measurement of the particle itself.

Concretely, this means that the state $\hat{\rho}_\textrm{S}(t)$ has the effective expression
\begin{align}
    \hat{\rho}_\textrm{S,red}(t)=
    \begin{pmatrix}
        \alpha+\frac{1}{2}(1-2\alpha)\lambda & \beta-(\lambda+\varphi)\beta-i\,h_2\\
        \beta^*-(\lambda+\varphi)\beta^*+i\,h_2 & (1-\alpha)-\frac{1}{2}(1-2\alpha)\lambda
    \end{pmatrix}.
\end{align}
We can now compute the probability $P_K(t):=\bra{K}\hat{\rho}_\textrm{S,red}(t)\ket{K}$ for the particle to be found at location $K$. We have
\begin{subequations}
    \begin{align}
        P_\textrm{L}(t):=&\alpha+(1-2\alpha)\epsilon^2\,Q(t)\,e^{-\sigma^2L^2},\\
        P_\textrm{R}(t):=&1-\alpha-(1-2\alpha)\epsilon^2\,Q(t)\,e^{-\sigma^2L^2}.
    \end{align}
\end{subequations}
We see here that the particle ``localizes'' as a function of time, that is, the probability of finding the particle on the left L or the right R changes. Interestingly, when $\alpha\neq\frac{1}{2}$ the particle does not tend to localize at the position where it initially has a higher probability to be found, but undergoes a process of classical diffusion, where the probability of finding it in any of the possible positions tends to equalize. In other words, the state tends to the most classical state available, that is, the maximally mixed state. Even more interestingly, when $\alpha=\frac{1}{2}$ the probabilities $P_K(t)$ do not change as a function of time. 

We also can compute the coherence in the state. In general, a quantum state of the form $\hat{\rho}=\sum_{jk} \rho_{jk}\ket{j}\bra{k}$, where $\ket{j}$ form an orthonormal basis, contains quantum coherence. This can be seen by using the \emph{ relative entropy of coherence} $C (\hat{\rho}):=S(\hat{\rho}_\textrm{diag})-S(\hat{\rho})$, where $S(\hat{\rho}):=-\textrm{Tr}(\hat{\rho}\ln\hat{\rho})$ is the von Neumann entropy of a state $\hat{\rho}$ and $\hat{\rho}_\textrm{diag}$ is the state obtained from $\hat{\rho}$ by removing all off-diagonal terms \cite{Zhang:Shao:2016}. In our case we have 
\begin{align}
    C (\hat{\rho}_\textrm{S,red}(t))=-\sum_{\sigma=\pm}\lambda^\textrm{diag}_\sigma\ln\lambda^\textrm{diag}_\sigma+\sum_{\sigma=\pm}\lambda_\sigma\ln\lambda_\sigma,
\end{align}
where we have the following eigenvalues
\begin{subequations}
    \begin{align}
        \lambda^\textrm{diag}_\sigma=&\frac{1}{2}\left[1+\sigma|1-2\alpha|\left(1-\frac{1}{2}\lambda\right)\right],\\
        \lambda_\sigma=&\frac{1}{2}\left[1+\sigma\sqrt{(1-2\alpha)^2\left(1-\lambda\right)+4\beta^2\left(1-2\lambda-2\varphi\right)}\right].
    \end{align}
\end{subequations}
which have the explicit expressions
\begin{subequations}
    \begin{align}
        \lambda^\textrm{diag}_\sigma=&\frac{1}{2}\left[1+\sigma|1-2\alpha|(1-\epsilon^2Q(t)\,e^{-\sigma^2L^2})\right],\\
        \lambda_\sigma=&\frac{1}{2}\left[1+\sigma\sqrt{(1-2\alpha)^2(1-2\epsilon^2Q(t)\,e^{-\sigma^2L^2})+4\beta^2(1-2(Q(t)-Q_{2L}(t)))}\right].
    \end{align}
\end{subequations}
It is immediate to see that $C (\hat{\rho}_\textrm{S,red}(t))\equiv0$ when $\beta=0$, which clearly shows that there is coherence in the system if and only if initially there was coherence in the system.

Together, these results can be seen as a corroboration of the fact that, from a gravitational perspective, a system tends toward the most classical state possible

%---------------------------%
\subsection{Timescales}
%---------------------------%
We now argue that there are two timescales intrinsic to the problem.

\vspace{0.2cm}

\noindent \textbf{Diffusion timescale}---The first timescale is the diffusion timescale $t_\textrm{diff}$ that has been already computed before, and is expected from standard quantum field theoretical arguments \cite{Ford:Oconnell:2002,Edery:2021}. Given that its specific form cannot be determined in a universal fashion, we write it more generally as
\begin{align}
    t_\textrm{diff}:=\kappa_\textrm{diff}\frac{\ell_0^2 m}{\hbar}=&\kappa_\textrm{diff}\,t_\textrm{P}\frac{\ell_0^2}{\ell_\textrm{P}^2}\frac{m}{m_P},
\end{align}
where $\kappa_\textrm{diff}$ is a constant that dependes on the specific details of the setup (such as the shape of the particle).

\vspace{0.2cm}

\noindent \textbf{Dephasing timescale}---We then note that, since we are working in the perturbative regime, all corrections to the lowest order must be small. The expressions \eqref{channel:parameters:appendix} for the coefficients of the channels inform us that, in order for $\lambda\epsilon^2\ll1$, and $\varphi\epsilon^2\ll1$ to be valid at all times, we must at least require that
\begin{align}
    Q(t)\epsilon^2\ll1.
\end{align}
This motivates us to define a \emph{dephasing} timescale $t_\textrm{deph}$ as the time for which this inequality becomes violated. Thus, we use $Q(t)\epsilon^2\approx1$ to implicitly define the timescale through
\begin{align}
    f_\textrm{deph}\left(\frac{t_\textrm{deph}}{t_\textrm{P}}\frac{\ell_\textrm{P}}{\ell_0}\right)=&\kappa_\textrm{deph}\frac{m_\textrm{P}^2}{m^2},
\end{align}
 Here, also $\kappa_\textrm{deph}$ is a constant that depends only on the specific details of the setup. In our case, $Q(\sigma\,t)\approx \ln(\sigma\,t)$ for $\sigma\,t\gg1$, and thus $ f_\textrm{deph}(x)\approx\ln x$ for the case under consideration.

 %---------------------------%
 %---------------------------%
\end{document}